\documentclass[aps,amsmath,amssymb, prc,twocolumn,superscriptaddress]{revtex4-2}

\usepackage{graphicx}
\usepackage{dcolumn}
\usepackage{bm}
\usepackage{siunitx}
\usepackage{longtable}
\usepackage{booktabs}
\usepackage{gensymb}
\DeclareSIUnit[number-unit-product = {}]{\inchQ}{\textquotedbl}
\DeclareSIUnit[number-unit-product = {\thinspace}]{\inch}{in}

\usepackage{hyperref}
\newcommand{\nuc}[2]{\hbox{$^{#1}$#2}}

\begin{document}


\title{Detailed experimental study of excited states in \nuc{50}{Ti} via the $(d,p)$ and $(d,p\gamma)$ reactions}

\author{B. Kelly}
 \email{bkelly4@fsu.edu}
\affiliation{Department of Physics, Florida State University, Tallahassee, Florida 32306, USA}

\author{M. Spieker}
 \email{Corresponding author: mspieker@fsu.edu}
\affiliation{Department of Physics, Florida State University, Tallahassee, Florida 32306, USA}

 \author{L.T. Baby}
\affiliation{Department of Physics, Florida State University, Tallahassee, Florida 32306, USA}

 \author{S. Baker}
\affiliation{Department of Physics, Florida State University, Tallahassee, Florida 32306, USA}

 \author{A.L. Conley}
\affiliation{Department of Physics, Florida State University, Tallahassee, Florida 32306, USA}

 \author{D. Houlihan}
\affiliation{Department of Physics, Florida State University, Tallahassee, Florida 32306, USA}

 \author{K.W. Kemper}
\affiliation{Department of Physics, Florida State University, Tallahassee, Florida 32306, USA}

\author{E. Litvinova}
\affiliation{Department of Physics, Western Michigan University, Kalamazoo, Michigan 49008, USA}

\author{N. Tsoneva}
\affiliation{Extreme Light Infrastructure (ELI-NP), Horia Hulubei National Institute of Physics and Nuclear Engineering (IFIN-HH), Bucharest-Magurele RO-077125, Romania}

\author{A. Volya}
\affiliation{Department of Physics, Florida State University, Tallahassee, Florida 32306, USA}

\date{\today}

\begin{abstract}
Excited states of semi-magic \nuc{50}{Ti} were studied up to the neutron-separation energy via the $(d,p)$ and $(d,p\gamma)$ reactions. In total, 82 excited states were identified based on the measurement of angular distributions with the Super-Enge Split-Pole Spectrograph (SE-SPS) at Florida State University. From the experimental data, sum rules related to vacancies were calculated for the $2p_{3/2}$, $2p_{1/2}$, $1f_{5/2}$, $1g_{9/2}$, and $2d_{5/2}$ neutron single-particle orbitals and compared to predictions obtained with the time-dependent continuum shell model (TDCSM), the quasiparticle-phonon model (QPM), and the relativistic equation of motion theory (REOM$^3$). A comparison for the $(d,p)$ data obtained for \nuc{51}{Ti} and \nuc{50}{Ti} is also presented, focusing on differences of the single-particle strength fragmentation in even-$A$ and odd-$A$ nuclei.


\end{abstract}

\maketitle


\section{\label{Introduction}Introduction}

Single-nucleon transfer reactions are a powerful tool to probe the single-particle structure of atomic nuclei. If performed systematically along isotopic, isotonic, and isobaric chains, these transfer reactions can provide information on the occupation of single-particle orbitals and the evolution of empirical shell gaps with changing nucleon number. The extracted experimental information is essential for benchmarking nuclear models, exploring the robustness of magic numbers far off stability, and for probing how single-particle configurations mix with more complex configurations. Such mixing could possibly influence the fragmentation of the associated single-particle, i.e., spectroscopic strength and contribute to the observed quenching \cite{Lap93a, Kay13a, SumRule_TransferReactions, Tos21a, Aum21a, Kay_A15_SF, Mac25a}, i.e., the reduction of single-particle strength in atomic nuclei.

The quenching of the spectroscopic strength by approximately 60\,$\%$ relative to independent-particle shell-model predictions has been established in systematic studies spanning a wide range of nuclei and using a suite of different single-nucleon transfer reactions \cite{Lap93a, Kay13a, SumRule_TransferReactions, Tos21a, Aum21a, Kay_A15_SF, Mac25a}. The latter include light-ion induced neutron-adding and neutron-removal reactions like the $(d,p)$, $(p,d)$, $(\nuc{3}{He},\alpha)$, and $(\alpha,\nuc{3}{He})$ reactions performed on nuclei in the $sd$ and $fp$ shells, as well as in heavier mass regions with typical beam energies of 2 to 8\,MeV/u \cite{Kay_A15_SF,Kay13a, SumRule_TransferReactions}. The consistency of the reduction (quenching) of the spectroscopic strength across proton- and neutron-transfer channels, for both adding and removal reactions, and for nucleons added or removed from different single-particle orbitals, i.e., for different angular momentum transfers underscores the universality of the observed quenching phenomenon \cite{Lap93a, Kay13a, SumRule_TransferReactions, Tos21a, Aum21a, Kay_A15_SF, Mac25a}. It has been argued that the reduction could be attributed to short-range correlations (SRC) between nucleons, see, {\it e.g.}, Ref.\,\cite{Mac25a} and references therein. These are typically not described correctly in nuclear models -- like configuration limited shell models or nuclear mean field models -- which assume that nucleons move in single-particle orbits in a nuclear mean field generated by all the other nucleons. However, theoretical work exists which questions that SRC alone can account for the observed quenching. Instead, the work of Barbieri showed that most of the quenching originated from a combination of configuration mixing at the Fermi surface and from coupling to collective vibrations, whose correct description requires large model spaces \cite{Bar09a}. These effects are typically referred to as long-range correlations (LRC) between nucleons. 

Both types of correlations, SRC and LRC, are defined in the model-independent many-body framework \cite{Dickhoff2005}, which departs from a picture where single nucleons move in a static mean-field potential. Instead, they are dynamic participants in the formation of a complex medium. In this framework, the leading beyond-mean-field contributions to the single-particle in-medium self-energy, characterizing nucleonic on-shell appearance, are determined by their coupling to the normal and pairing vibrations. The latter can be interpreted as a dynamical Brueckner G-matrix \cite{RingSchuck}, mostly associated with SRC, while the former is dominated by LRC originating from the coupling of the single-particle states to surface vibrations. In Bogoliubov's quasiparticle basis, both contributions are unified and can be viewed as coupling to superfluid phonons \cite{Litvinova2021a}. The pole structure of the dynamical self-energy causes fragmentation of the mean-field states in both superfluid and non-superfluid regimes.

Theoretical studies of doubly-magic (non-superfluid) nuclei showed that single-particle states close to the Fermi energy appeared to be ``good'' quasiparticle states, see, {\it e.g.}, Refs. \cite{Bortignon1981a,MahauxBortignonBrogliaEtAl1985,LitvinovaRing2006}. This was supported by the presence of a fragment with a significant spectroscopic factor, $S$. However, exceptions exist as, {\it e.g.}, discussed in the recent work of Ref.\,\cite{Vaquero2020}. Theoretically, it is expected that away from the Fermi energy the fragmentation becomes stronger. For deep hole states, one would anticipate wide distributions of the spectroscopic strengths. In open-shell nuclei, superfluidity affects the fragmentation of the strength further. In particular, a strong fragmentation is already obtained at the Fermi surface, see Refs. \cite{Soloviev,VanderSluys1993,Mishev2010,Litvinova2012,Afanasjev2015} for a theoretical discussion of spherical medium-heavy nuclei. Deformed systems display an even more pronounced fragmentation of the single-particle strength, both near and far from the Fermi surface \cite{Elbek1969, Malov1976, satchler, Zhang2022}, thereby questioning the very notion of well-defined quasiparticle states. A large fraction of the expected spectroscopic strength can be recovered though if the many small strength fragments are collected \cite{Elbek1969, Fre17a}. In general, appreciable strength fragmentation complicates the discussion of single-particle strength severely. The fragmentation of strength might significantly contribute to the observed quenching phenomenon as both SRC and LRC induced fragmentation push strength to higher excitation energies. Because of that, even for well-bound orbitals some spectroscopic strength can be pushed above particle-emission thresholds, possibly escaping experimental detection.

The discussion and comparison between theory and experiment is further complicated when data obtained with different probes and under different kinematic conditions are compared as it becomes difficult to disentangle structure and reaction effects; see, {\it e.g.}, the discussion in Refs.\,\cite{Muk10a, Dug12a, Kay13a, Dug15a, Ata18a, Aum21a, Poh23a, Jia25a}. In this context, the impact of uncertainties coming from optical model parameters and the single-particle potential used was also discussed\,\cite{Heb23a, Heb25a}. We note that the recent theoretical studies of Ref.\,\cite{Din26a} showed that traditional nuclear shell structure emerges from a realistic nuclear force based on chiral effective field theory with decreasing resolution scale, putting the statements made in, {\it e.g.}, Refs.\,\cite{Dug12a, Dug15a} regarding the non-observability of effective single-particle energies and their connection to the ``true'' ({\it ab-initio}) nuclear potential into further context. Given the intricacies of the discussion around nuclear shell structure and single-particle motion, it is, thus, helpful to provide high-quality data for different nuclei and establish systematics obtained under comparable kinematic conditions as the real experimental observables, i.e., cross sections will be consistent.

\begin{figure}[t]
    \centering
    \includegraphics[width=\linewidth]{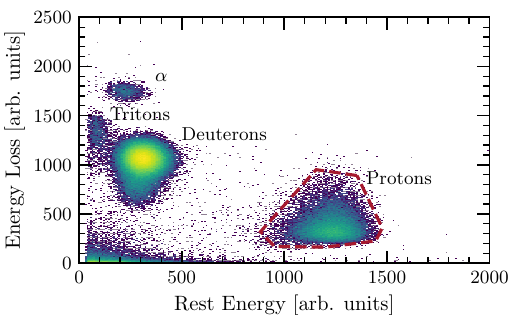}
    \caption{Particle identification plot for the SE-SPS at a magnetic field setting of 8.6 kG and $\theta_{\mathrm{SE-SPS}}$ = 25$\degree$ obtained with its focal-plane detector. Energy loss is measured by the rear anode wire of the proportional-counter section, and the rest energy is measured by the plastic scintillator placed behind that section. The proton group corresponding to the $(d,p)$ reaction is shown within the red-dashed line. Other particle groups are listed and correspond to the $(d,d')$, $(d,t)$, and $(d,\alpha)$ reactions.}
    \label{fig:PID_plot}
\end{figure}

The work presented here is a continuation of our efforts to map out the single-neutron adding strengths around the $N=28$ shell closure for the $f$ and $p$ orbitals, as well as for the $1g_{9/2}$ and $2d_{5/2}$ orbitals by systematically performing $(d,p)$ single-neutron transfer reactions at the Super-Enge Split-Pole Spectrograph (SE-SPS) of the John D. Fox Accelerator Laboratory at Florida State University under the same kinematic conditions \cite{Riley_51Ti, Riley_55Fe, Riley_53Cr, 61Nidp_Mark, Hay_52V, spi24a, Spi25a}. In this manuscript, we present new data for the $N=28$ isotone \nuc{50}{Ti} ($Z=22$) obtained from a $\nuc{49}{Ti}(d,p)\nuc{50}{Ti}$ one-neutron transfer experiment at the SE-SPS with $E_d = 16$\,MeV and a complementary $\nuc{49}{Ti}(d,p\gamma)\nuc{50}{Ti}$ particle-$\gamma$ coincidence experiment performed with the CeBr$_3$ Array (CeBrA) demonstrator \cite{CeBrA_NIM}. The latter allowed us to identify which states of \nuc{50}{Ti} were populated in the $(d,p)$ reaction, if there were any ambiguities to start with, and to identify contaminants in our $(d,p)$ spectra. 

We note that our previous $(d,p)$ studies on $fp$ shell nuclei provided a consistent picture of $40 - 60$\,$\%$ quenching relative to the sum rules for the well-bound $fp$ orbitals \cite{Riley_51Ti, Riley_55Fe, Riley_53Cr, 61Nidp_Mark, Hay_52V, spi24a, Spi25a}. For the even angular momentum, $\ell$, transfers, i.e., neutron transfers to the $1g_{9/2}$ and $2d_{5/2}$ orbitals, a more pronounced reduction of the spectroscopic strength was observed though. At first glance, the observation appeared in conflict with the systematic analysis of Kay {\it et al.}, who demonstrated that quenching effects are largely independent of the orbital the nucleon is added to or removed from\,\cite{Kay13a, Kay_A15_SF}. Reconciling this apparent discrepancy and whether it reflects an underlying structural effect in these $fp$-shell nuclei or whether it is an artifact of reaction selectivity, including momentum matching as discussed in, {\it e.g.}, Ref.\,\cite{Sch13a}, forms one of the central questions addressed in our previous \cite{Riley_51Ti, Riley_55Fe, Riley_53Cr, 61Nidp_Mark, Hay_52V, spi24a, Spi25a} and the present work.

\begin{figure*}[t]
\includegraphics[width=\linewidth]{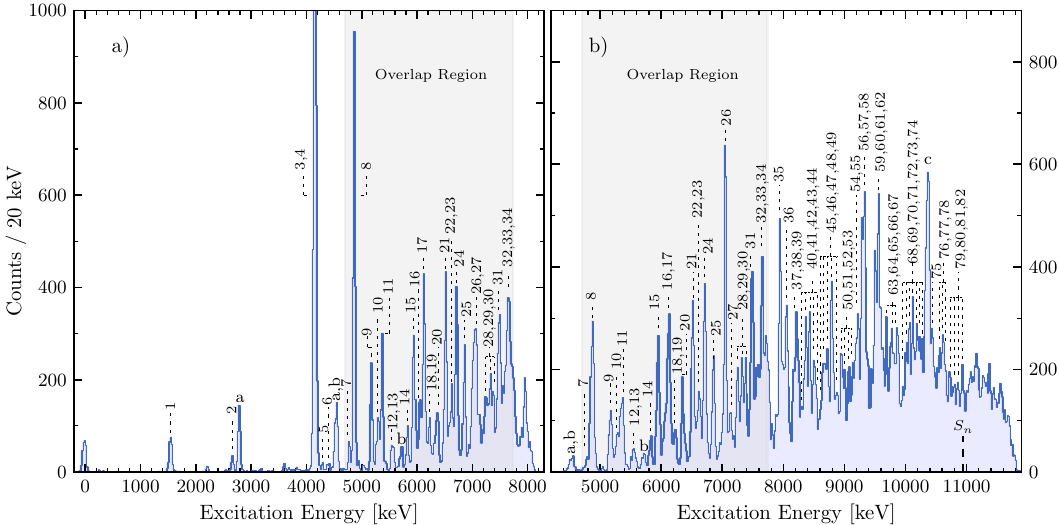}
\caption{\label{fig:EnergySpectra} $(d,p)$ spectra measured with the position-sensitive part of the SE-SPS focal-plane detector and the SE-SPS placed at a laboratory scattering angle of $\theta_{\mathrm{SE-SPS}}$ = 25$\degree$ for two different magnetic field settings, (a) 8.6\,kG and (b) 7.6\,kG. The shaded region in both panels highlights the overlap between the two magnetic field settings. Observed states from the \nuc{49}{Ti}$(d,p)$\nuc{50}{Ti} reaction are labeled with numbers. The corresponding excitation energies are listed in Table\,\ref{tab:EnergyTable}. Contaminants [a,b,c] originate from the \nuc{48}{Ti}$(d,p)$\nuc{49}{Ti}, \nuc{50}{Ti}$(d,p)$\nuc{51}{Ti}, and \nuc{12}{C}$(d,p)$\nuc{13}{C} reactions on target contaminants, respectively.}
\end{figure*}

As we will show in this manuscript, the neutron-adding (spectroscopic) strength is strongly fragmented up to the neutron-separation energy, $S_n$, in \nuc{50}{Ti} for all observed angular momentum transfers, i.e., $\ell = 1$ through 4. In addition, the spectroscopic strength is fragmented among significantly more states than in the $N=29$ isotones. In order to better understand the mechanisms that fragment spectroscopic strengths in atomic nuclei, it is instructive to study the fragmentation in both the even-$A$ and odd-$A$ isotopes. In this manuscript, the new high-quality $(d,p)$ data for \nuc{50}{Ti} allowed us to compare the fragmentation of the spectroscopic strengths for the $fp$ orbitals, and the $1g_{9/2}$ and $2d_{5/2}$ orbitals in even-even \nuc{50}{Ti} and even-odd \nuc{51}{Ti}. To better understand the fragmentation and possible quenching of spectroscopic strengths in \nuc{50}{Ti}, we compared our new data to predictions coming from the time-dependent continuum shell model (TDCSM) \cite{Vol09a, Vol14a} with the FSU cross-shell interaction \cite{Lub19a, Lub20a}, from the energy-density functional plus quasiparticle-phonon model approach (EQPM) \cite{Tsoneva2016, Soloviev}, and from the relativistic equation of motion theory including the coupling of two quasiparticles with up to two phonons (REOM$^3$) \cite{Lit19a, Lit22a, Lit23a}.

\section{Experimental Details}

Both the \nuc{49}{Ti}$(d,p)$\nuc{50}{Ti} and \nuc{49}{Ti}$(d,p\gamma)$\nuc{50}{Ti} experiments were conducted at the John D. Fox Superconducting Linear Accelerator Laboratory at Florida State University\,\cite{spi24a}. A 16-MeV deuteron beam was accelerated by the 9-MV Super-FN Tandem Van-de-Graaff accelerator and impinged on a self-supporting metal foil highly enriched in \nuc{49}{Ti} and with a nominal areal density of 413 $\mu$g/cm$^2$, which had to be corrected as will be described in Sec.\,\ref{sec:Contaminant_States}. Protons were detected by the position-sensitive focal-plane detector of the Super-Enge Split-Pole Spectrograph (SE-SPS)\,\cite{spi24a}. The SE-SPS has two dipole magnets that bend reaction products towards the magnetic focal plane. The angular acceptance of the SE-SPS was set to $\Delta\Omega$ = 4.6 msr for the experiments discussed in this work.  The focal-plane detector is comprised of a position-sensitive proportional counter and a large plastic scintillator placed behind it \cite{spi24a}. The latter is used to measure the rest energy of particles after they pass through the proportional counter, where particles continuously lose energy in the isobutane gas. For this experiment, the proportional counter was filled with isobutane gas at a pressure of around 160 Torr. The energy loss in the proportional counter, along with the rest energy measured by the plastic scintillator, allows the identification of different reaction products, thus aiding the selection of different reactions in the offline analysis. An example of a particle identification (PID) plot is shown in Fig.\,\ref{fig:PID_plot}. Each particle group corresponds to another deuteron-induced reaction on the target highly enriched in \nuc{49}{Ti}.

The position in the focal plane is reconstructed using the two position-sensitive sections of the focal-plane detector\,\cite{spi24a}, each equipped with pick-up pads and associated delay lines. Particles are dispersed across the length of the focal plane based on their magnetic rigidity, B$\rho$. Two magnetic field settings of 8.6\,kG and 7.6\,kG were chosen to ensure overlap between excitation energy ranges and detect excited states in \nuc{50}{Ti} up to the neutron-separation energy, $S_n = 10939.19(4)$ keV \cite{Mass50DataSheet}. Examples of $(d,p)$ spectra collected for \nuc{50}{Ti} are shown in Fig.\,\ref{fig:EnergySpectra}. The focal-plane position was converted to excitation energies in \nuc{50}{Ti} using well-known excited states and information from a previous $(d,p)$ experiment\,\cite{Barnes1}. Angular distributions were measured by positioning the SE-SPS at ten different laboratory scattering angles between $10^{\circ}$ to $60^{\circ}$. Ultimately, these allow for the determination of the angular momentum transfer, $\ell$, and, thus, provide information for the spin-parity assignment, $J^{\pi}$, to excited states. In order to calculate absolute differential cross sections, the number of incoming deuterons was determined through continuous beam current integration using a Faraday cup positioned downstream of the reaction target. All cross sections determined in this work include a systematic uncertainty of 15\%, accounting for contributions from charge integration, target thickness, and solid-angle acceptance. 

For the $\nuc{49}{Ti}(d,p\gamma)\nuc{50}{Ti}$ coincidence experiment, the SE-SPS was used in conjunction with the CeBrA demonstrator, an array of five CeBr$_3$ scintillators recently commissioned at the John D. Fox Accelerator Laboratory \cite{CeBrA_NIM}. The data were collected as part of the commissioning runs with the magnetic field set to 8.6 kG and the SE-SPS positioned at a laboratory scattering angle of $37^{\circ}$. More details about the setup and its capabilities can be found in Refs. \cite{CeBrA_NIM, spi24a, Spi25a}. In this work, the particle-$\gamma$ coincidence data were used to identify the states of \nuc{50}{Ti} populated in the $(d,p)$ reaction, to identify contaminants, and to determine the isotopic composition of the target.  

\section{Experimental Results}

\begin{figure}[t]
\includegraphics[width=\linewidth]{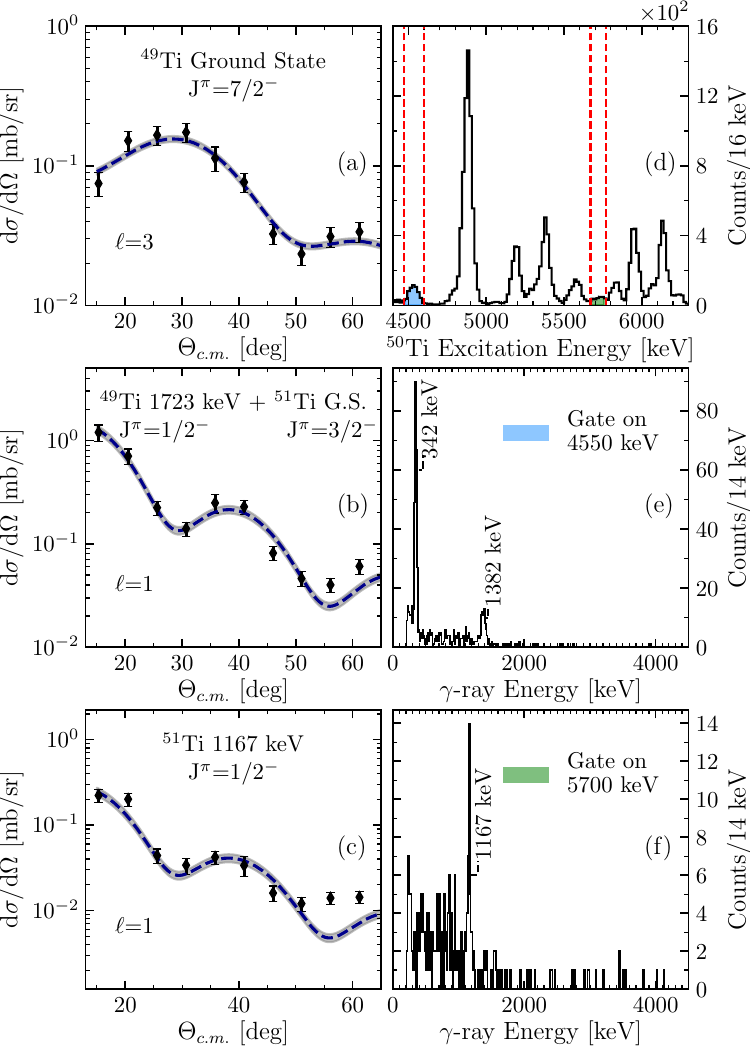}
\caption{(a), (b), and (c) $(d,p)$ angular distributions of states identified to belong to other Ti isotopes. (d) SE-SPS focal-plane spectrum if $\gamma$ rays were detected in coincidence with CeBrA. (e) $\gamma$-ray spectrum of the 1723-keV excited state in \nuc{49}{Ti}, obtained when a gate is applied around 4.5\,MeV of excitation energy in \nuc{50}{Ti} [highlighted in blue in panel (d)]. (f) $\gamma$-ray spectrum of the 1167-keV state of \nuc{51}{Ti}, obtained when a gate is set on the excitation energy range highlighted in green in panel (d). See text for more information.}
\label{fig:contaminant_plot}
\end{figure}

\subsection{\label{sec:Contaminant_States}Determination of target composition and identification of contaminants}

The detection of $\gamma$ rays with CeBrA in coincidence with light ions detected in the focal plane of the SE-SPS allows us to identify excited states populated in a specific reaction, to identify contaminants, and to effectively suppress contaminants. See Refs.\,\cite{CeBrA_NIM, spi24a, Spi25a} for examples. In this section, we will highlight how we used the coincidence data to identify contaminants in the focal-plane spectrum coming from $(d,p)$ reactions on other Ti isotopes present in the target and how we ultimately used that information to determine the target composition. The states used for this analysis are shown in Fig.\,\ref{fig:contaminant_plot}.

A previous study of the \nuc{49}{Ti}$(d,p)$\nuc{50}{Ti} reaction can be found in Ref.\,\cite{Barnes1} and a re-analysis of the data was reported in Ref. \cite{Barnes2}. In these previous $(d,p)$ experiments, two states at 4536 and 4576 keV were identified as excited states of \nuc{50}{Ti}. The states are currently adopted \cite{Mass50DataSheet}. No unique $\ell$ transfer could be determined in Refs.\,\cite{Barnes1, Barnes2} though. As briefly mentioned in Ref.\,\cite{CeBrA_NIM}, our $p\gamma$-coincidence data established that these states are not excited states of \nuc{50}{Ti}. Instead, the peak in the proton spectrum results partly from the 1723-keV, $J^{\pi} = 1/2^-_1$ excited state of \nuc{49}{Ti} populated in the $(d,p)$ reaction on a \nuc{48}{Ti} target contaminant. Fig.\,\ref{fig:contaminant_plot}\,(d) shows protons detected in coincidence with $\gamma$ rays. The 4.5-MeV structure is clearly visible. Due to the excellent particle energy resolution of the SE-SPS, a narrow excitation-energy gate can be applied around that peak, enabling the study of $\gamma$ rays emitted in coincidence. By applying this gate -- indicated by the red dashed line in Fig.\,\ref{fig:contaminant_plot}\,(d) around the 4.5-MeV peak highlighted in blue -- the $\gamma$-ray spectrum shown in Fig.\,\ref{fig:contaminant_plot}\,(e) is obtained. No known $\gamma$ rays of \nuc{50}{Ti} are observed. Instead, we observe the 342-keV, $1/2^-_1 \rightarrow 3/2^-_1$ transition from the 1723-keV state of \nuc{49}{Ti} and the 1382-keV, $3/2^-_1 \rightarrow 7/2^-_1$ transition following it \cite{Mass50DataSheet}. The measured $(d,p)$ angular distribution in Fig.\,\ref{fig:contaminant_plot}\,(b) follows the one expected for populating a $J^{\pi} = 1/2^-$ state, i.e., it shows the features of an $\ell = 1$ angular momentum transfer. We note that the $\nuc{51}{Ti}$, $J^{\pi} = 3/2^-$ ground state is also part of the peak structure at 4.5\,MeV of excitation energy in \nuc{50}{Ti}. That is why two labels are shown at this energy in Fig.\,\ref{fig:EnergySpectra}. It was taken into account in the fit shown in Fig.\,\ref{fig:contaminant_plot}\,(b). We propose to reevaluate the assignment of the two states at 4.5\,MeV to \nuc{50}{Ti}, and to identify them as states of \nuc{49}{Ti} and \nuc{51}{Ti} instead.

We found evidence for another probably incorrect assignment. The previous $(d,p)$ experiments \cite{Barnes1, Barnes2} reported the observation of a state at an excitation energy of 5717 keV and assigned it to \nuc{50}{Ti}. Following the same strategy as described above, the $\gamma$-ray spectrum shown in Fig.\,\ref{fig:contaminant_plot}\,(f) was obtained. The proton-gated spectrum returns a single $\gamma$ ray with an energy of 1167\,keV. When considering the different $Q$ values for $(d,p)$ reactions on the Ti isotopes, it is clear that this peak in the proton spectrum corresponds to the first excited state of \nuc{51}{Ti} at 1167\,keV with a spin-parity assignment of $J^{\pi} = 1/2^-$. The observed $(d,p)$ angular distribution in Fig.\,\ref{fig:contaminant_plot}\,(c) is consistent with the population of the 1167-keV state of \nuc{51}{Ti}, i.e., it follows an $\ell$ = 1 angular momentum transfer. We, thus, propose that also the assignment of the adopted 5717-keV state to \nuc{50}{Ti} should be revisited.

Having established \nuc{48}{Ti} and \nuc{50}{Ti} contaminants in our target, we looked for the \nuc{49}{Ti}, $J^{\pi} = 7/2^-$ ground state in our proton spectrum. As expected from $Q$-value considerations, it was observed right above the 2$^{\mathrm{nd}}$ excited state of \nuc{50}{Ti} ($E_x = 2675$\,keV) in our proton spectrum, see Fig.\,\ref{fig:EnergySpectra}. The angular distribution clearly follows the expected $\ell = 3$ angular momentum transfer, see Fig.\,\ref{fig:contaminant_plot}\,(a). As expected, no $\gamma$ rays were detected in coincidence with this state.

We used the information obtained from this analysis and combined it with known $\gamma$-decay branching ratios and spectroscopic factors from previous $(d,p)$ experiments \cite{Barnes2, Ball_49Ti_dp_study} to determine the isotopic composition of the target; 83(3) \% \nuc{49}{Ti}, 13(2)\% \nuc{48}{Ti}, and 4.1(7)\% \nuc{50}{Ti}. The effective areal density of \nuc{49}{Ti} nuclei in the target was, thus, reduced to 344(12)\,$\mu$g/cm$^2$ compared to the reported thickness of 413\,$\mu$g/cm$^2$. Differential cross sections for states of \nuc{50}{Ti} were calculated using the corrected target thickness. Note that the corrected target thickness was also used in Ref.\,\cite{Kel26a}.

\onecolumngrid
\LTcapwidth=\textwidth
\renewcommand{\arraystretch}{1.2}
\begin{longtable}[T]{lcccccccccc}
\caption{Experimental data for states populated via the $^{49}$Ti$(d,p)$$^{50}$Ti reaction. Reported excitation energies (level energies), $J^{\pi}$ assignments, angular momentum, $\ell$, transfers, transfer configurations, and spin-weighted spectroscopic factors, $S'$ [see Eq.\,(\ref{eq:specfac})], are given. The values obtained for $S'$ are compared to those listed in Ref.\,\cite{Mass50DataSheet}. The angle-integrated cross sections, $\sigma_{total}$ [see Eq.\,(\ref{eq:angintcs}) and text], are also listed. The uncertainties for $\sigma_{\mathrm{total}}$ include a 15\% systematic uncertainty arising from beam-current integration, while the uncertainties in the spin-weighted spectroscopic factors, $S'$, reflect the 1$\sigma$ errors obtained from the $\chi^2$ minimization procedure used to fit the experimental $(d,p)$ angular distributions. Spin ranges for states are given for the transfer configuration listed. Only the states observed in this experiment are compared with those reported in Ref.\,\cite{Barnes2}.} \label{tab:EnergyTable}\\ 
\midrule
\midrule

    &
  \multicolumn{2}{c}{Level Energy (keV)} &
  \multicolumn{2}{c}{$J^{\pi}_f$} &
  \multicolumn{2}{c}{$\ell$ transfer} &
  Transfer &
  \multicolumn{2}{c}{$S^{'}$} &
  $\sigma_{total}$ \\ \cmidrule(lr){2-3} \cmidrule(lr){4-5} \cmidrule(lr){6-7} \cmidrule(lr){9-10}
  $\#$ & This Work & Ref. \cite{Mass50DataSheet} & This Work & Ref. \cite{Mass50DataSheet} & This work & Ref. \cite{Mass50DataSheet} & Configuration  & This work  & Ref. \cite{Mass50DataSheet} & (mb)      \\ 
\midrule
\endfirsthead
\multicolumn{11}{c}%
{{\tablename\ \thetable{}: Continued from previous page}}\\ 
\midrule
\midrule

    &
  \multicolumn{2}{c}{Level Energy (keV)} &
  \multicolumn{2}{c}{$J^{\pi}_f$} &
  \multicolumn{2}{c}{$\ell$ transfer} &
  Transfer &
  \multicolumn{2}{c}{$S^{'}$} &
  $\sigma_{total}$ \\ \cmidrule(lr){2-3} \cmidrule(lr){4-5} \cmidrule(lr){6-7} \cmidrule(lr){9-10}
  $\#$ & This Work & Ref. \cite{Mass50DataSheet} & This Work & Ref. \cite{Mass50DataSheet} & This work & Ref. \cite{Mass50DataSheet} & Configuration  & This work  & Ref. \cite{Mass50DataSheet} & (mb)      \\ 
\midrule
\endhead

\hline
\endfoot
\hline
\hline

\multicolumn{10}{l}{${}^{a}$ Contaminant from $\nuc{48}{Ti}(d,p)\nuc{49}{Ti}$, see section \ref{sec:Contaminant_States} for further discussion.} \\
\multicolumn{10}{l}{${}^{b}$ Contaminant from $\nuc{50}{Ti}(d,p)\nuc{51}{Ti}$, see section \ref{sec:Contaminant_States} for further discussion.} \\
\multicolumn{10}{l}{${}^{c}$ Not observed in $(\gamma,\gamma')$ experiment of Ref.\,\cite{Kel26a}, which does not support the
   adopted $J^{\pi} = 1^+$ assignment.} \\
\multicolumn{10}{l}{${}^{d}$ Assigned $J^{\pi} = 1^+$ based on complementary data. See discussion in Ref.\,\cite{Kel26a}.} \\
\multicolumn{10}{l}{$^e$ Ref.\,\cite{50Ti_ee'} reported evidence for possible doublet of states with $J^{\pi} = 1^+$ and $3^-$.} \\
\multicolumn{10}{l}{$^f$ Data of Ref.\cite{50Ti_ee'} preferred $J^{\pi} = 1^-$ assignment.} \\
\multicolumn{10}{l}{$^g$ State was not reported as $J^{\pi} = 1^+$ state in Refs.\,\cite{50Ti_pp'_201MeV, 50Ti_pp', 50Ti_ee'}.} \\

\endlastfoot

1  & 1555(6)         & 1553.794(8)    & $2^+$         & $2^+$           & 1+3  & 1+3     & $2p_{3/2}$    & 0.038(8)      & 0.08       & 0.286(13)  \\
   &                 &                &               &                 &      &         & +$1f_{7/2}$   & +0.120(23)    & 0.32       &            \\
2  & 2673(10)        & 2674.932(10)   & $4^+$         & $4^+$           & 1    & 1       & $2p_{3/2}$    & 0.0189(27)    & 0.04       & 0.098(5)   \\
3  & 4152(7)         & 4147.210(13)   & $4^+$         & $4^+$           & 1    & 1       & $2p_{3/2}$    & 0.43(4)       & 0.74       & 1.98(10)   \\
4  & 4178(8)         & 4172.003(19)      & $3^+$         & $3^+$           & 1    & 1       & $2p_{3/2}$    & 0.58(4)       & 0.93       & 2.49(13)   \\
   &                 & 4172.8(3)         &  & $(2^+)$  &     &       &    &        &  & \\
5  & 4309(13)        & 4309.86(11)    & $2^+$         & $2^+$           & 1    & 1       & $2p_{3/2}$    & 0.0128(15)    & 0.02       & 0.059(3)   \\
6  & 4403(12)        & 4410.02(3)     & $3^-$         & $3^-$           & 2+4  & 0+2     & $2d_{5/2}$    & 0.0035(11)    & 0.003      & 0.0533(29) \\
   &                 &                &               &                 &      &         & +$1g_{9/2}$   & +0.018(7)     & 0.009      &            \\
   &                 & 4536(20)$^{a}$ &               &                 &      &         &               &               &            &            \\
   &                 & 4576(20)${^a}$ &               &                 &      &         &               &               &            &            \\
7  & 4790(6)         & 4789.97(6)     & $2^+$         & $2^+$           & 1    & 1       & $2p_{3/2}$    & 0.027(3)      & 0.05       & 0.141(8)   \\
8  & 4879(6)         & 4880.705(15)   & $5^+$         & $5^+$           & 1    & 1       & $2p_{3/2}$    & 0.45(3)       & 0.81       & 1.69(9)    \\
9  & 5189(3)         & 5186.103(18)   & ($3,4$)$^+$   & ($3,4$)$^+$     & 1    & 1       & $2p_{3/2}$    & 0.108(14)     & 0.2        & 0.451(23)  \\
10 & 5313(11) & 5334(5)   & ($4-6$)$^-$ & ($4-6$)$^-$ & 2 & & $2d_{5/2}$ & 0.0158(21) &            & 0.173(16) \\
11 & 5376(9)         & 5379.942(19)   & $4^+$         & $4^+$           & 1    & 1       & $2p_{3/2}$    & 0.122(14)     & 0.23       & 0.63(3)    \\
12 & 5548(17)        & 5547.81(4)        &    $2^+-5^+$           & $(4^+)$     & 1    &        & $2p_{3/2}$    & 0.0135(14)    &        & 0.083(5)   \\
   &                 & 5561(6)         &  & ($2-5$)$^+$  &     &   1    &    &       & 0.03 & \\
13 & 5588(17) & 5600(6)   & $1^+-6^+$   & ($2-5$)$^+$ & 3 & 1+3     & $1f_{5/2}$ & 0.0294(21) & 0.01+0.08  & 0.073(5)  \\
   &                 & 5717(6)$^{b}$  &               &                 &      &  &               &               &            &            \\
14 & 5835(9)  & 5837.2(6) & $(3-5)^-$   & ($2^+-5^+$) & 2 & 1+3     & $2d_{5/2}$ & 0.0176(14) & 0.004+0.08 & 0.144(7)  \\
15 & 5955(4)         & 5946.479(22)   & $3^+,4^+$     & $3^+,4^+$       & 1    & 1       & $2p_{3/2}$    & 0.176(14)     & 0.29       & 0.86(4)    \\
16 & 6066(7)         & 6072(15)       & $(2)^+$         & $(2)^+$         & 1    & 1       & $2p_{3/2}$    & 0.068(5)      & 0.02       & 0.357(17)  \\
17 & 6133(11)        & 6123.15(4)  & $(4)^+$     & ($4^+$)         & 1+3  & 1+3     & $2p_{3/2}$    & 0.047(12)     & 0.15       & 0.79(4)    \\
   &                 &                &               &                 &      &         & +$1f_{5/2}$   & +0.31(4)      & 0.5        &            \\
18 & 6209(20)        & 6212(5)        & $2^+-5^+$     & ($1^--6^-$)     & 1    & (2)     & $2p_{3/2}$    & 0.0179(21)    & 0.04       & 0.088(6)   \\
19 & 6249(13)        & 6249(6)        & $1^+-6^+$     & ($0-7$)$^+$     & 3    &         & $1f_{5/2}$    & 0.068(5)      &            & 0.120(7)   \\
20 & 6387(9)         & 6392(6)        & $2^+-5^+$     & ($2-5$)$^+$     & 1    & 1       & $2p_{3/2}$    & 0.0294(21)    & 0.07       & 0.158(8)   \\
21 & 6524(7)         & 6521.41(4)     & $3^+,4^+$     & $3^+,4^+$       & 1+3  & 1+3     & $2p_{3/2}$    & 0.0263(21)    & 0.09       & 0.73(3)    \\
   &                 &                &               &                 &      &         & +$1f_{5/2}$   & +0.28(5)      & 0.47       &            \\
22 & 6590(23)        & 6583(10)       & $1^--6^-$     & ($1-6$)$^-$     & 2    &         & $2d_{5/2}$    & 0.0126(9)     &            & 0.124(11)  \\
23 & 6634(13)        & 6636(6)        & $1^+-6^+$     & ($0-7$)$^+$     & 3    & 3       & $1f_{5/2}$    & 0.087(8)      & 0.26       & 0.212(11)  \\
24 & 6721(6)         & 6710.570(24)   & $4^+$         & $4^+$           & 1    & 1       & $2p_{3/2}$    & 0.189(14)     & 0.25       & 0.95(4)    \\
25 & 6864(4)         & 6864(5)        & ($5$)$^+$     & ($5$)$^+$       & 1+3  & 1+3     & $2p_{3/2}$    & 0.043(7)      & 0.09       & 0.505(23)  \\
   &                 &                &               &                 &      &         & +$1f_{5/2}$   & +0.132(17)    & 0.41       &            \\
26 & 7029(7)         & 7029.39(25)    & $2^+-4^+$     & $2^+-4^+$       & 1    & 1       & $2p_{3/2}$    & 0.056(5)      & 0.05       & 0.325(26)  \\
27 & 7132(18)  & 7132(20)       & $3^-$         & $3^-$           & 2    & 2       & $2d_{5/2}$    & 0.0176(14)    & 0.05       & 0.166(10)  \\
28 & 7249(4)         & 7249(6)        & $2^+-5^+$     & ($2-5$)$^+$     & 1    & 1       & $2p_{3/2}$    & 0.066(5)      & 0.09       & 0.445(22)  \\
29 & 7325(8)         & 7335(10)       & $1^--6^-$     & $(2)^+$         & 2    &         & $2d_{5/2}$    & 0.135(15)     &            & 0.321(16)  \\
   &                 &                & or  $1^+-6^+$ & $(2)^+$         & or 3 &         & or $1f_{5/2}$ & or 0.0368(23) &            &            \\
30 & 7394(8)         & 7382(9)        & $(3)^-$         & $(3)^-$           & 2    & 2+4     & $2d_{5/2}$    & 0.034(4)      & 0.005+0.56 & 0.301(14)  \\
31 & 7493(3)   &      & $1^--6^-$         &         & 2    &        & $2d_{5/2}$    & 0.105(11)     &        & 0.97(4)    \\
32 & 7587(7)         & 7577(10)       & $1^+-6^+$     & $0^+,1^+$$^c$                & 3    &         & $1f_{5/2}$    & 0.198(14)     &            & 0.461(24)  \\
33 & 7662(4)         & 7667(15)       & $1^--6^-$     & $(2)^+$         & 2+4  & (3)     & $2d_{5/2}$    & 0.068(15)     & 0.8        & 0.88(4)    \\
   &                 &                &               &                 &      &         & +$1g_{9/2}$   & +0.23(8)      &            &            \\
   &                 &                & or  $1^+-6^+$ &                 & or 3 &         & or $1f_{5/2}$ & or 0.44(3)    &            &            \\
34 & 7739(11)  & 7734(15)       & $1^--6^-$         &                 & 2    &         & $2d_{5/2}$    & 0.061(4)      &            & 0.601(26)  \\
35 & 7951(4)         & 7941(15)       & $2^+-5^+$     &                 & 1    &         & $2p_{3/2}$    & 0.143(15)     &            & 0.92(4)    \\
36 & 8046(4)         & 8034(10)       & $(4)^+$     & ($4$)$^+$       & 3    &         & $1f_{5/2}$    & 0.338(27)     &            & 0.88(4)    \\
37 & 8178(10)        &                & $1^--8^-$     &                 & 4    &         & $1g_{9/2}$    & 0.179(11)     &            & 0.390(17)  \\
38 & 8226(10)  & 8241(10)       & $1^+ - 6^+$     & $0^+,1^+$$^c$       & 3    &         & $1f_{5/2}$    & 0.185(14)     &            & 0.493(24)  \\
39 & 8280(14)  & 8290(10)       & $(3)^-$         & $(3)^-$         & 2    &         & $2d_{5/2}$    & 0.0216(18)    &            & 0.209(11)  \\
40 & 8321(16)        &                & $1^--8^-$     &                 & 4    &         & $1g_{9/2}$    & 0.063(8)      &            & 0.163(7)   \\
41 & 8373(6)         &                & $1^--8^-$     &                 & 4    &         & $1g_{9/2}$    & 0.210(21)     &            & 0.477(19)  \\
42 & 8438(8)   &  8444(10)              & $1^- - 6^-$         &   $0^+,1^+$              & 2    &         & $2d_{5/2}$    & 0.0410(23)    &            & 0.401(18)  \\
43 & 8503(10)        &                & $1^--6^-$     &                 & 2    &         & $2d_{5/2}$    & 0.0221(18)    &            & 0.226(12)  \\
44 & 8551(21)$^{d}$  & 8560(20)       & $1^+$         & $1^+$           & 3    &         & $1f_{5/2}$    & 0.0231(21)    &            & 0.065(6)   \\
45 & 8611(12)        &    8606(10)            & $2^+-5^+$     &     $(1)^+$            & 1    &         & $2p_{3/2}$    & 0.0173(23)    &            & 0.103(7)   \\
46 & 8661(22)        & 8640(20)       & $2^-$     & $2^-$           & 2    &         & $2d_{5/2}$    & 0.0330(15)    &            & 0.313(17)  \\
47 & 8726(8)   & 8725(10)       & $(2)^-$         & $(2^-)$         & 2+4  &         & $2d_{5/2}$    & 0.016(6)      &            & 0.406(18)  \\
   &                 &                &               &                 &      &         & +$1g_{9/2}$   & +0.10(3)      &            &            \\
48 & 8799(5)$^{d}$   & 8810(20)       & $1^+$         & $1^+$           & 3+4  &         & $1f_{5/2}$    & 0.113(27)     &            & 0.653(28)  \\
   &                 & 8815(10)       & $(3)^-$       & $(3)^-$         &      &         & +$1g_{9/2}$   & +0.13(3)      &            &            \\
49 & 8876(14)        & 8870(20)       & $(2)^+$     & $(2^+)$         & 3    &         & $1f_{5/2}$    & 0.038(4)      &            & 0.126(6)   \\
50 & 8952(7)         & 8973(10)       & $1^--6^-$     & $(3)^-$         & 2    &         & $2d_{5/2}$    & 0.0248(14)    &            & 0.277(13)  \\
51 & 9005(7)$^{d}$   & 9030(20)       & $1^+$       & $(1)^+$           & 3    &         & $1f_{5/2}$    & 0.061(6)      &            & 0.197(10)  \\
52 & 9077(9)         &     9050(20)           & $(2)^-$     &    $2^-$             & 2    &         & $2d_{5/2}$    & 0.0165(8)     &            & 0.166(10)  \\
53 & 9145(10)        &    9127(10)            & $1^+-6^+$     &                 & 3    &         & $1f_{5/2}$    & 0.049(3)      &            & 0.166(8)   \\
54 & 9191(12)  &    9188(15)            & $1^- - 6^-$         &                 & 2    &         & $2d_{5/2}$    & 0.0188(15)    &            & 0.194(10)  \\
55 & 9229(7)$^d$         & 9210(20)       & $1^+$         & $1^+$           & 3    &         & $1f_{5/2}$    & 0.098(8)      &            & 0.335(16)  \\
56 & 9288(5)         & 9282(10)       & $(6)^+$     & $(5^-,6^+)$     & 3    &         & $1f_{5/2}$    & 0.135(15)     &            & 0.507(23)  \\
57 & 9333(4)         & 9339(10)       & $1^+-6^+$     & $(3)^-$         & 3    &         & $1f_{5/2}$    & 0.203(15)     &            & 0.71(3)    \\
58 & 9380(6)         & 9391(10)       & $2^+-5^+$     & $(4)^+$         & 1+3  &         & $2p_{3/2}$    & 0.023(8)      &            & 0.259(12)  \\
   &                 &                &               &                 &      &         & +$1f_{5/2}$   & +0.038(15)    &            &            \\
59 & 9452(9)         & 9442(10)               & $1^+-6^+$     &                 & 3    &         & $1f_{5/2}$    & 0.037(3)      &            & 0.094(6)   \\
60 & 9510(4)         & 9508(10)       & $(6)^+$     & $(5^-,6^+)$     & 3    &         & $1f_{5/2}$    & 0.158(11)     &            & 0.576(24)  \\
61 & 9562(2)   & 9550(10)               & $1^+ - 6^+$       &                 & 3    &         & $1f_{5/2}$    & 0.151(8)      &            & 0.587(26)  \\
62 & 9615(9)         &  9614(10)               & $1^+-6^+$     & $(1)^+$$^c$                & 3    &         & $1f_{5/2}$    & 0.069(4)      &            & 0.262(12)  \\
63 & 9691(9)         &                & $1^--6^-$     &                 & 2    &         & $2d_{5/2}$    & 0.0189(11)    &            & 0.216(12)  \\
64 & 9746(14)        & 9752(10)       & $(3)^-$     & $(3)^-$         & 2    &         & $2d_{5/2}$    & 0.0147(21)    &            & 0.156(11)  \\
65 & 9785(18)  &  9790(20)              & $1^- - 6^-$         &    $0^+,1^+$             & 2    &         & $2d_{5/2}$    & 0.0242(11)    &            & 0.251(14)  \\
66 & 9828(18)        &  9842(10)              & $(4)^-$     &     $(4^-,5^+)$            & 2    &         & $2d_{5/2}$    & 0.0126(11)    &            & 0.138(10)  \\
   &                 &                & or $1^+-6^+$  &                 & or 3 &         & or $1f_{5/2}$ & or 0.034(4)   &            &            \\
67 & 9876(9)         &                & $1^+-6^+$     &                 & 3    &         & $1f_{5/2}$    & 0.069(3)      &            & 0.287(14)  \\
68 & 9959(16)$^{d}$  & 9957(14)       & $1^+$         & $1^+$           & 3    &         & $1f_{5/2}$    & 0.017(4)      &            & 0.054(6)   \\
69 & 10014(20) &    10000(20)            & $(2)^-$         &     $(2^-,1^+)$            & 2    &         & $2d_{5/2}$    & 0.0210(21)    &            & 0.209(13)  \\
70 & 10067(11)       &   10045(10)             & $1^--6^-$     &    $(1)^+$$^e$             & 2    &         & $2d_{5/2}$    & 0.033(3)      &            & 0.347(18)  \\
71 & 10123(8)        & 10140(20)      & $(2)^-$     & $(1^+,2^-)$     & 2    &         & $2d_{5/2}$    & 0.0315(21)    &            & 0.346(17)  \\
72 & 10193(12)       &    10206(10)            & $1^--6^-$     &  $(1)^+$$^f$               & 2    &         & $2d_{5/2}$    & 0.0200(21)    &            & 0.184(10)  \\
73 & 10240(17)       & 10240(20)      & $(2,3)^-$         & $(1^+,2^-,3^-)$ & 2    &         & $2d_{5/2}$    & 0.023(3)      &            & 0.140(8)   \\
74 & 10286(19)       &                & $1^+-6^+$     &                 & 3    &         & $1f_{5/2}$    & 0.042(5)      &            & 0.186(12)  \\
75 & 10440(11)       &                & $1^+-6^+$     &                 & 3    &         & $1f_{5/2}$    & 0.070(6)      &            & 0.324(18)  \\
76 & 10556(15)$^{d}$ & 10540(20)      & $1^+$         & $(1^+,2^-)$           & 3    &         & $1f_{5/2}$    & 0.054(5)      &            & 0.271(14)  \\
77 & 10616(16)$^{d}$       & 10580(20)      & $1^+$     & $1^+$                 & 3    &         & $1f_{5/2}$    & 0.066(5)      &            & 0.318(19)  \\
78 & 10663(24)$^{d}$ & 10660(20)      & $1^+$         & $1^+$           & 3    &         & $1f_{5/2}$    & 0.026(3)      &            & 0.118(12)  \\
79 & 10723(15)       &                & $1^--6^-$     &                 & 2    &         & $2d_{5/2}$    & 0.0368(21)    &            & 0.174(10)  \\
   &                 &                & or $1^+-6^+$  &                 & or 3 &         & or $1f_{5/2}$ & or 0.0168(11) &            &            \\
80 & 10805(20)       &      10800(20)           & $1^- - 6^-$         &     $1^+$$^{g}$           & 2    &         & $2d_{5/2}$    & 0.0126(21)    &            & 0.125(8)   \\
81 & 10867(12)       &    10870(20)            & $(1,2)^-$         &  $(1,2^-)$               & 2    &         & $2d_{5/2}$    & 0.0168(11)    &            & 0.172(10)  \\
82 & 10939(13)       &                & $1^- - 6^-$         &                & 2    &         & $2d_{5/2}$    & 0.0179(21)    &            & 0.154(8)  
\end{longtable}
\twocolumngrid

\subsection{Results from $\nuc{49}{Ti}(d,p)\nuc{50}{Ti}$ and $\nuc{49}{Ti}(d,p\gamma)\nuc{50}{Ti}$}

\begin{figure}[t]
    \centering
    \includegraphics[width=\linewidth]{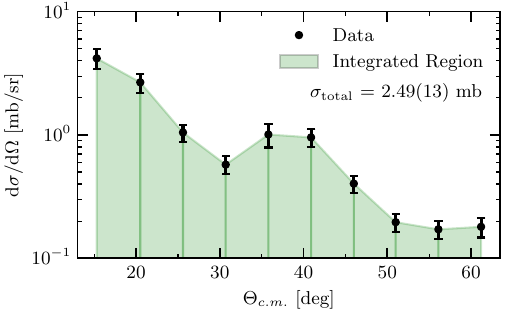}
    \caption{Calculation of angle-integrated cross section for the 4172-keV state of \nuc{50}{Ti}. The area under the experimentally measured angular distribution (shaded green) is integrated using a trapezoidal sum, with each point weighted by its corresponding solid-angle element (shown in discrete steps with green vertical lines), to obtain the model-independent angle-integrated cross section, $\sigma_{\mathrm{total}}$ of 2.49(13)\,mb.}
    \label{fig:Integrated_cross_section}
\end{figure}

A total of 82 states were identified in \nuc{49}{Ti}$(d,p)$\nuc{50}{Ti} up to $S_n = 10939.19(4)$\,keV \cite{Mass50DataSheet}. This number excludes the ground state as it was by choice placed at the edge of the focal-plane detector acceptance where cross sections cannot be reliably determined. The observed excited states of \nuc{50}{Ti} are highlighted in the example excitation spectra in Fig.\,\ref{fig:EnergySpectra}. Information on the observed states is listed in Table\,\ref{tab:EnergyTable} with the numbers used to label them in Fig.\,\ref{fig:EnergySpectra}. Angle-integrated cross sections were calculated for the finite set of measured angles using the trapezoidal rule for integration:

\begin{equation}
\sigma_{\mathrm{total}} = \int \left( \frac{d\sigma}{d\Omega} \right) d\Omega = 2\pi \sum_{i} \left( \frac{d\sigma}{d\Omega} \right)_i\,\sin\theta_i\,\Delta\theta_i
\label{eq:angintcs}    
\end{equation}
 
\begin{figure*}[t]
\includegraphics[width=\linewidth]{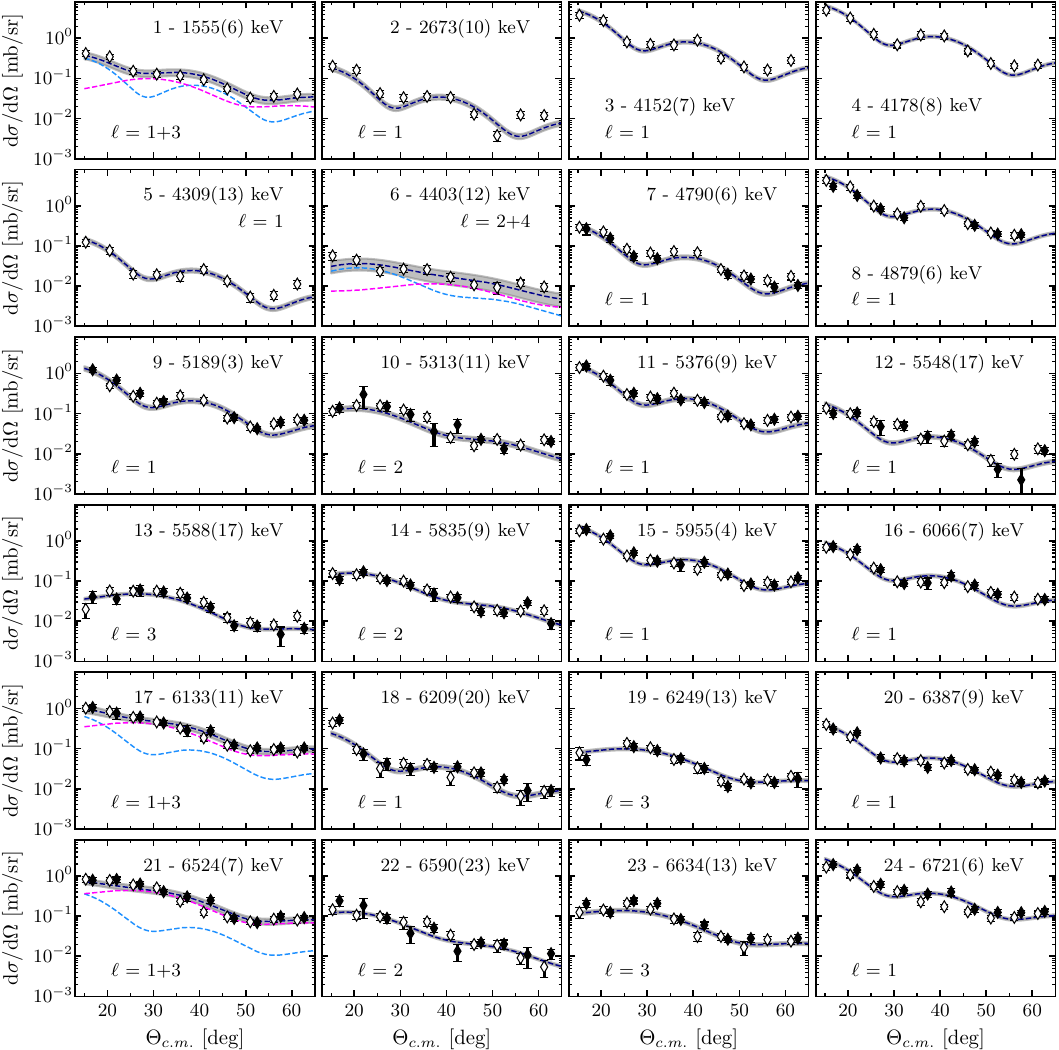}
\caption{\label{fig:AngDist_1stplot}Angular Distributions for states observed in \nuc{50}{Ti} populated via \nuc{49}{Ti}$(d,p)\nuc{50}{Ti}$. Experimental differential cross sections (symbols) are compared to Adiabatic Distorted Wave Approximation (ADWA) calculations performed with \textsc{fresco} (lines). The calculated lines were scaled to data through a $\chi^2$ minimization to obtain the spectroscopic factor, $S$, and assuming a specific $J$. The 1$\sigma$ uncertainty band is shown (gray shaded band). State labels, excitation energies, and preferred angular momentum transfer are given and also listed in Table\,\ref{tab:EnergyTable}. For states observed in both the 8.6-kG and 7.6-kG magnetic settings, the data are presented using open and closed symbols, respectively. For states with mixed transfer configurations, their individual contributions are highlighted with blue and pink dashed lines.}
\end{figure*}

\begin{figure*}[t]
\includegraphics[width=\linewidth]{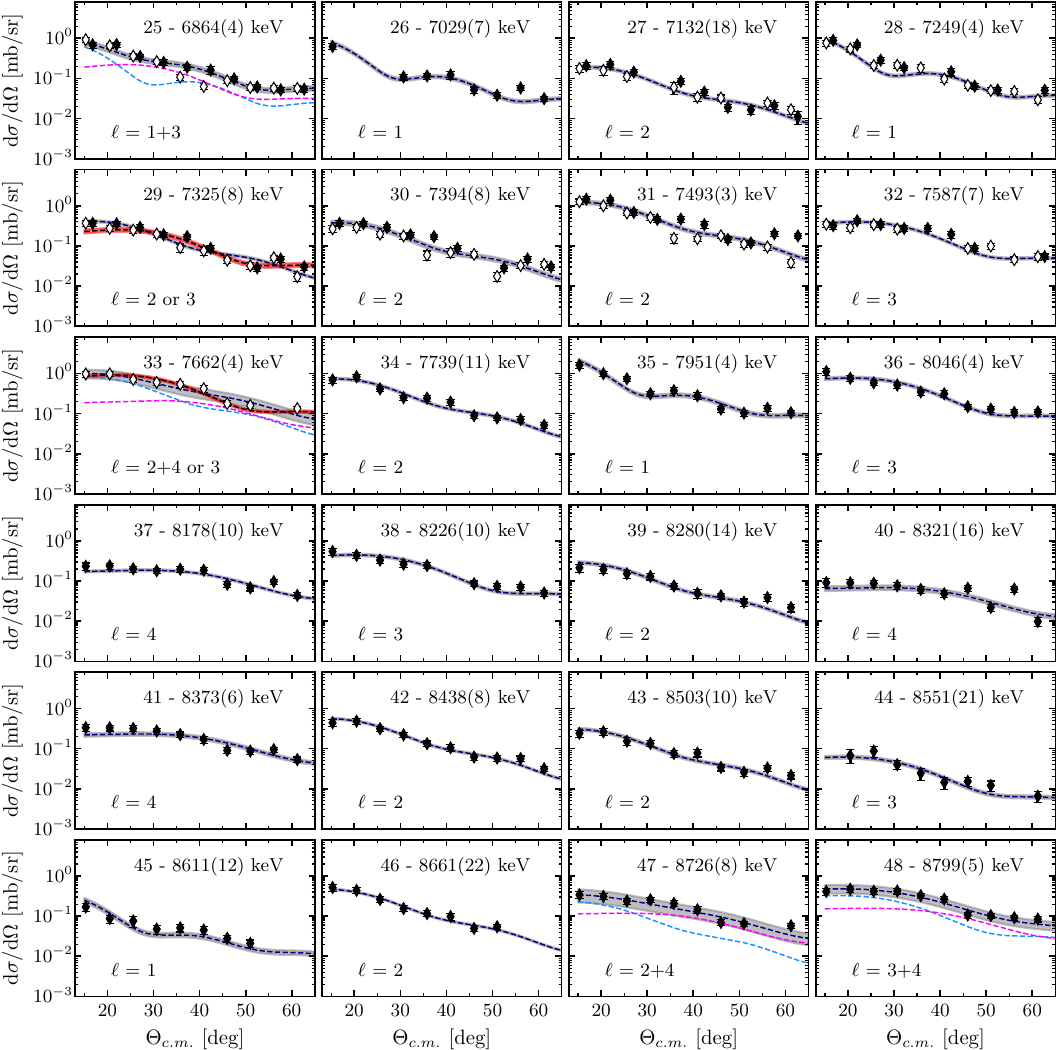}
\caption{\label{fig:AngDist_2ndplot}Same as Fig.\,\ref{fig:AngDist_1stplot} but for the second set of angular distributions. For state 29, the predicted distribution for both $\ell=2$ (gray band) and $\ell=3$ transfers (red band) led to equally good descriptions of the measured distribution. For state 33, the predicted distribution for both a mixed $\ell=2+4$ (gray band) and single $\ell=3$ transfers (red band) led to equally good descriptions. The uncertainty for the latter is smaller resulting in a tighter uncertainty band.}
\end{figure*}

 \noindent An example for the 4172-keV state is shown in Fig.\,\ref{fig:Integrated_cross_section}. The measured $(d,p)$ angular distributions, i.e., differential cross sections measured at the different laboratory angles are presented in Figs.\,\ref{fig:AngDist_1stplot} to \ref{fig:AngDist_4thplot}. For each state, the excitation energy determined from this work, proposed spin-parity, $J^{\pi}$, assignment based on the observed angular momentum transfer, $\ell$, and the model-independent angle-integrated cross section, $\sigma_{total}$, are included in Table\,\ref{tab:EnergyTable}. To extract the model-dependent spectroscopic factors, $S$, the transfer configurations given in Table\,\ref{tab:EnergyTable} were used in our reaction calculations with \textsc{fresco} \cite{fresco}. They were determined from a $\chi^2$ minimization of the theoretical distributions to the $(d,p)$ angular distributions (shown as lines in Figs. \ref{fig:AngDist_1stplot}--\ref{fig:AngDist_4thplot}). The calculated $(d,p)$ cross sections do, however, depend on the total angular momentum, $J_f$, assumed for the state to which the neutron transfer leads as it contributes with the weight $(2J_f+1)$, see, {\it e.g.}, Refs.\,\cite{Mac25a, Kunz1993}.  Since for even-$A$ nuclei, the spin, $J_f$, of a populated state cannot be uniquely determined from the $(d,p)$ angular distribution alone, see also the comments in Refs.\,\cite{Hay_52V, 61Nidp_Mark, spi24a}, we decided to list the ``spin-weighted'' spectroscopic factor, $S'$, in Table\,\ref{tab:EnergyTable}. It is defined as,

\begin{equation}
  S' = S \cdot\frac{(2J_f(\nuc{50}{Ti})+1)}{(2J_i(\nuc{49}{Ti})+1)}  
  \label{eq:specfac}
\end{equation}
 
 \noindent where $J_f$ is the spin of the final state in \nuc{50}{Ti} and $J_i = 7/2^-$ is the ground-state spin of \nuc{49}{Ti}. Rather then just separating the spectroscopic factor out, we then get the following relation between the experimental and theoretical cross section:

 \begin{equation}
     \left(\frac{d\sigma}{d\Omega}\right)_{\mathrm{exp.}} = \frac{2J_f+1}{2J_i+1}  S \left(\frac{d\sigma}{d\Omega}\right)_{\mathrm{th.}}
 \end{equation}
 
$S'$ is a quantity, also sometimes referred to as ``spectroscopic strengths'' \cite{Bertulani}, that can be calculated without knowing $J_f$. Values of $S'$ can, thus, be compared to previous measurements if these reported the quantity or the spin assumed in their reaction calculations needed for us to calculate $S'$ retrospectively. As will be discussed in Sec.\,\ref{sec:50Ti_SF_strength}, $S'$ is used to evaluate the sum rule for the neutron-adding strength to the different single-particle orbitals. We will also show that $S'$ corresponds to the neutron-removal spectroscopic factor $S_j^-(\nuc{50}{Ti}(E_x) \to \nuc{49}{Ti}(\mathrm{g.s.}))$, if we were to remove a neutron from an excited state in \nuc{50}{Ti}, in Sec.\,\ref{sec:Theory_comparison} and Appendix\,\ref{appendix}.

The theoretical reaction cross sections were calculated with the Adiabatic Distorted Wave Approximation (ADWA) using the coupled-channels code \textsc{fresco} \cite{fresco}. To determine the deuteron entrance channel, the optical model parameters were calculated using the approach of Ref. \cite{Wales_Johnson}. The optical model parameters for the proton and neutron were calculated from the global optical model parameters of Koning and Delaroche \cite{Koning_OMPS}. As in our previous work, the overlaps between the \nuc{50}{Ti} and \nuc{49}{Ti} + $n$ system were calculated using binding potentials of Woods-Saxon form, whose depth was varied to reproduce the given state's binding energy, with geometry parameters of $r_0$ = 1.25 fm and $a_0$ = 0.65 fm, and a Thomas spin-orbit term with a strength $V_{so} = 6$ MeV that was not varied. Most state's measured angular distributions could be described using a single transfer configuration, i.e., one angular momentum transfer. For some states, as shown in Table\,\ref{tab:EnergyTable} and Figs.\,\ref{fig:AngDist_1stplot} to \ref{fig:AngDist_4thplot}, the superposition of two angular momentum transfers was necessary to describe the measured distribution. No multi-step excitation was considered and no coupled-channels calculations were performed to describe the experimental data.

 In the following sections, we will discuss implications of our new $(d,p)$ and $(d,p\gamma)$ data for adopted and previously observed states. With one exception, we will go in order of increasing excitation energy. The highest-lying state populated in the previous $(d,p)$ measurement was at an excitation energy of 7.67\,MeV \cite{Barnes1, Barnes2} (state number 33 in Table\,\ref{tab:EnergyTable} and Fig.\,\ref{fig:AngDist_2ndplot}). Above this energy, 49 additional states were identified in the present work. Any observed discrepancies in spin-parity assignments will be addressed. 

  \subsubsection{$J^{\pi}=1^+$ states}

 We discussed $J^{\pi} = 1^+$ states populated in the $\nuc{49}{Ti}(d,p)\nuc{50}{Ti}$ reaction and the implications of our findings for the spin-flip $M1$ strength in Ref.\,\cite{Kel26a}. The experimental information for these states, obtained from our $(d,p)$ experiment, is also listed in Table\,\ref{tab:EnergyTable}. The $J^{\pi} = 1^+$ states of \nuc{50}{Ti} were populated because of $(1f_{7/2})^{-1}(1f_{5/2})^{+1}$ neutron 1p-1h excitations contributing to their wavefunctions, i.e., through $\ell =3$ transfers. Previously, some of the $1^+$ states had been observed in $(e,e')$ \cite{50Ti_ee'} and $(p,p')$ \cite{50Ti_pp'_201MeV, 50Ti_pp'}. Assignments were confirmed based on a comparison of our $(d,p)$ data to these earlier data and new $(\gamma,\gamma')$ data \cite{Kel26a}.

 \subsubsection{4403-keV level}

The first discrepancy between this work and previous $(d,p)$ experiments is observed for the level adopted at 4410 keV \cite{Mass50DataSheet} (state number 6 in Table\,\ref{tab:EnergyTable} and Fig.\,\ref{fig:AngDist_1stplot}). Earlier $(d,p)$ studies reported a mixed $\ell=0+2$ transfer to a weakly populated state, which is the only instance of an $\ell=0$ transfer in their data \cite{Barnes2}. In this work, no evidence of $\ell=0$ strength is observed, which is consistent with findings in previous $(d,p)$ experiments of our group on \nuc{62}{Ni}\,\cite{61Nidp_Mark} and \nuc{53}{Cr}\,\cite{Spi25a}. For the state observed at 4403(12)\,keV in our work, a mixed $\ell=2+4$ transfer provides a good description of the measured angular distribution and is consistent with the adopted $J^{\pi} = 3^-$ assignment. The mixed transfer is also consistent with the presence of $\ell=4$ strength near this excitation energy, as observed in all $N=29$ isotones studied at the SE-SPS \cite{Riley_51Ti, Riley_55Fe, Riley_53Cr, Hay_52V, Spi25a}. Note that the values of $S'$, given in Table\,\ref{tab:EnergyTable}, are not directly comparable for this state. The value of $S' = 0.003$ corresponds to the $\ell=0$ component and of $0.009$ to the $\ell=2$ component \cite{Mass50DataSheet}. For the latter, we get $S' = 0.0035(11)$. As can be seen in Table\,\ref{tab:EnergyTable} and in agreement with previous work \cite{Barnes2}, the spectroscopic factors are comparably small for this state confirming its weak population in the $(d,p)$ reaction.

\subsubsection{5588-keV level}

 We observe a state at 5588(17)\,keV (state 13 in Table\,\ref{tab:EnergyTable} and Fig.\,\ref{fig:AngDist_1stplot}). In terms of energy, this state could either correspond to the states adopted around 5560\,keV or the state currently adopted at 5600 \,keV\,\cite{Mass50DataSheet}. Previous $(d,p)$ experiments suggested the existence of two states at 5561 and 5600\,keV\,\cite{Barnes2}. Both states were weakly populated via $\ell=1$ and a mixed $\ell=1+3$ angular momentum transfer, respectively. This means that the states should have a positive parity quantum number, which excluded the possibility that the adopted 5560-keV, $J^{\pi} = (3)^-$ state\, was observed in the $(d,p)$ reaction \cite{Mass50DataSheet}. In our work, the angular distribution, obtained with both magnetic field settings, is described best by a pure $\ell=3$ transfer, see Fig.\,\ref{fig:AngDist_1stplot}. The $\ell=3$ transfer also indicates that the state's parity quantum number is $\pi = +1$ and that the 5588-keV state likely corresponds to the level previously reported at 5600\,keV. The complementary $(d,p\gamma)$ data confirms that this state belongs to \nuc{50}{Ti} as the $4^+_1\rightarrow2^+_1$ and $2^+_1\rightarrow0^+_1$ $\gamma$-ray transitions are observed when gating on that state. No primary $\gamma$-ray transitions could be established though. Given the higher excitation energy of the state, we assume that it was populated in the $(d,p)$ reaction because of a $(1f_{7/2})^{-1}(1f_{5/2})^{+1}$ 1p-1h component in its wavefunction. This gives a spin range of $J=1 - 6$.

\begin{figure*}[t]
\includegraphics[width=\linewidth]{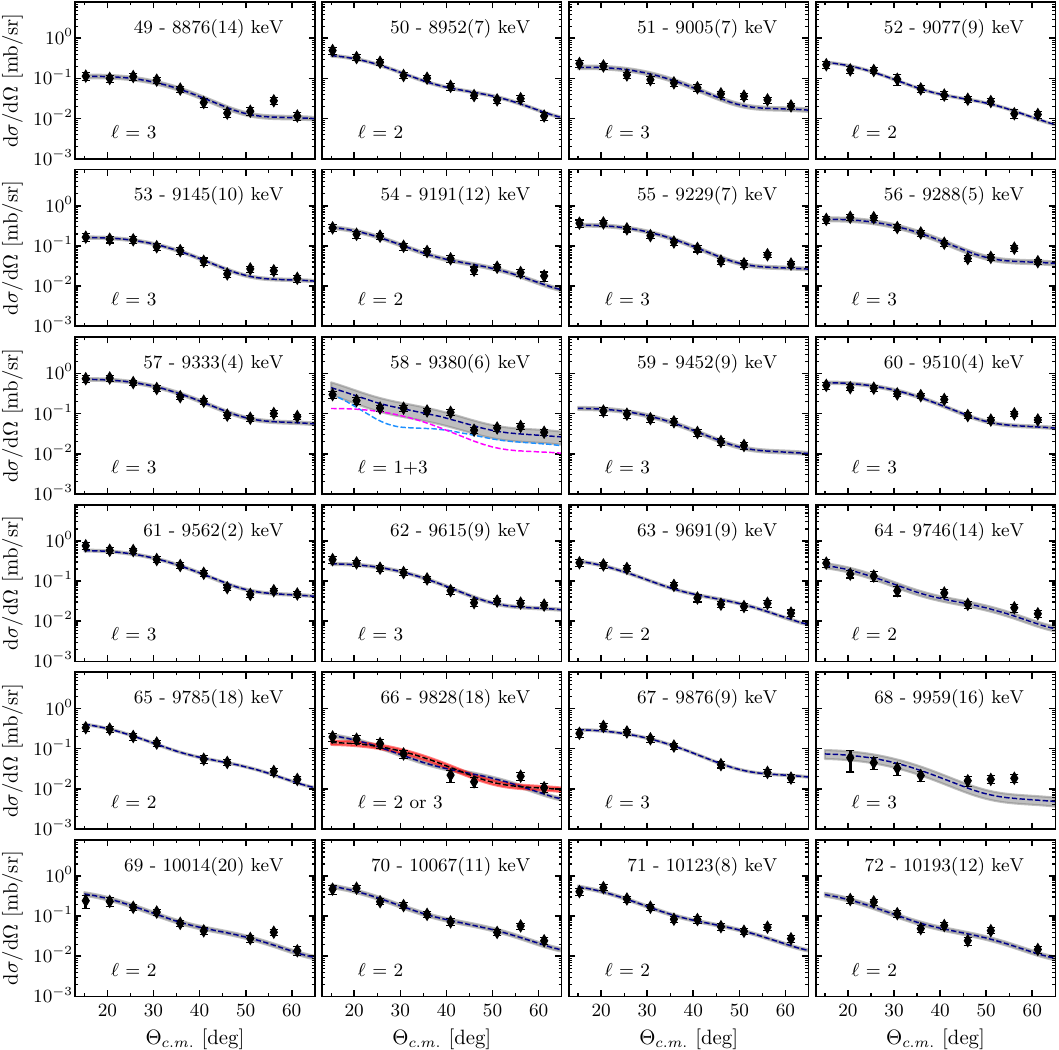}
\caption{\label{fig:AngDist_3rdplot}Same as Fig.\,\ref{fig:AngDist_2ndplot} but for the third set of angular distributions. For state 66, the predicted distribution for both $\ell=2$ (gray band) and $\ell=3$ transfers (red band) led to equally good descriptions of the measured distribution.}
\end{figure*}

\subsubsection{5835-keV level}

A state is adopted at 5837.2(6) keV with a tentative spin-parity assignment of $J^{\pi} = (2-5)^{(+)}$. We observe a state at 5835(9)\,keV (see state 14 in Table\,\ref{tab:EnergyTable} and Fig.\,\ref{fig:AngDist_1stplot}). Its $(d,p)$ angular distribution, measured again with both field settings as it is within the overlap region, is best described by an $\ell=2$ transfer. This points to a spin-parity assignment of $J^{\pi} = 1^--6^-$. Available $\gamma$-ray data allowed us to provide a narrower spin range. Previously published $(d,p\gamma)$ \cite{50Ti_dpgamma} data suggest transitions with $\gamma$-ray energies of $E_{\gamma} = 1690$ and 3162\,keV to originate from the 5837-keV state\,\cite{Mass50DataSheet}. These primary transitions lead to the $4^+_2$ and $4^+_1$ states, respectively. The primary transitions to the $4^+$ states, likely of $E1$ character, suggest that the possible spin range for this state can be narrowed down to $J=3-5$ but that the parity quantum number should be $\pi = -1$ instead.

\begin{figure}[t]
\includegraphics[width=\linewidth]{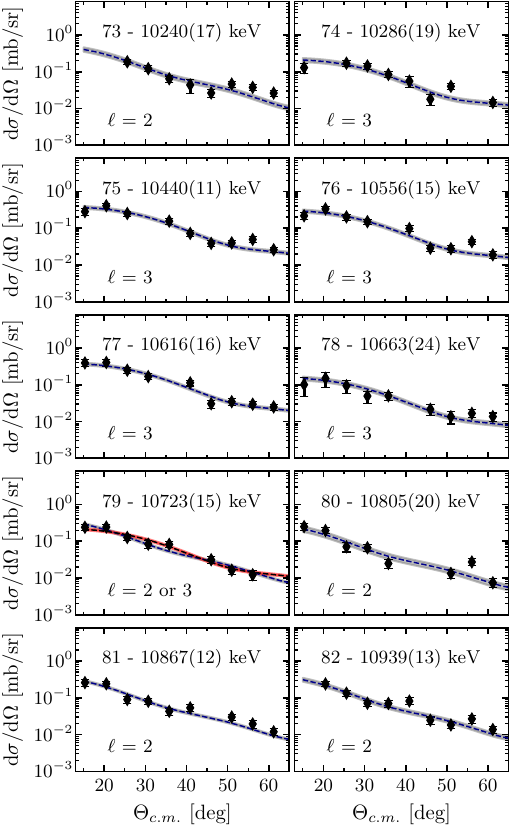}
\caption{\label{fig:AngDist_4thplot}Same as in Fig.\,\ref{fig:AngDist_3rdplot} but for the fourth set of angular distributions. For state 79, the predicted distribution for both $\ell=2$ (gray band) and $\ell=3$ transfers (red band) led to equally good descriptions of the measured distribution.}
\end{figure}

\subsubsection{6133-keV level}

The previous $(d,p)$ study reported a level at 6138 keV. The data suggested a mixed $\ell=1+3$ transfer \cite{Barnes1,Barnes2}. Our data confirm the previously observed mixed angular momentum transfer, see Fig.\,\ref{fig:AngDist_1stplot}, and place the level, populated in $(d,p)$, at 6133(11)\,keV. However, the adopted -- yet tentative -- spin-parity assignment of $J^{\pi} = (7)^+$ for the 6136-keV state is at odds with the observations \cite{Mass50DataSheet}. Unless this is an unresolved doublet, one cannot couple to $J=7$ if the state has $(1f_{7/2})^{-1}(2p_{3/2})^{+1}$, $(1f_{7/2})^{-1}(2p_{1/2})^{+1}$ ($\ell = 1$), and $(1f_{7/2})^{-1}(1f_{5/2})^{+1}$ ($\ell = 3$) 1p-1h components in its wavefunction.

\begin{figure}[t]
\includegraphics[width=\linewidth]{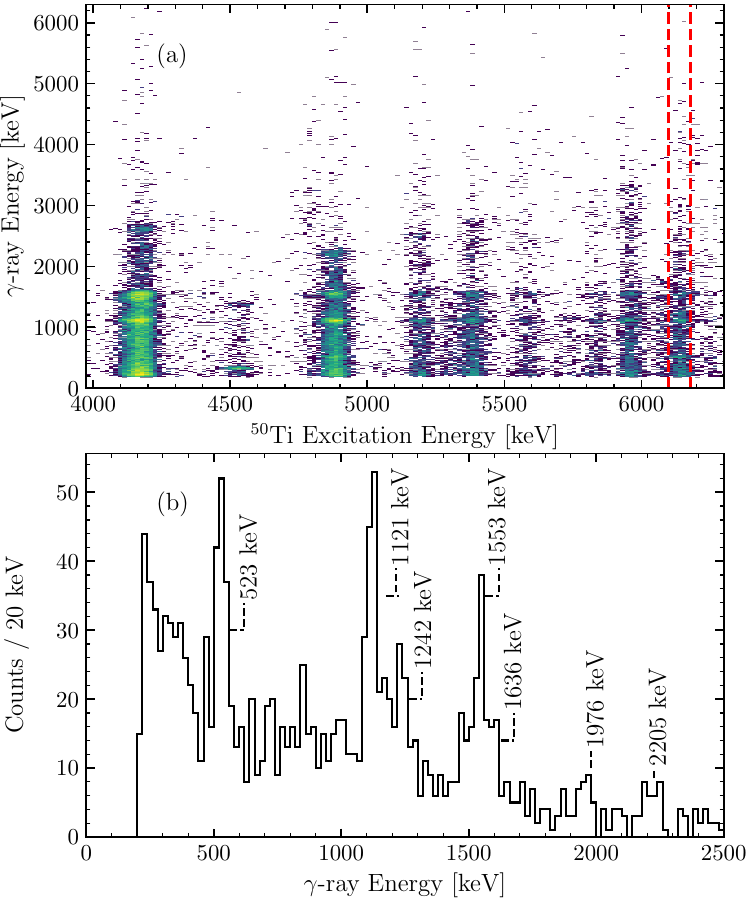}
\caption{(a) $p\gamma$-coincidence matrix for the \nuc{49}{Ti}$(d,p\gamma)$\nuc{50}{Ti} reaction for an excitation energy range of $4000 \ \mathrm{keV} < E_x < 7000 \ \mathrm{keV}$. (b) $\gamma$-ray spectrum for the 6123-keV level obtained if proton gate shown in panel (a) is applied (red dashed lines). The $\gamma$ rays with $E_{\gamma} = 1242$, 1636, and 1976\,keV correspond to known primary transitions coming from this level\,\cite{Mass50DataSheet}. Secondary transitions are also highlighted.}
\label{fig:50Ti_Matrix_Projections}
\end{figure}

To investigate further, we gated on the state in our $p\gamma$-coincidence matrix, shown in Fig.\,\ref{fig:50Ti_Matrix_Projections}\,(a), which returned the $\gamma$-ray spectrum of Fig.\,\ref{fig:50Ti_Matrix_Projections}\,(b). As can be seen, the known primary $\gamma$ rays of the 6123-keV state are observed\,\cite{Mass50DataSheet}. This suggests that it is not the 6136-keV but the 6123-keV state which is populated in the $(d,p)$ reaction. The state has as tentative spin-parity assignment of $J^{\pi} = (4^+)$\,\cite{Mass50DataSheet}. The combination of our $(d,p)$ and $(d,p\gamma)$ data clearly fixes the parity quantum number to be $\pi = +1$. We, thus, propose to reevaluate which state was populated in the $(d,p)$ reaction in the currently adopted data\,\cite{Mass50DataSheet}.

\subsubsection{6209-keV level}

In the previous $(d,p)$ work by Barnes \textit{et al.}\cite{Barnes1, Barnes2}, an excited state at 6210 keV was identified. A tentative angular momentum transfer of $\ell=2$ was reported. The state is currently adopted at 6212(5)\,keV with a spin-parity assignment of $J^{\pi} = 1^- - 6^-$\,\cite{Mass50DataSheet}, listing the mentioned $(d,p)$ experiment as source for the $J^{\pi}$ assignment. The $(t,p)$ experiment by Hinds and Middleton\,\cite{50Ti_tp}, also listed in the adopted data, observed the state at 6207(15) keV. This experiment suggests a $J^{\pi}=2^+$ assignment for that level based on the observation of an $\ell =2$ angular momentum transfer in $(t,p)$ \cite{50Ti_tp}. In the present work, the angular distribution, also measured with both magnetic field settings, is best described by an $\ell=1$ transfer, suggesting a positive parity quantum number for the state, see Fig.\,\ref{fig:AngDist_1stplot}. This observation is consistent with the previous $(t,p)$ experiment and suggests that the currently adopted assignment of $J^{\pi}=1^- - 6^-$ should be revised to $J^{\pi}=2^+$.

\subsubsection{7325-keV level}

A level at 7325(8) keV was observed in this work (see state 29 in Table\,\ref{tab:EnergyTable} and Fig.\,\ref{fig:AngDist_2ndplot}). Its measured $(d,p)$ angular distribution is equally well described by either an $\ell=2$ or $\ell=3$ transfer. Both fits yielded similar $\chi^2$ values. A state at 7335 keV was previously populated in a $(p,p')$ experiment\,\cite{50Ti_pp'}. Based on the observed $\ell =2$ transfer, $J^{\pi}=2^+$ was adopted for the state \cite{50Ti_pp', ensdf}. As the excitation energies overlap, the $\ell = 3$ transfer in the $(d,p)$ reaction would be consistent if the same state was populated. Previous work did not report the population of the state in the $(d,p)$ reaction\,\cite{Barnes2}.

\subsubsection{7662-keV Level}

We identified an excited state at 7662(4) keV. The previous $(d,p)$ study reported a state at 7667(15) keV and tentatively assigned it to have been populated through an $\ell=3$ transfer. However, a mixed $\ell=2+4$ transfer also provided a reasonable description of the data. As shown in Fig.\,\ref{fig:AngDist_2ndplot}, the $\ell=3$ calculation (red curve) yields a better fit to our data. The corresponding $\chi^2$ value is smaller by approximately a factor of two compared to the mixed $\ell=2+4$ case (gray curve). This state was also observed in the aforementioned $(t,p)$ study\,\cite{50Ti_tp}, where the observation of an $\ell=2$ transfer led to a spin-parity assignment of $J^{\pi}=2^+$ for the state. Taken together, the available data support the $\ell=3$ transfer observed in the present $(d,p)$ measurement, which is in agreement with the proposed $J^{\pi} = 2^+$ assignment.

\subsubsection{8799-keV level}

Two excited states are currently adopted at 8810(20) and 8815(10) keV with spin-parity assignments of $J^{\pi} = 1^+$ and $J^{\pi} = (3)^-$, respectively \cite{Mass50DataSheet}. This doublet could not be resolved in our $(d,p)$ experiment. However, the measured $(d,p)$ angular distribution is best described by a superposition of $\ell=3$ and $\ell=4$ transfers, see state 48 in Fig.\,\ref{fig:AngDist_2ndplot}, indicating the population of both states. As mentioned earlier, levels with such high excitation energies were not studied in previous $(d,p)$ experiments.

\subsubsection{9288- \& 9510-keV levels}

Two excited states were observed at 9288(5) and 9510(4) keV, with angular distributions best described by $\ell=3$ transfers. Corresponding levels were previously reported in an inelastic proton scattering study at nearly identical energies of 9282 and 9508 keV, with tentative spin–parity assignments of either $J^{\pi}=5^-$ or $6^+$ for both states\,\cite{50Ti_pp'}. Given the $\ell=3$ transfer observed in the present $(d,p)$ measurement, which unambiguously leads to a positive parity quantum number for both states, we suggest to revise the spin-parity assignments for both states to $J^{\pi}=6^+$.

\subsubsection{10014-, 10123-, \& 10240-keV levels}

We identified levels at 10014(20)\,keV, 10123(8)\,keV, and 10240(17)\,keV (see Table\,\ref{tab:EnergyTable} and Figs.\,\ref{fig:AngDist_3rdplot} and \ref{fig:AngDist_4thplot}). For the 10014-keV and 10123-keV states, the angular distributions are best described by $\ell=2$ transfers, implying a possible spin range of $J=1 - 6$ and negative parity for these states. An inelastic electron-scattering experiment reported states at 10000(20)\,keV and 10140(20)\,keV, excited either via $M1$ or $M2$ transitions\,\cite{50Ti_ee'}. Spin-parity assignments of $J^{\pi}=1^+$ or $2^-$ are then possible. Both scenarios were considered equally probable. If we populated the same levels, the only consistent spin-parity assignment would be $J^{\pi}=2^-$ for both states. The excitation energies agree within uncertainties.

 The $(d,p)$ angular distribution of the 10240-keV level is best reproduced assuming the population through an $\ell=2$ transfer. States at excitation energies of 10210(20) and 10250(20)\,keV were also observed in the mentioned $(e,e')$ experiment, where $E1$ and $E3$ excitations were listed, respectively\,\cite{50Ti_ee'}. This would lead to $J^{\pi}=1^-$ and $3^-$ assignments for the two states. The $\ell=2$ transfer observed here supports that a state with negative parity was populated. We cannot say for certain whether it is the $J^{\pi}=1^-$ or $3^-$ state. Our data does, however, support the two most probable scenarios presented in Ref.\,\cite{50Ti_ee'} for the excitation of these states in $(e,e')$.

\section{Discussion}

In this section, we will discuss the experimental results obtained for \nuc{50}{Ti} focusing on the fragmentation of the single-particle strengths up to $S_n$. In section \ref{sec:50Ti_SF_strength}, we will first summarize our results for the different $\ell$ transfers comparing integrated cross sections to the spin-weighted spectroscopic factors, $S'$, from which vacancies can be calculated. In section \ref{sec:comp_51Ti}, we will then compare the fragmentation of the single-particle strengths in even-$A$ \nuc{50}{Ti} $(N=28)$ and odd-$A$ \nuc{51}{Ti} $(N=29)$. Finally, we will confront three different nuclear structure models with our new data for \nuc{50}{Ti} in section \ref{sec:Theory_comparison}. These are the time-dependent continuum shell model (TDCSM) \cite{Vol09a, Vol14a} with the FSU cross-shell interaction \cite{Lub19a, Lub20a}, the energy-density functional plus quasiparticle-phonon model approach (EQPM) \cite{Tsoneva2016}, and the relativistic equation of motion theory including the coupling of two quasiparticles with up to two phonons (REOM$^3$) \cite{Lit19a, Lit22a, Lit23a}.

\subsection{Spectroscopic strengths in \nuc{50}{Ti}}
\label{sec:50Ti_SF_strength}

\begin{figure*}[t!]
\includegraphics[width=\linewidth]{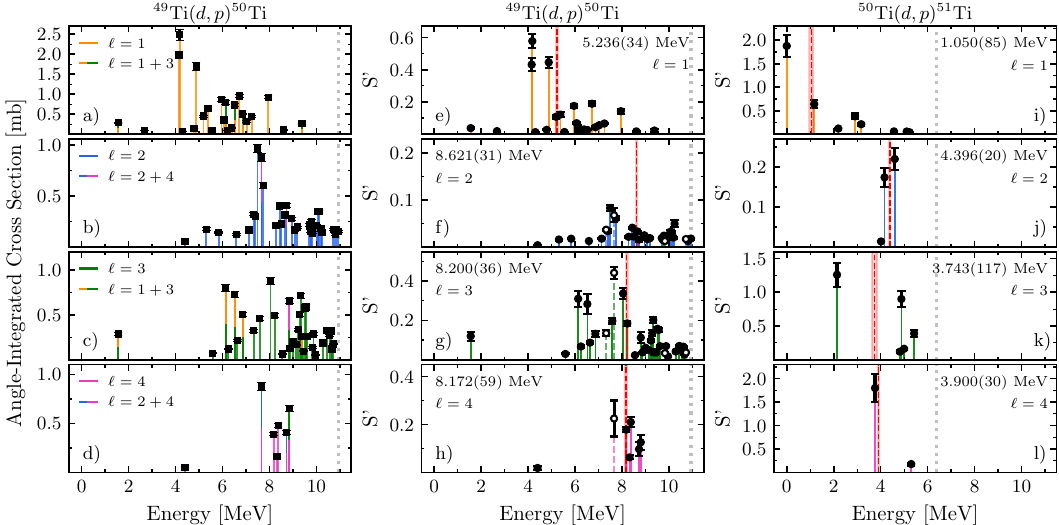}
\caption{(a)–(d) Angle-integrated cross sections for \nuc{50}{Ti}, shown separately for each orbital angular momentum transfer $\ell$. States with mixed transfers are displayed in both corresponding panels using split colors. (e)–(h) spin-weighted spectroscopic factors for \nuc{50}{Ti} and (i)–(l) for \nuc{51}{Ti}\,\cite{Riley_51Ti}. For \nuc{50}{Ti}, states for which more than one $\ell$ transfer could equally well describe the angular distribution are indicated by dashed lines with open symbols in the relevant panels. Vertical red dashed lines highlight the centroid energies for each orbital, with uncertainties depicted by the lighter, red shaded band. The centroid energies are also given in the panels. The vertical gray dotted line in each panel corresponds to the neutron separation energy, $S_n$.}
\label{fig:50Ti_51Ti_centroids}
\end{figure*}

The experimentally measured and very distinct $(d,p)$ angular distributions, see Figs.\,\ref{fig:AngDist_1stplot} to \ref{fig:AngDist_4thplot}, clearly highlight that excited states of \nuc{50}{Ti} were populated via different angular momentum, $\ell$, transfers in the $(d,p)$ reaction. As described above, these angular momentum transfers were identified through comparison to single-step ADWA calculations. For states, which were observed at several scattering angles, model-independent angle-integrated cross sections were determined. They are presented in Figs.\,\ref{fig:50Ti_51Ti_centroids}\,(a)-(d). Spectroscopic factors $S = (d\sigma/d\Omega)_{exp.}/(d\sigma/d\Omega)_{ADWA}$ were determined by fitting the calculated angular distribution to the measured one using $\chi^2$ minimization as also done in our previous studies \cite{Riley_51Ti, Riley_55Fe, Riley_53Cr, 61Nidp_Mark, Hay_52V, spi24a, Spi25a}. Since the total angular momentum quantum number of the states, $J$, cannot be unambiguously determined from the $(d,p)$ data if an unpolarized beam is used, we chose to calculate the spin-weighted spectroscopic factors $S'$ as defined in Eq.\,(\ref{eq:specfac}). These are shown in Figs.\,\ref{fig:50Ti_51Ti_centroids}\,(e)-(h) for the different $\ell$ transfers. In contrast to the angle-integrated cross sections, where mixed transfers could not be separated and where we chose to present them with associated color coding in Figs.\,\ref{fig:50Ti_51Ti_centroids}\,(a)-(d), separate spin-weighted spectroscopic factors could be determined for states (or doublets) that were populated through mixed angular momentum transfers (see Table\,\ref{tab:EnergyTable} and Figs.\,\ref{fig:AngDist_1stplot}-\ref{fig:AngDist_4thplot}). As outlined in Ref.\,\cite{satchler}, these spin-weighted spectroscopic factors are part of the sum rule for stripping (nucleon adding) reactions like the $(d,p)$ reaction:

\begin{equation}
    G_{lj,+}=\sum_k S_{lj,k} \frac{2J_{f,k}+1}{2J_i+1} = (2j+1)-n(l,j)
\label{eq:stripping}
\end{equation}

In Eq.\,(\ref{eq:stripping}), $S_{l,j}$ is the nucleon single-particle spectroscopic factor for the stripping (nucleon adding) reaction, $J_f$ the spin of the final state, $J_i$ the spin of the target ground state, $(2j+1)$ the occupancy (degeneracy) of a nucleon single-particle orbit, and $n(l,j)$ measures how many nucleons already occupy the single-particle orbit $(l,j)$ in the target ground state. The summation index $k$ indicates summation over all possible final states after populating those states in the stripping reaction because of contributions of the single-particle configuration $(l,j)$, i.e., the corresponding nucleon 1p-1h component to its wavefunction. Generally, $G_{lj,+}$ measures the average number of holes in the single-particle orbit $(l,j)$ of the target ground state, i.e., its vacancy. The combination of pick-up (nucleon removal) and stripping (nucleon adding) reactions provides the means to determine the occupancy of single-particle orbitals \cite{satchler, SumRule_TransferReactions}. We acknowledge that the model dependency of occupation numbers -- and spectroscopic factors -- has been and continues to be discussed controversially, see, {\it e.g.}, Refs.\,\cite{SumRule_TransferReactions, Fur02a, Fur10a, Muk10a, Dug15a}. However, as has been undoubtedly demonstrated, they provide important insights for interpreting a variety of nuclear-structure phenomena.

A polarized deuteron beam is not available at the Fox Laboratory. Therefore, we cannot distinguish between spin-orbit partners since both are populated through the same angular momentum transfer, $\ell$, in the $(d,p)$ reaction. As shown in Refs.\,\cite{spi24a, Spi25a}, particle-$\gamma$ angular correlations possibly provide the means to determine $j$ and not just $\ell$ even when using an unpolarized deuteron beam. The CeBrA demonstrator was not set up to measure these correlations in the present work. In the absence of information to determine the quantum numbers $(l,j)$ uniquely, we chose to report results on the $\ell = 1$ strength in Fig.\,\ref{fig:50Ti_51Ti_centroids} rather than for the $2p_{3/2}$ and $2p_{1/2}$ orbitals separately. A similar approach was also adopted in, {\it e.g.}, Ref.\,\cite{SumRule_TransferReactions}. To first order, both orbitals are expected to be unoccupied (vacant). We will refer to them as $2p$ orbitals in the following. For \nuc{50}{Ti} ($N=28$), we expect that little $1f_{7/2}$ transfer strength is left as the \nuc{49}{Ti}, $J^{\pi} = 7/2^-$ ground state should be dominated by a simple $(1f_{7/2})^{-1}$ neutron hole configuration. So, naively, only one spot should be vacant in the $1f_{7/2}$ neutron orbital. In this scenario, the remaining $1f_{7/2}$ neutron strength should only lead to the population of the $J^{\pi} = 0^+$ ground state of \nuc{50}{Ti}. By choice, the ground state was placed at the acceptance limits of the focal-plane detector preventing us from reliably determining differential cross sections. The $J^{\pi} = 2^+_1$ state was populated through a mixed $\ell = 1+3$ transfer (see Table\,\ref{tab:EnergyTable} and Fig.\,\ref{fig:AngDist_1stplot}). As a Gedankenexperiment, we can assume that there is more than one vacant spot in the $1f_{7/2}$ orbital and that the observed $\ell =3$ component for $2^+_1$ state results from transferring a neutron into the $1f_{7/2}$ orbital. Then, the spin-weighted spectroscopic factor for the $\ell = 3$ component is $S'=0.12(2)$. The previous study of Ref.\,\cite{Barnes2} reported $S'=0.32$. This level of reduction compared to the previous work of Refs.\,\cite{Barnes1, Barnes2} was also observed in Ref.\,\cite{Riley_51Ti} for \nuc{51}{Ti}. The ``normalization'' (reduction) factor would be $R= 0.38(4)$. Applying this normalization to the previously reported spin-weighted spectroscopic factor for the \nuc{50}{Ti} ground state \cite{Barnes2} yields $S'= 0.29(2)$. Only considering these two states, one then computes $G_{1_{f_{7/2}},+} = 0.40(4)$ for the average number of holes in the $1f_{7/2}$ orbital for the \nuc{49}{Ti} ground state, which is, of course, subject to quenching. In our $(d,p)$ experiment, we observed a large energy gap between the $2^+_1$ state and the next state populated through an $\ell = 3$ transfer [see Figs.\,\ref{fig:50Ti_51Ti_centroids}\,(c) and (g)]. Based on this observation, we will in the following assume that the bulk of the $\ell = 3$ strength at higher energies can be attributed to transferring a neutron into the $1f_{5/2}$ orbital. We will come back to this discussion when comparing to theoretical predictions in Sec.\,\ref{sec:Theory_comparison}. As mentioned earlier, for $\ell = 4$ and $\ell = 2$ transfers we assumed that neutrons were transferred into the $1g_{9/2}$ and $2d_{5/2}$ orbitals. This will also be backed up by the theoretical calculations discussed in Sec.\,\ref{sec:Theory_comparison}.

As can be seen in Figs.\,\ref{fig:50Ti_51Ti_centroids}\,(a)-(h), the spectroscopic strengths are strongly fragmented in \nuc{50}{Ti}. By comparing the model-independent angle-integrated cross sections in Fig.\,\ref{fig:50Ti_51Ti_centroids}\,(a)-(d) to the model-dependent spin-weighted spectroscopic factors in Figs.\,\ref{fig:50Ti_51Ti_centroids}\,(e)-(h), it is quite apparent that the same strength fragmentation pattern is observed. The sum rule $G_{lj,+}$, i.e., the number of holes in the \nuc{49}{Ti} ground state for the associated orbitals were determined to be 2.7(3) [45(4)\% relative to 6] for $\ell=1$, 0.83(8) [14(2)\% relative to 6] for $\ell=2$, 3.9(4) [65(7)\% relative to 6] for $\ell=3$, and 0.9(2) [9(2)\% relative to 10] for $\ell=4$ according to Eq.\,(\ref{eq:stripping}). Note that the value stated for $\ell =3$ includes the $2^+_1$ state. If we include the scaled value for the \nuc{50}{Ti} ground-state spin-weighted spectroscopic factor, see discussion above, then the total vacancy is 4.2(4) for $\ell=3$. We observe that the vacancies for the well-bound odd-$\ell$ neutron orbitals are consistent with the quenching of single-particle strength expected in the limit of the independent single-particle model. As in our previous work \cite{Riley_53Cr, Hay_52V, 61Nidp_Mark, Spi25a}, significantly less strength is collected for the $1g_{9/2}$ and $2d_{5/2}$ orbitals, which makes them appear less vacant. Possible explanations were provided in Refs.\,\cite{Riley_53Cr, Hay_52V, Spi25a}. We will get back to this discussion when comparing to model predictions. However, we already point out that the strong fragmentation of the spectroscopic strengths clearly shows that it is not sufficient to just study single fragments and that instead states in \nuc{50}{Ti} had to be studied to high excitation energy to accumulate the strength. The running sums for $G_{lj,+}$ are presented in Fig.\,\ref{fig:50Ti_dp_theory_comparison}.

\subsection{Comparison to \nuc{51}{Ti}}
\label{sec:comp_51Ti}

The data to calculate the vacancies for \nuc{51}{Ti} were reported in Ref.\,\cite{Riley_51Ti}. The vacancies are 3.4(5) [56(8)\% relative to 6] for $\ell=1$, 0.41(5) [7(1)\% relative to 6] for $\ell=2$, 2.8(4) [47(6)\% relative to 6] for $\ell=3$, and 2.0(3) [20(3)\% relative to 10] for $\ell=4$. For the two Ti isotopes, the vacancies for the $2p$ orbitals ($\ell=1$) agree well within uncertainties even though the strength fragmentation is very different, see Figs.\,\ref{fig:50Ti_51Ti_centroids}\,(e) and (i). For $\ell=3$, the larger value in \nuc{50}{Ti} could possibly be explained by vacancies in the $1f_{7/2}$ orbit for the \nuc{49}{Ti} ground state ($N=27$), which would lead to additional $1f_{7/2}$ ($\ell=3$) neutron strength in \nuc{50}{Ti} in contrast to \nuc{51}{Ti} (see discussion in Sec.\,\ref{sec:50Ti_SF_strength}). However, weaker $\ell =3$ strength associated with placing a neutron in the $1f_{5/2}$ orbital might also have been missed in Ref.\,\cite{Riley_51Ti}. As for the $\ell = 1$ strength, the $\ell = 3$ strength in \nuc{50}{Ti} is much more fragmented than in \nuc{51}{Ti}, see Figs.\,\ref{fig:50Ti_51Ti_centroids}\,(g) and (k). The same is observed for the $\ell =2$ and $\ell =4$ strengths.

We use the example of the $\ell = 4$ strength to highlight how differently single-particle strengths fragment between even-$A$ \nuc{50}{Ti} $(N=28)$ and odd-$A$ \nuc{51}{Ti} $(N=29)$. In \nuc{51}{Ti}, one strong fragment was observed below $S_n$\,\cite{Riley_51Ti} [see Fig.\,\ref{fig:50Ti_51Ti_centroids}\,(l)], while several fragments share the strength in \nuc{50}{Ti} [see Fig.\,\ref{fig:50Ti_51Ti_centroids}\,(h)]. To some extent, this is expected from simple angular momentum coupling considerations. For the $(1f_{7/2})^{-1}(1g_{9/2})^{+1}$ neutron 1p-1h configuration, one expects final states with $J^{\pi} = 1^- - 8^-$. Each one of these final states will contribute to the vacancy of the $1g_{9/2}$ orbital, i.e., the sum rule defined in Eq.\,(\ref{eq:stripping}). More specifically, partial vacancies can be attributed to the different $J^{\pi}$, i.e., 0.38 for $J^{\pi} = 1^-$, 0.63 for $J^{\pi} = 2^-$, 0.88 for $J^{\pi} = 3^-$, 1.13 for $J^{\pi} = 4^-$, 1.38 for $J^{\pi} = 5^-$, 1.63 for $J^{\pi} = 6^-$, 1.88 for $J^{\pi} = 7^-$, and 2.13 for $J^{\pi} = 8^-$ according to Eq.\,(\ref{eq:stripping}). See a similar discussion in Ref.\,\cite{Spi25b}. Consequently, we expect the strength to at least fragment between eight different states with different $J^{\pi}$ in \nuc{50}{Ti}. We observe seven fragments with one being tentatively assigned as $\ell=4$. Knowing that $G_{1g_{9/2},+}=2.0(3)$ for the $\ell=4$ strength in \nuc{51}{Ti}, we can use this to scale the partial vacancies in \nuc{50}{Ti}. If no further fragmentation takes place, then the $7^-$ and $8^-$ states should be the strongest populated states with $S' = 0.38(6)$ and 0.43(6), respectively. The two strongest fragments are observed at 8178(10)\,keV and 8373(6)\,keV with $S'$ values of  0.179(11) and 0.21(2), respectively. The ratio 0.85(10) between these two $S'$ values is surprisingly similar to the expected ratio 0.9(2). However, as for the sum rule, the spin-weighted spectroscopic factors appear to be further quenched by 50\,$\%$. Thus, it seems likely that other mechanisms must be at play, too. The position of the $\ell =4$ strength centroid relative to $S_n$ is about the same in both isotopes though, i.e., roughly 2\,MeV below $S_n$ [see Figs.\,\ref{fig:50Ti_51Ti_centroids}\,(h) and (l)]. Consequently, one wonders if continuum effects should contribute similarly to the two isotopes and, therefore, whether they should not be a major factor here. A word of caution is necessary. If such a small fraction of the expected single-particle strength is collected, as is the case for the $\ell = 4$ strength, one should not interpret the centroid energy as an effective single-particle energy. We will, however, show in the following that also the comparison of ``centroid'' energies determined for the $\ell =4$ and $\ell =2$ strengths in \nuc{50}{Ti} and \nuc{51}{Ti} gives access to another fundamental quantity which can be deduced from our data, i.e., the two-neutron pairing gap, $2\Delta_n$, or two-quasiparticle (2qp) energy. 

Before we come to that discussion, let us have a look at the combined centroid energies for the $2p$ orbitals and the centroid energies for the other orbitals. They are compiled in Table\,\ref{tab:pairing_centroids} and also shown in Fig.\,\ref{fig:50Ti_51Ti_centroids}. The combined centroid energy for the $2p$ orbitals is the lowest of the orbitals studied for both Ti isotopes. The centroid energy in \nuc{51}{Ti} is significantly lower than in \nuc{50}{Ti}. The energy difference is $\Delta E_c = 4.19(9)$\,MeV. For \nuc{50}{Ti}, all other orbitals have centroid energies in excess of 8\,MeV with the caveat of collecting significantly less strength for $\ell = 2$ and $\ell =4$ as for the $fp$ orbitals. Still, the energy differences between the centroids of \nuc{50}{Ti} and \nuc{51}{Ti} for these orbitals are remarkably similar to the one stated for the $2p$ orbitals. On average, the energy difference -- or this energy gap -- between the corresponding centroids in \nuc{50}{Ti} and \nuc{51}{Ti} is 4.29(4)\,MeV as also shown in Table\,\ref{tab:pairing_centroids}.

\begin{table}[t]
\centering
\caption{Centroid energies, $E_c$, for each $\ell$ transfer in \nuc{50}{Ti} and \nuc{51}{Ti}, and the corresponding energy difference $\Delta E_c = E_c(\nuc{50}{Ti}) - E_c(\nuc{51}{Ti})$. The average of these differences is given at the bottom of the table.}
\begin{tabular}{
>{\centering\arraybackslash}p{0.8cm} 
>{\centering\arraybackslash}p{2.4cm} 
>{\centering\arraybackslash}p{2.4cm} 
>{\centering\arraybackslash}p{2.4cm} 
}
\hline\hline
$\ell$  & $E_{\text{c}}$(\nuc{50}{Ti}) [MeV] & $E_{\text{c}}$(\nuc{51}{Ti}) [MeV] & $\Delta E_c$ [MeV] \\
\hline
1 & 5.24(3) & 1.05(9) & 4.19(9) \\
2 & 8.62(3) & 4.40(2) & 4.23(4) \\
3 & 8.20(4) & 3.74(12) & 4.46(12) \\
4 & 8.17(6) & 3.90(3) & 4.27(7) \\
\hline
\textbf{Average} &  &  & \textbf{4.29(4)} \\
\hline\hline
\end{tabular}
\label{tab:pairing_centroids}
\end{table}

Typically, one observes an energy gap between the ground state and the bulk of excited states in even-even nuclei, with only collective states dropping to energies lower than that gap. This gap corresponds to the energy required to create configurations with one broken pair and is referred to as two-quasiparticle (2qp) energy. For breaking a neutron pair, we identify this energy with the two-neutron pairing gap, $2\Delta_{n}$. The excitation-energy spectrum of odd-$A$ nuclei does typically not show this gap as particles can be easily excited to other orbits without having to break a pair. The neutron pairing gap, $\Delta_n$, can be calculated using the empirical three-point formula \cite{Bohrbook, Cha15a}:

\begin{equation}
    \Delta_n(N) = -1/2[S_n(N+1, Z) - S_n(N,Z)]
    \label{eq:gap}
\end{equation}

 In Eq.\,(\ref{eq:gap}), $S_n$ denotes the neutron separation energy. For \nuc{50}{Ti} ($S_n=10.939$ MeV \cite{Mass50DataSheet}) and \nuc{51}{Ti} ($S_n=6.372$ MeV \cite{Mass51DataSheet}), this yields $\Delta_n = 2.28$ MeV and, consequently, $2\Delta_n=4.57$ MeV for the empirical two-neutron pairing gap in \nuc{50}{Ti}. This number is close to the centroid-energy differences calculated for the different $\ell$ transfers. In a nutshell, one indeed needs to break a pair in the \nuc{50}{Ti} ground state first to create a neutron 1p-1h excitation to any of the orbitals above the $N=28$ shell closure, which requires an energy of $2\Delta_n$. The small difference between the empirical value of $2\Delta_n$ and the centroid-energy differences $\Delta E_c$ -- in addition to the differences between the $\Delta E_c$ values themselves -- might either hint at a variable gap or, more probably, the scenario that the position of the single-particle orbits is slightly different between \nuc{50}{Ti} and \nuc{51}{Ti}.

Finally, we look at the energy differences between the different orbitals in \nuc{50}{Ti} and compare these to the corresponding differences in \nuc{51}{Ti}. Similar comparisons were presented in our previous work for the $N=29$ isotones\,\cite{Riley_51Ti, Riley_55Fe, Riley_53Cr, spi24a, Spi25a}.  In \nuc{51}{Ti}, the $N=32$ subshell gap, which corresponds to the separation between the spin-orbit partners $2p_{3/2}$ and $2p_{1/2}$, was determined to be $\Delta SO = 1459(365)$\,keV \cite{Riley_51Ti}. The rather large uncertainty arises from several fragments at higher excitation energy that could not be assigned unambiguously to either $p$ orbital due to the lack of firm $J^{\pi}$ assignments. Since the $2p$ spin–orbit partners could not be separated experimentally in \nuc{50}{Ti}, the spin-orbit splitting determined in Ref.\,\cite{Riley_51Ti} for \nuc{51}{Ti} was adopted and applied to the measured $2p$ centroid in \nuc{50}{Ti} to calculate the centroid energies for the $2p_{3/2}$ and $2p_{1/2}$ orbitals. This assumes that the splitting is the same for \nuc{50}{Ti} and \nuc{51}{Ti}. In doing so, centroid energies of $4505(217)$\,keV for the $2p_{3/2}$ orbital and $5965(217)$\,keV for the $2p_{1/2}$ orbital were obtained. Having these values, we can now compare the gaps between the pseudo-spin partners, i.e., $2p_{3/2}$ and $1f_{5/2}$, and between the $2p_{1/2}$ and $1f_{5/2}$ orbitals. The resulting energy gaps are $\Delta_{pf} = E(1f_{5/2})-E(2p_{1/2})=2031(217)$ keV and $\Delta PS = E(1f_{5/2})-E(2p_{3/2})=3491(217)$ keV. These gaps agree, within uncertainties, with those determined for \nuc{51}{Ti}. The gaps in \nuc{51}{Ti} are $\Delta_{pf} = 1712(216)$\,keV and $\Delta PS = 3171(149)$\,keV \cite{Riley_51Ti}. The gaps determined for \nuc{50}{Ti} leave room for the scenario mentioned above that the orbitals' energies actually change slightly, with the gaps getting smaller with increasing neutron number.

\subsection{Comparison to Theoretical Models}
\label{sec:Theory_comparison}

Being aware of the ongoing discussion of the basis dependence and suggested non-observability of spectroscopic factors, calculations were performed with three independent, state-of-the-art nuclear-structure models. As mentioned earlier, these were the time-dependent continuum nuclear shell model (TDCSM) \cite{Vol09a, Vol14a} using the recently developed cross-shell FSU interaction \cite{Lub19a, Lub20a}, the energy-density functional theory plus quasiparticle-phonon model approach (EQPM) including up to three-phonon contributions \cite{Tsoneva2016}, and the also recently developed relativistic equation of motion theory including the coupling of two quasiparticles with up to two phonons (REOM$^3$) \cite{Lit19a, Lit22a, Lit23a}. All three approaches are beyond mean-field theories even though all of them start from a mean-field potential. One of the main differences between the shell model and the density functional theory based models, i.e., the EQPM and REOM$^3$, are the configuration spaces considered. For the purpose of this work, the current shell-model results are effectively restricted to the $fp$ valence space. The use of the FSU interaction implies, however, that the TDCSM configuration space could in principle include the full $s$–$p$–$sd$–$fp$ valence space \cite{Lub19a, Lub20a}. Consequently, particle-hole excitations involving these shells are accessible. In this work, cross-shell mixing with shells below the $fp$ shells was not considered though. In contrast, the relativistic mean field basis for the REOM$^3$ spans the single-quasiparticle states with angular momenta up to 41/2 in both the Fermi (particle) and Dirac (antiparticle) sectors\,\cite{Lit19a}. Consequently, and of interest to this work, single-particle orbitals above the $fp$ shell are part of the model space. It must be stated though that, unlike the TDCSM, both the EQPM and REOM$^3$ do not include ground-state correlations. This means that the \nuc{50}{Ti} ground state has no vacancy in the neutron $1f_{7/2}$ orbital in these models and that, consequently, the \nuc{49}{Ti} ground state is described as a pure $(1f_{7/2})^{-1}$ neutron-hole configuration. In the following, we will briefly comment on how spectroscopic factors were calculated in the EQPM, the REOM$^3$, and the TDCSM.

For the QPM, the one-, two-, and three-phonon excitations are built from a set of phonons calculated at the level of the quasiparticle random phase approximation (QRPA)\,\cite{Tsoneva2016, Soloviev}. The QRPA phonons are defined by Eq.\,(\ref{eq:qrpa_phonon}):

\begin{equation}
    Q^+_{\lambda \mu i } = \frac{1}{2}\sum_{j_1j_2}\left[\psi^{\lambda i}_{j_1 j_2}A^+_{\lambda \mu}(j_1j_2) - \varphi^{\lambda i}_{j_1j_2}\tilde{A}_{\lambda \mu}(j_1j_2) \right]
    \label{eq:qrpa_phonon}
\end{equation}

In Eq.\,(\ref{eq:qrpa_phonon}), $i$ labels the number of the QRPA state, $j \equiv (nljm\tau)$ corresponds to a single-particle proton or neutron state, and $A^+_{\lambda \mu}$ and $\tilde{A}_{\lambda \mu}$ are time-forward and time-backward operators. These operators couple 2qp creation and annihilation operators to a total angular momentum $\lambda$ with projection $\mu$ by means of the Clebsch-Gordon coefficients $C^{\lambda \mu}_{j_1 m_1 j_2 m_2}= \langle j_1 m_1 j_2 m_2 | \lambda \mu \rangle$. The excitation energies of the QRPA phonons and the time-forward $\psi^{\lambda i}_{j_1 j_2}$ and time-backward $\varphi_{j_1 j_2}^{\lambda i}$ amplitudes are determined by solving QRPA equations\,\cite{Soloviev}. 

For spherical even-even nuclei, the QRPA phonons defined by Eq.\,(\ref{eq:qrpa_phonon}) build the basis of the QPM wavefunction, $\Psi_\nu(JM)$: 

\begin{align}
\label{eq:QPM_wavefunction}
\Psi_\nu(JM) = 
\Biggl\{ &
\sum_{i} R_i(J_{\nu}) Q^{+}_{JMi} 
\\
&+ \sum_{\substack{\lambda_1 i_1 \\ \lambda_2 i_2}}
P^{\lambda_1 i_1}_{\lambda_2 i_2}(J_{\nu})
\left[ 
Q^{+}_{\lambda_1 \mu_1 i_1} 
\otimes 
Q^{+}_{\lambda_2 \mu_2 i_2} 
\right]_{JM} \nonumber \\ 
&+ \sum_{\substack{\lambda_1 i_1 \lambda_2 i_2 \\ \lambda_3 i_3 I}}
T^{\lambda_1 i_1 \lambda_2 i_2 I}_{\lambda_3 i_3}(J_{\nu}) \Biggl[
\Big[ 
Q^{+}_{\lambda_1 \mu_1 i_1} \nonumber \\
&\otimes 
Q^{+}_{\lambda_2 \mu_2 i_2} \Big]_{IK}
\otimes 
Q^{+}_{\lambda_3 \mu_3 i_3}
\Biggr]_{JM}
\Biggr\} \Psi_0 \nonumber
\end{align}

In Eq.\,(\ref{eq:QPM_wavefunction}), $R$, $P$, and $T$ are the one-, two-, and three-phonon amplitudes, $\nu$ labels the individual excited states, and $\Psi_0$ is the phonon vacuum state \cite{Tsoneva2016, Soloviev}. As outlined in Ref.\,\cite{Sol81a}, the 2qp component, i.e., the 1p-1h component $(j_1,j_2)$ coupling to spin $J$ in $\Psi_\nu(JM)$ is defined by:

\begin{equation}
    \Phi_{j_1 j_2}(J;\eta_{\nu})= \frac{1}{2} \left| \sum_i R_i(J_{\nu})\psi^{\lambda i}_{j_1 j_2}\right|^2,
    \label{eq:2qp-comp}
\end{equation}

\noindent where $\eta_{\nu}$ corresponds to the energy of the QPM state. For the $(d,p)$ reaction placing a neutron in the orbital $j$ and if we identify the hole of the odd-$A$ target nucleus in orbital $j_0$, the spectroscopic factor is given by \cite{Sol81a}:

\begin{equation}
    S_{jj_0}(J;\eta_{\nu}) = C^2_{j_0 \nu_0} U^2_j \Phi_{jj_0}(J;\eta_{\nu}),
\end{equation}

\noindent where $U_j$ is a Bogolubov transformation coefficient and $C_{j_0 \nu_0}$ the one-quasiparticle amplitude for the odd-$A$ target nucleus wavefunction neglecting quasiparticle-plus-one and two-phonon terms. For \nuc{49}{Ti}, this is equivalent to assuming that its ground state can be considered as a pure neutron hole in the $1f_{7/2}$ orbital. As mentioned in Ref.\,\cite{Wei21a}, some of these factors can be combined into one-quasiparticle occupation numbers for a given orbit and be calculated by solving BCS equations. As shown in Ref.\,\cite{Sol81a}, a 2qp state obtained at the QRPA level typically fragments over several QPM states. For more details, see Refs.\,\cite{Soloviev, Sol81a}.


For the REOM$^3$, the spectroscopic factors constitute the residues of the two-body Green's function, which describes the propagation of two nucleons within the correlated medium. The two-time two-fermion particle-hole propagator, associated with the response function, is comprised of the time-dependent nucleonic field operators shown as \cite{Lit19a},

\begin{align}
\label{response_t}
R(12,1'2') &\equiv R_{12,1'2'}(t-t') \\
           &= -i\langle T(\psi_1^\dagger\psi_2)(t) (\psi_{2'}^\dagger\psi_{1'})(t')\rangle \nonumber \\
& = -i\langle T\,\psi^\dagger(1)\psi(2)\psi^\dagger(2')\psi(1')\rangle \nonumber,
\end{align}

\noindent where $t_1=t_2=t$, $t_1'=t_2'=t'$, and the
subscripts label the complete set of single-nucleon
quantum numbers in an arbitrary representation for the one-nucleon fields $\psi_1$ and $\psi_2$. The Fourier transformation to the energy domain leads to the spectral expansion \cite{Lit19a},

\begin{equation}
R_{12,1'2'}(\omega)
= \sum_{\nu > 0}
\left[
\frac{\rho_{21'}^\nu \rho_{2'1}^{\nu *}}{\omega - \omega_\nu + i\delta}
-
\frac{\rho_{12'}^{\nu *} \rho_{1'2}^\nu}{\omega + \omega_\nu - i\delta}
\right] ,
\label{response_E}
\end{equation}
where the matrix elements
\begin{equation}
    \rho_{12}^{\nu} = \langle 0|\psi_2^\dagger \psi_1| \nu \rangle = \langle \nu|\psi_1^\dagger \psi_2| 0 \rangle^{\ast}
    \label{TrDen}
\end{equation}
are, by construction, the transition densities between the correlated, formally exact, ground $|0\rangle$ and excited $|\nu\rangle$ states, and the summation runs over all excited states, enabling non-zero values of $\rho_{12}^{\nu}$.
The latter represent the weights of the particle-hole configurations, built on top of the ground state $|0\rangle$, in the excited state $|\nu\rangle$, i.e., has the meaning of the two-fermion spectroscopic factors in the basis associated with the number indices $\{1\}$. The poles of $R(\omega)$ are the energies of the excited states, $\omega_{\nu} = E_{\nu} - E_0$. In the context of this work, we track the specific components of $\rho_{12}^{\nu}$ associated with the neutron hole in the $1f_{7/2}$ shell represented by $\psi_2$ and transferring a neutron to the orbits above the Fermi energy by $\psi^{\dagger}_1$ \cite{Lit19a}, while $|0\rangle$ and $|\nu\rangle$ quantify the many-body states of the reference nucleus $^{50}$Ti. In this work, $|0\rangle$ is approximated by the relativistic Hartree-Bogoliubov superfluid mean field \cite{Ring1996}, and the states $|\nu\rangle$ are represented by superpositions of two-quasiparticle plus two-phonon (2qp$\otimes$2phonon) configurations on top of $|0\rangle$.
Superfluid pairing is included in the proton open-shell subsystem of $^{50}$Ti, to which a superfluid version of REOM$^3$ is applied. In this case, Eqs. (\ref{response_t}) through (\ref{TrDen}) are reformulated in the basis of Bogoliubov quasiparticles \cite{Lit22a}.


\begin{figure*}[t!]
\includegraphics[width=\linewidth]{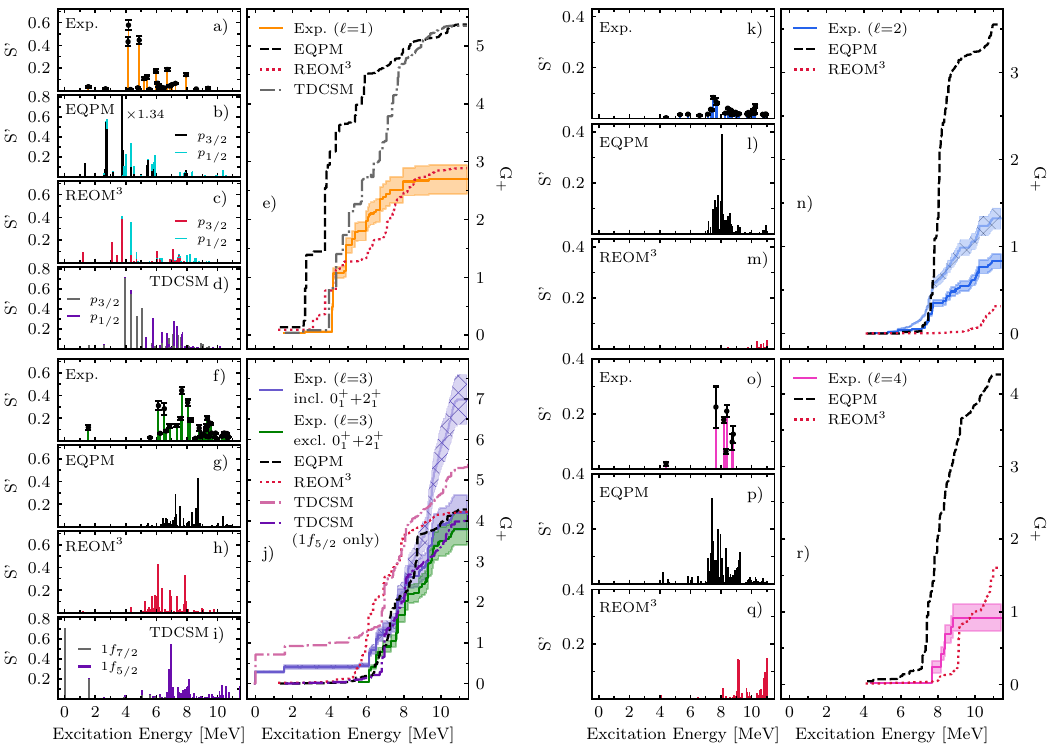}
\caption{Spin-weighted spectroscopic factors, $S'$, and running sum of these, as defined in Eq.\,(\ref{eq:stripping}) and yielding the observed vacancy of a single-particle orbital. Each block of four panels corresponds to one specific angular momentum, $\ell$, transfer. (a–e) $\ell=1$, (f–j) $\ell=3$, (k–n) $\ell=2$, and (o–r) $\ell=4$. For the odd-$\ell$ transfers, the left four panels display the fragmentation of strength observed in $(d,p)$ [symbols with error bars] and predictions from the EQPM, REOM$^3$, and TDCSM. The right panel in each block displays the corresponding running sums of the spin-weighted spectroscopic factors [vacancies/sum rule] (lines with associated color coding). For the TDCSM, the running sum for the combined $1f_{7/2} + 1f_{5/2}$ strength (dash-dotted dark pink) and the $1f_{5/2}$-only contribution (dash-dotted purple) are shown in panel (j).(k-r) same comparison but for even-$\ell$ transfers. Only predictions from the EQPM and REOM$^3$ are compared to data. The associated running sums are shown in panels (n) and (r). In panels (e), (j), (n), and (r), experimental uncertainties are presented with shaded bands. The hatched band for the $\ell=2$ and $\ell=3$ transfers in panels (j) and (n) include strength from unresolved strength, see discussion in Section \ref{sec:unresolved}. The running sums are extended just past $S_n$ up to an excitation energy of $E_{x}=11.5$ MeV.}
\label{fig:50Ti_dp_theory_comparison}
\end{figure*}

Some additional words of caution are necessary before continuing with the discussion. In the current implementation, the EQPM and REOM$^3$ understand the ``spectroscopic factors'' as the weights of specific neutron 1p-1h excitations with a hole in the $1f_{7/2}$ neutron orbital built on top of the \nuc{50}{Ti} ground state, as highlighted in the discussion pertaining to Eq.\,(\ref{TrDen}). Specifically, the spectroscopic factor of Eq.\,(\ref{TrDen}) might be understood as a two-nucleon spectroscopic factor. In principle, this is fundamentally different from the one-nucleon spectroscopic factors calculated with the TDCSM, where the overlap of the excited state in \nuc{50}{Ti} with the $\nuc{49}{Ti}+n$ system is calculated after removing a neutron from the considered single-particle orbital. Here, the spectroscopic factor for neutron removal, $S_j^{-}(J_i \rightarrow J_f)$, is calculated according to

\begin{align}
    S_j^{-}(J_i \rightarrow J_f) &= \frac{1}{2J_i+1}\sum_{m,M_f,M_i} \left| \langle J_fM_f|a_{jm}|J_iM_i \rangle\right|^2 \nonumber \\
    &= \frac{1}{2J_i+1}\left| \langle J_f \|a_{j} \| J_i \rangle\right|^2
    \label{eq:tdcsm_s}
\end{align}

\noindent where the initial state $|i,J_iM_i\rangle$ is an excited state in \nuc{50}{Ti}, the final state $|f,J_fM_f\rangle$ corresponds to the \nuc{49}{Ti} ground state, and where $a_{jm}$ is an annihilation operator removing a neutron from an orbital with quantum numbers $(jm)$ in \nuc{50}{Ti}. Experimentally, the neutron addition spectroscopic factor, $S^+_j(J_i(\nuc{49}{Ti}) \to J_f(\nuc{50}{Ti}))$, was determined. The two quantities are connected to each other as more explicitly discussed in Appendix \ref{appendix}, 

\begin{align}
S_j^+(\nuc{49}{Ti} \to \nuc{50}{Ti}) = \frac{2J_f(\nuc{50}{Ti})+1}{2J_i(\nuc{49}{Ti})+1} S^-_j(\nuc{50}{Ti} \to \nuc{49}{Ti}),
\label{eq:text_connection}
\end{align}

\noindent where $J_i(\nuc{49}{Ti)} = 7/2^-$ for the ground state of \nuc{49}{Ti} and $J_f(\nuc{50}{Ti)}$ corresponds to the total angular momentum of the excited state considered (or calculated) in \nuc{50}{Ti}. As can be seen from Eq.\,(\ref{eq:text_connection}), $S^{-}_j\left(\nuc{50}{Ti} \to \nuc{49}{Ti}\right)$ is equivalent to the spin-weighted spectroscopic factor, $S'$, defined in Eq.\,(\ref{eq:specfac}) and which is part of the sum rule of Eq.\,(\ref{eq:stripping}).

Even though, spectroscopic factors are calculated differently in the models, the comparison between the models still leads to meaningful results since, as we will highlight, also the TDCSM expects the \nuc{49}{Ti} ground state to be dominated by a $(1f_{7/2})^{-1}$ neutron-hole configuration. Then, spectroscopic factors for neutron 1p-1h configurations like $(1f_{7/2})^{-1}(2p_{3/2})^{+1}$ predicted by the EQPM and REOM$^3$ can be compared to the predictions of the TDCSM for the overlap $\langle \nuc{49}{Ti}(\mathrm{g.s})+(\nu:2p_{3/2})|\nuc{50}{Ti}(E_x,J^{\pi}) \rangle$.

The theoretical predictions for the spin-weighted spectroscopic factors, $S'$, and vacancies (sum rule), $G_{lj,+}$, are compared to the experimentally determined quantities in Fig.\,\ref{fig:50Ti_dp_theory_comparison} and Table\,\ref{tab:occupancies}. In the TDCSM, EQPM, and REOM$^3$, spectroscopic factors were calculated for all possible final states $J_f$ in \nuc{50}{Ti} that could be reached in the $(d,p)$ reaction by populating a specific neutron 1p-1h component in their wavefunction. All models calculated the $\ell =1$ strength for placing a neutron into either the $2p_{3/2}$ or $2p_{1/2}$ orbital, and the $\ell =3$ strength for placing a neutron into the $1f_{5/2}$ orbital. For the TDCSM calculations, we also considered $1f_{7/2}$ ($\ell =3$) neutron-adding strength as, in principle, the $1f_{7/2}$ could have more than one vacancy. Unlike the TDCSM with the FSU interaction, the EQPM and REOM$^3$ were not restricted to the $fp$ valence space. Therefore, comparisons for the $(1f_{7/2})^{-1}(1g_{9/2})^{+1}$ ($\ell =4$) and $(1f_{7/2})^{-1}(2d_{5/2})^{+1}$ ($\ell =2$) neutron 1p-1h strengths were also possible. However, only states with $J^{\pi} \leq 6^-$ could be calculated in the EQPM because of the restricted phonon basis.

As we were not able to discriminate between spin-orbit partners experimentally, the neutron $2p_{3/2}$ and $2p_{1/2}$ neutron-adding strengths (spin-weighted spectroscopic factors) were grouped together as $\ell =1$, and the $1f_{7/2}$ and $1f_{5/2}$ neutron-adding strengths as $\ell =3$ strength in Figs.\,\ref{fig:50Ti_dp_theory_comparison}\,(a)-(j). However, one can make an argument that most of the $\ell=3$ strength at higher excitation energies should be due to the $(1f_{7/2})^{-1}(1f_{5/2})^{+1}$ neutron 1p-1h configuration, i.e., due to transferring a neutron into the $1f_{5/2}$ orbital. In fact, and, thus, partly supporting the experimentally based arguments made above, the TDCSM predicts that most of the remaining $1f_{7/2}$ neutron strength is concentrated in the $0^+_1$ ground state $(S' = 0.71)$ and the $2^+_1$ state at 1.59\,MeV $(S' = 0.21)$. Only considering these two states leads to $G_{1f_{7/2},+} = 0.92$. Compared to the experimental value of 0.40(4), this corresponds to a quenching of 43(4)\,\% in agreement with the empirically determined quenching of single-particle strength \cite{Kay13a, Kay_A15_SF, Tos21a, Aum21a, Mac25a}. For the total vacancy of the $1f_{7/2}$ neutron orbital, the TDCSM predicts $G_{1f_{7/2},+}=1.32$, meaning that there is more than one vacancy but that it is not significantly more. The corresponding additional spectroscopic strength is fragmented over many excited states. The predicted spin-weighted spectroscopic factors, corresponding to the neutron-removal spectroscopic factors as shown in the discussion around Eq.\,(\ref{eq:text_connection}) and in Appendix\,\ref{appendix}, are very small. If this prediction is correct, then we would not observe this remaining $1f_{7/2}$ strength in resolved peaks. In addition and in agreement with the experimental observations, the predicted neutron-addition spectroscopic factors, associated with placing a neutron in the $1f_{7/2}$ orbital, are small for the $4^+_1$ and $6^+_1$ states.

\begin{table}[t]
\centering
\caption{
Vacancies, $G_{lj,+}$, for the neutron orbitals in \nuc{50}{Ti} discussed in the text. Experimentally determined vacancies are compared to predictions from EQPM, REOM$^3$, and TDCSM up to $E_x = 11.5$\,MeV, i.e., to an energy slightly above $S_n$. For the TDCSM predictions, contributions of both the $1f_{7/2}$ $(G_+ = 1.3)$ and $1f_{5/2}$ $(G_+ = 4.2)$ orbitals to the observed $\ell =3$ strength (vacancy) are considered. The predicted vacancies for the EQPM and REOM$^3$ only consider the $1f_{5/2}$ contribution. For the $p$ orbitals, contributions from the $2p_{1/2}$ (first summand) and $2p_{3/2}$ (second summand) to the overall vacancy are given for all models.}
\label{tab:occupancies}
\renewcommand{\arraystretch}{1.25}

\begin{tabular}{
>{\centering\arraybackslash}p{1.55cm}
>{\centering\arraybackslash}p{1.6cm}
>{\centering\arraybackslash}p{1.6cm}
>{\centering\arraybackslash}p{1.6cm}
>{\centering\arraybackslash}p{1.6cm}
}
\hline\hline
Orbital & Exp. & EQPM & REOM$^3$ & TDCSM \\
\hline
$2p$          & $2.7(3)$ & $5.3$ & $2.9$ & $5.3$ \\
 & & $(1.7+3.6)$ & $(1.0+1.9)$ & $(1.8+3.5)$ \\
$1f$    & $4.2(4)$ & $4.3$ & $4.3$ & $5.5$ \\
 & & $(1f_{5/2})$ & $(1f_{5/2})$ & $(1.3+4.2)$ \\
$1g_{9/2}$    & $0.9(2)$ & $4.3$ & $1.6$ & \multicolumn{1}{c}{--} \\
$2d_{5/2}$    & $0.83(8)$ & $3.6$ & $0.3$ & \multicolumn{1}{c}{--} \\
\hline\hline
\end{tabular}
\end{table}

\begin{figure}
    \centering
    \includegraphics[width=0.9\linewidth]{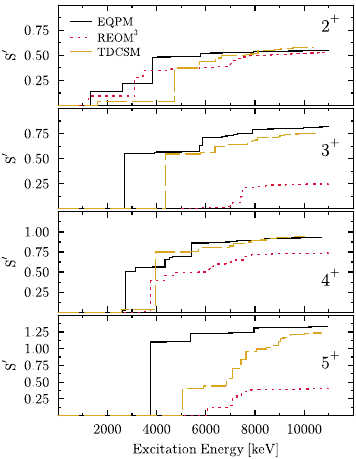}
    \caption{Spin-weighted spectroscopic factors, $S'$, as defined in Eq.\,(\ref{eq:specfac}) for the $2p_{3/2}$ neutron-adding strength. States with total angular momentum $J^{\pi} = 2^+, 3^+, 4^+$, and $5^+$ can be reached in the $(d,p)$ reaction for the $(1f_{7/2})^{-1}(2p_{3/2})^{+1}$ neutron 1p-1h configuration. Their individual contributions to the total predicted $2p_{3/2}$ vacancy are shown.}
    \label{fig:sf_frag}
\end{figure}

\begin{figure*}[t]
    \centering
    \includegraphics[width=1.0\linewidth]{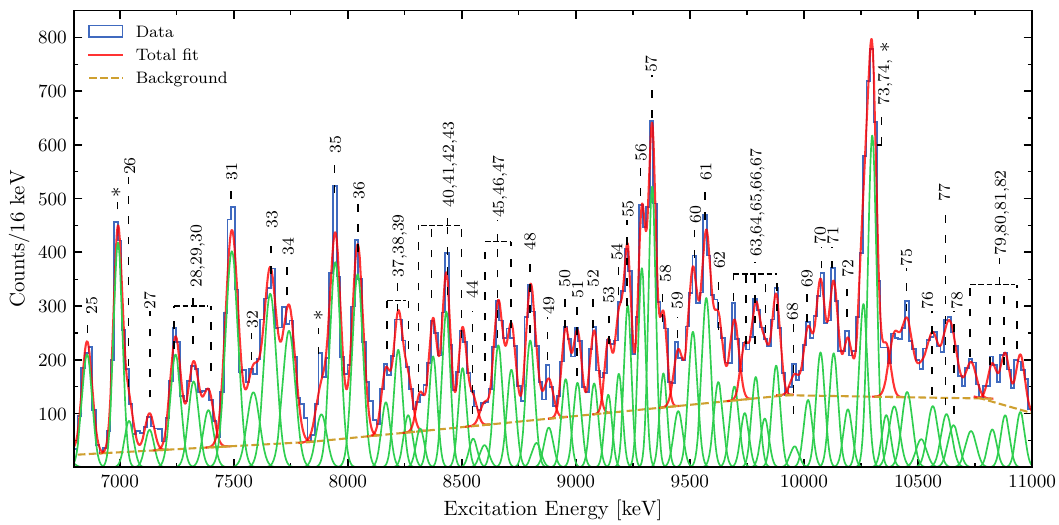}
    \caption{$(d,p)$ spectrum measured at $\theta_{SE-SPS} = 20^{\circ}$. The excitation-energy range between 6750\,keV and 11000\,keV is shown. The total fit (red line) resulting from the superposition of fits to individual peaks (green lines) and a smooth background (orange, dashed line) is shown. Individual peaks were fitted with a Gaussian fit template. The width of the peaks remained nearly constant over the length of the focal plane. States are highlighted with the numbering presented in Table\,\ref{tab:EnergyTable} and Figs.\,\ref{fig:AngDist_1stplot}-\ref{fig:AngDist_4thplot}. Contaminants  are presented with asterisks.}
    \label{fig:spectrum_02}
\end{figure*}

We now focus further on the $\ell =1$ and $\ell =3$ strength, see Fig.\,\ref{fig:50Ti_dp_theory_comparison}\,(a)-(j). All three models predict a high level of fragmentation of the spectroscopic strengths in agreement with experiment. In addition, all model predictions agree reasonably well with the overall strength distribution including the location of where most of the strength is observed. For $\ell =1$, the REOM$^3$ captures the observed quenching of the spectroscopic strength, here presented as vacancies in Fig.\,\ref{fig:50Ti_dp_theory_comparison}\,(e) and Table\,\ref{tab:occupancies}. The shell model and EQPM overshoot the experimental data significantly, i.e., both predict a vacancy of 5.3 leading to the usual quenching factor of 0.51(6) compared to the experimental data for states up to $E_x = 11.5$\,MeV. For the $\ell =3$ strength, a quenching factor of 0.76(7) is obtained when comparing experimental data to shell-model predictions up to $E_x = 11.5$\,MeV. This is less quenching than for the $\ell =1$ strength, which appears consistent with observations in the $N=29$ isotones \cite{Spi25a}. Subtracting the spin-weighted spectroscopic factors for the $0^+_1$ and $2^+_1$ states from the entire experimentally determined $\ell =3$ sum rule gives $G_+ = 3.8(4)$. Note that this is strength coming from experimentally resolved states, i.e., states that showed up as discrete states (peaks) in our $(d,p)$ spectra. We will discuss unresolved strength shortly. When comparing Figs.\,\ref{fig:50Ti_dp_theory_comparison}\,(f) and (i), we see that this resolved $\ell =3$ strength is expected to result mostly from placing a neutron in the $1f_{5/2}$ orbital. Comparing this value to the predicted vacancy for the $1f_{5/2}$ orbital, when considering states up to $E_x = 11.5$\,MeV, leads to quenching factors of 0.90(10) for the TDCSM and 0.88(9) for the EQPM and REOM$^3$. Surprisingly, that would mean that we picked up almost all predicted $1f_{5/2}$ neutron-adding strength up to $E_x = 11.5$\,MeV and that different from the $\ell=1$ strength all three model predictions agree in terms of the accumulated strength. Note though that the experimental data for \nuc{50}{Ti} might still include some strength coming from transferring a neutron into the $1f_{7/2}$ orbital.

The findings mentioned above beg the question why we observe agreement for the $\ell =3$ strength but not for the $\ell=1$ strength between the models. We, thus, decided to take a closer look at the individual contributions of the $2p_{3/2}$ and $2p_{1/2}$ strengths to the overall $2p$ vacancy. The predicted vacancies were added to Table\,\ref{tab:occupancies} for all three models and the spin-weighted spectroscopic factors were added to Figs.\,\ref{fig:50Ti_dp_theory_comparison}\,(b) to (c) with associated color coding. As can be seen in Table\,\ref{tab:occupancies}, the model predictions are quite different for both orbitals. For the $2p_{1/2}$ orbital, the REOM$^3$ predicts 50\,\% of the expected strength to be collected up to 11.5\,MeV, the EQPM 85\,\%, and the TDCSM 90\,\%. For the $2p_{3/2}$ orbital, the TDCSM and EQPM predict that 88\,\% and 90\,\% of the strength gets collected up to 11.5\,MeV, while the REOM$^3$ expects 48\,\%. Differences between the model predictions, thus, originate from varying predictions for both $2p$ orbitals, with the TDCSM and EQPM expecting that almost all strength gets picked up.

\begin{figure}[t!]
\includegraphics[width=\linewidth]{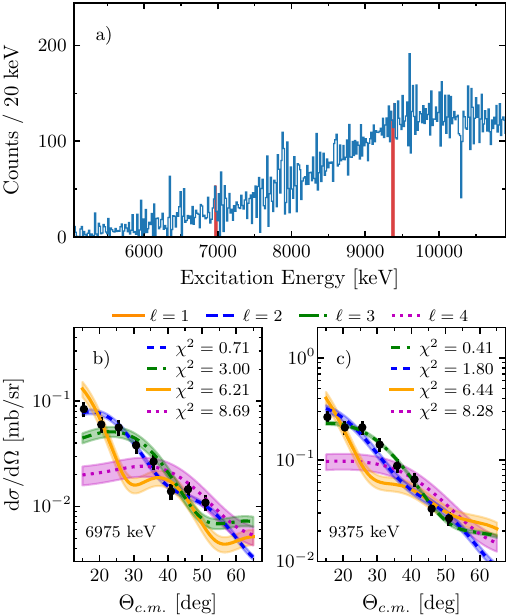}
\caption{(a) Extracted ``background'' spectrum for the $(d,p)$ spectrum taken at $\theta_{\mathrm{SPS}}=25^{\circ}$, with regions of interest highlighted by the red-shaded areas. (b) and (c) two examples of $(d,p)$ angular distributions extracted from the background at the indicated bin centers, with best-fit $\chi^2$ values shown. The color scheme used to identify the different $\ell$ transfers is indicated. See text for further discussion.}
\label{fig:Bg_AngDist}
\end{figure}

As mentioned throughout this manuscript, we cannot distinguish between spin-orbit partners. However, we can compare details of the $2p$ strength distribution to the predictions at lower energies where the density of states is lower. With respect to $\ell =1$ transfers, the three most strongly populated states are $J^{\pi} = 3^+$, $4^+$, and $5^+$ states at $E_x = 4147$, 4172, and 4881\,keV, respectively, see Fig.\,\ref{fig:50Ti_dp_theory_comparison}\,(a) and Table\,\ref{tab:EnergyTable}. The TDCSM predicts a $3^+$ state at $4358$\,keV with $S'_{2p_{3/2}}= 0.55$, a $4^+$ state at $3968$\,keV with $S'_{2p_{3/2}}= 0.71$, and a $5^+$ state at $5050$\,keV with $S'_{2p_{3/2}}= 0.41$ in remarkable agreement with experiment. For the $3^+$ and $5^+$ states, i.e., the unnatural-parity states, the predicted values agree with the experimentally determined values within uncertainties. For the $4^+$ state, the strength is quenched (reduced), i.e., $R = 0.43(4)/0.71 = 0.61(6)$. The other excited state with an appreciable $S'_{2p_{3/2}}= 0.33$ value is a $2^+$ state predicted at 4713\,keV. We did not observe a single $2^+$ state with such a large spectroscopic factor. The $2^+$ state, which we observed at 4790\,keV, has a spin-weighted spectroscopic factor of $S' = 0.027(3)$. It appears that for firmly assigned $2^+$ states the $2p_{3/2}$ spectroscopic strength is significantly more fragmented (or quenched) than predicted by the TDCSM. To evaluate whether the full strength is quenched, we would need to identify more $2^+$ states experimentally. The EQPM and REOM$^3$ also predict another $2^+$ state with a sizable spectroscopic factor of $S' = 0.27$ at 3.8\,MeV and $S' = 0.19$ at 3.1\,MeV, respectively, in addition to the $2^+_1$ state. As stated, there is no experimental equivalent though. Furthermore, it needs to be stressed that our data do not support a large $S'_{2p_{3/2}}$ value for the $2^+_1$ state. In agreement with experiment, the TDCSM does not predict a large value for the $2^+_1$ state. In conflict with experiment, the REOM$^3$ does not predict a single $3^+$ or $5^+$ state with sizable spectroscopic factors at energies close to the experimental states. For the case of the REOM$^3$, an excited $4^+$ state with $S'_{2p_{3/2}} = 0.38$ is, however, predicted at 3.76\,MeV in excellent agreement with experiment. The EQPM predicts the $4^+$ state with the largest spin-weighted spectroscopic factor, $S'_{2p_{3/2}}= 0.33$, at 2.7\,MeV, i.e., too low in energy when compared to experiment. The experimentally observed $4^+$ state at 2.7\,MeV has $S'_{2p_{3/2}}= 0.0189(27)$ (see Table\,\ref{tab:EnergyTable}). A $4^+$ state with $S_{2p_{1/2}}'=0.24$ is predicted at 4.33\,MeV by the EQPM in reasonable agreement with the experimentally observed state at 4147\,keV. However, the proximity to the 4172-keV $3^+$ state suggests that the two states should have a similar structure, i.e., a strong $(1f_{7/2})^{-1}(2p_{3/2})^{+1}$ component. For the neutron-adding strength to the $2p_{3/2}$ orbital, the predictions for the different values of total angular momentum, $J$, and their contributions to the total $2p_{3/2}$ vacancy are summarized in Fig.\,\ref{fig:sf_frag}. As can be seen, the main contributor to the differences between the TDCSM, the EQPM, and the REOM$^3$, seen in Fig.\,\ref{fig:50Ti_dp_theory_comparison} and Table\,\ref{tab:occupancies}, are indeed the unnatural-parity states. It is also obvious that the strongest fragments for $J= 3$, 4, and 5 are systematically weaker in the REOM$^3$ than in the TDCSM or EQPM. The strongest fragments for the unnatural-parity states are too high in energy for the REOM$^3$ and too low in energy for the EQPM, see Fig.\,\ref{fig:sf_frag}, when compared to experiment.

To conclude this part of the discussion, while the REOM$^3$ might appear to provide a better description of the accumulated spectroscopic strength for the $2p$ orbitals, it lacks at the moment the level of detail that the TDCSM and EQPM provide for the specific single-particle structure of some selected states discussed here. Thus, although the REOM$^3$ reproduces the reduction of the spectroscopic strength associated with the $2p$ neutron orbitals below 11.5 MeV, substantial discrepancies in the detailed structure persist, and the apparent agreement in total strength should be interpreted with caution. Further investigations are clearly needed, which should also focus on how spectroscopic strength gets distributed among states with different total angular momenta (spins).

In the last part of this discussion, we turn to the $\ell = 4$ and $\ell =2$ strengths and their comparison to predictions made with the EQPM and REOM$^3$, see Figs.\,\ref{fig:50Ti_dp_theory_comparison}\,(k)-(r). As mentioned in the earlier sections, experimentally, significantly less of the expected sum rule was accumulated up to $S_n$ than for the $fp$ orbitals. Furthermore, the predictions for the $\ell = 4$ and $\ell = 2$ strengths are remarkably different. For instance, the EQPM predicts that 60\,$\%$ of the sum rule is collected for the $(1f_{7/2})^{-1}(2d_{5/2})^{+1}$ neutron 1p-1h configuration up to energy slightly above $S_n$, while the REOM$^3$ expects only about 5\,$\%$, see Fig.\,\ref{fig:50Ti_dp_theory_comparison}\,(n) and Table\,\ref{tab:occupancies}. Similar observations are made for the $(1f_{7/2})^{-1}(1g_{9/2})^{+1}$ ($\ell =4$) strength, even though the REOM$^3$ prediction for the exhaustion of the sum rule agrees reasonably with the experimentally determined vacancy, see Fig.\,\ref{fig:50Ti_dp_theory_comparison}\,(r) and Table\,\ref{tab:occupancies}. In the REOM$^3$, a substantial fraction of the strength is shifted to higher excitation energies, with approximately 40\,\% of the expected $2d_{5/2}$ strength being collected up to 15\,MeV, which is still less than predicted by the EQPM. For the $1g_{9/2}$ orbit, around 40\,$\%$ of the sum rule is reached at about 20\,MeV. Given the obvious differences between the models and the significantly lower exhaustion of the sum rule, we cannot comment on the quenching of the strength for the $1g_{9/2}$ and $2d_{5/2}$ neutron orbitals. However, as commented on in our previous work \cite{Riley_53Cr, Hay_52V, Spi25a}, it appears likely that much of the $\ell =4$ and $\ell =2$ strength indeed resides above $S_n$ and is significantly fragmented. To support this statement and, especially, with the stark difference between the EQPM and REOM$^3$ for the $\ell =2$ strength in mind, we decided to analyze the unresolved strength hidden in the background, i.e., strength that did not show up in the form of discrete peaks in our $(d,p)$ spectra.

\subsection{Analysis of unresolved strength}
\label{sec:unresolved}

To determine the unresolved strength, discrete peaks were subtracted from the $(d,p)$ spectra using fit parameters determined from modeling their Gaussian-like shapes on top of a smooth background distribution, see Fig.\,\ref{fig:spectrum_02}. This procedure allowed us to obtain a smooth background spectrum, as shown in Fig.\,\ref{fig:Bg_AngDist}\,(a), for all scattering angles covered in our experiment. From these background spectra, containing possibly unresolved strength, angular distributions were generated in 50-keV bins and analyzed from 5\,MeV to 11\,MeV analogously to the ones obtained for resolved states. Two examples for energy bins centered on 6975 and 9375\,keV, respectively, are presented in Figs.\,\ref{fig:Bg_AngDist}\,(b) and (c). While the angular distribution of the 6975-keV energy bin is better described by an $\ell =2$ transfer, the distribution obtained for the 9375-keV energy bin follows the one predicted for an $\ell = 3$ transfer more closely. All unresolved strength hidden in the background was determined to be either of $\ell =2$ or of $\ell =3$ character. No additional $\ell =1$ and $\ell=4$ strengths were detected. The additional strengths obtained from the background analysis were added to the sum rules. For $\ell =2$ and $\ell =3$, their contribution is highlighted in Fig.\,\ref{fig:50Ti_dp_theory_comparison} with lighter shading and hatching. The $\ell=2$ vacancy increases to 1.32(12), which corresponds to only 22\% of the expected vacancy. This increase does, therefore, not change the hypothesis that a significant fraction of the $2d_{5/2}$ strength is likely located above $S_n$. For $\ell=3$, the inclusion of unresolved strength increases the vacancy significantly from 4.2(4) to 7.0(6), i.e., by almost three particles. This number is larger than the number of neutrons which the $1f_{5/2}$ orbital alone could accommodate. The problem could be solved if both the $1f_{7/2}$ and $1f_{5/2}$ orbitals contributed to unresolved strength in the background. If we assume that we have a vacancy of one nucleon in the neutron $1f_{7/2}$ orbital for \nuc{49}{Ti}, we could in principle accommodate a total of seven neutrons consistent with the experimental upper limit for the $\ell =3$ strength. However, then there would be appreciable tension with the empirically determined quenching of the single-particle strength in nuclei. We, thus, suggest to interpret the vacancies obtained from our background analysis as an upper limit. Likely, some of the $\ell =3$ strength, hidden in the background, is in fact $\ell =2$ or mixed $\ell = 2+4$ strength, as in some cases comparable $\chi^2$ values were obtained. There is also a small probability that some of the unresolved strength does not belong to $\nuc{49}{Ti}(d,p)\nuc{50}{Ti}$. Given that we are dealing with unresolved strength here, it will be challenging to quantify and characterize it more precisely. Experimentally, the only avenues to tackle this challenge that one could imagine right now are significantly improved particle energy resolution or detailed, high-statistics $(d,p\gamma)$ measurements as presented in Ref. \cite{Spi25a} to possibly resolve additional, weakly populated states.

\section{Summary}

In this article, we presented new data for excited states of \nuc{50}{Ti}, which were obtained from $\nuc{49}{Ti}(d,p)\nuc{50}{Ti}$ and $\nuc{49}{Ti}(d,p\gamma)\nuc{50}{Ti}$ experiments performed at the FSU John D. Fox Accelerator Laboratory. In total, 82 excited states of \nuc{50}{Ti} were identified up to the neutron-separation threshold, $S_n$, through the measurement of detailed $(d,p)$ angular distributions with the Super-Enge Split-Pole Spectrograph. Additionally, some incorrect assignments of states to \nuc{50}{Ti}, based on previous $(d,p)$ data, could be identified thanks to the complementary $\gamma$-decay information obtained with the CeBrA demonstrator and revisions of the adopted data were suggested.

Up to $S_n$, model-dependent spectroscopic factors and associated sum rules related to vacancies for the $2p_{3/2}$, $2p_{1/2}$, $1f_{5/2}$, $1g_{9/2}$, and $2d_{5/2}$ neutron single-particle orbitals were determined from the model-independent $(d,p)$ cross sections. For the well-bound $fp$ orbitals, the new data for \nuc{50}{Ti} and previously reported data for \nuc{51}{Ti} \cite{Riley_51Ti} are consistent, with the centroids of the spectroscopic strength in \nuc{50}{Ti} shifted up in energy by twice the value of the neutron-pairing gap, $\Delta_n$, compared to \nuc{51}{Ti}. While this energy shift was also observed for the strength associated with placing the neutron in the $1g_{9/2}$ and $2d_{5/2}$ orbitals, significantly less spectroscopic strength was observed for these orbitals than for the $fp$ orbitals. This observation is consistent with our previous work on the $N=29$ isotones \cite{Riley_51Ti, Riley_55Fe, Riley_53Cr, Hay_52V, spi24a, Spi25a}. However, the comparison between \nuc{50}{Ti} and \nuc{51}{Ti} revealed that significantly less $\ell =4$ strength was observed in the even-$A$ than in the odd-$A$ titanium isotope. For now, this is an observation without an explanation.

We also compared our new data to three leading nuclear-structure models. These were the time-dependent continuum shell model (TDCSM) with the FSU interaction, the energy-density functional theory plus quasiparticle-phonon model approach (EQPM) including up to three-phonon contributions, and the relativistic equation of motion theory including the coupling of two-quasiparticles with up to two phonons (REOM$^3$). A detailed comparison of the predicted strength fragmentation and associated vacancies for the $fp$ orbitals to data revealed that all models provided an excellent description of the observed $1f_{5/2}$ neutron strength, also including the correct amount of quenching up to $S_n$. However, stark differences were observed for the predicted $2p_{3/2}$ and $2p_{1/2}$ spectroscopic strengths. While both the TDCSM and EQPM overestimated the experimentally determined vacancy for the combined $2p$ orbitals, the REOM$^3$ largely agreed with the data. A closer look at the details of the strength fragmentation and comparing the strongly populated states in experiment to the theoretical predictions uncovered deficiencies for the REOM$^3$ though. Specifically, the REOM$^3$ did not fragment any appreciable $2p_{3/2}$ or $2p_{1/2}$ strength to unnatural-parity states at lower excitation energies which were, however, strongly populated in the $(d,p)$ reaction. In contrast, the TDCSM and EQPM predicted sizable spectroscopic factors for these states. Thus, although the REOM$^3$ reproduces the reduction of the spectroscopic strength associated with the $2p$ neutron orbitals below 11.5 MeV, substantial discrepancies in the detailed structure persist, and the apparent agreement in total strength should be interpreted with caution. The comparison suggests that further investigations are needed to better understand how strength fragmentation depends on the total angular momentum, $J$. For instance, while the TDCSM got the $S'$ values for the $J^{\pi} = 3^+$ and $5^+$ states expected for the $(1f_{7/2})^{-1}(2p_{3/2})^{+1}$ neutron 1p-1h excitation spot on, it overestimated the strength for the strongest $J=4$ fragment by about 40\,\%. The differences between predicted and experimentally observed $2^+$ states was even more striking. Experimentally, no $2^+$ state with a significant $\ell =1$ spectroscopic factor was observed, while all models predicted at least one $2^+$ state with appreciable strength.

For the strength associated with the $1g_{9/2}$ and $2d_{5/2}$ neutron orbitals, conflicting predictions were obtained as well. For both single-particle orbitals, the EQPM predicted significantly more spectroscopic strength below $S_n$ than the REOM$^3$. While for the $\ell =4$ strength, the data clearly favor the REOM$^3$ predictions, which expect the bulk of the strength above $S_n$ and to be strongly fragmented, the data lies between the EQPM and REOM$^3$ predictions for the $\ell =2$ strength. However, also in this case, an additional analysis of unresolved strength in the background points to a scenario where significantly less $2d_{5/2}$ neutron strength is found below $S_n$ than presently predicted by the EQPM. For a more definite answer, also regarding the strength quenching for these orbitals, additional data above $S_n$ would be needed. As pointed out in Ref.\,\cite{Spi25a}, experiments with isotopically-pure targets without significant carbon or oxygen contamination would be needed to study additional states and strength fragmentation above $S_n$. Coincident $\gamma$-ray detection might provide a path towards resolving weaker fragments as suggested in Ref.\,\cite{Spi25a}. One might also consider neutron-transfer reactions like $(\alpha,\nuc{3}{He})$, as presented in Ref.\,\cite{Sch13a}, since the momentum matching is more favorable for the population of states through larger $\ell$ transfers.

\appendix

\section{Comments on Spectroscopic Factors}

\label{appendix}

We defined the spectroscopic factor for nucleon removal in Eq.\,(\ref{eq:tdcsm_s}). The nucleon-addition spectroscopic factor, $S_j^{+}(J_i \rightarrow J_f)$, is given by,

\begin{align}
    S_j^{+}(J_i \rightarrow J_f) &= \frac{1}{2J_i+1}\sum_{m,M_f,M_i} \left| \langle J_fM_f|a^{\dagger}_{jm}|J_iM_i \rangle\right|^2 \nonumber \\
    &= \frac{1}{2J_i+1}\left| \langle J_f\|a^{\dagger}_{j}\| J_i \rangle\right|^2,
    \label{eq:tdcsm_s+}
\end{align}

\noindent where the Wigner-Eckart theorem with the following convention,

\begin{equation}
\langle J_f M_f | a^{\dagger}_{jm} | J_i M_i\rangle
=
\frac{\langle J_f\|a^{\dagger}_j\|J_i\rangle}{\sqrt{2J_f+1}}
\langle J_i M_i\, j m | J_f M_f\rangle ,
\end{equation}

\noindent and

\begin{equation}
    \sum_{m,M_f,M_i}\left|\langle J_i M_i\, j m | J_f M_f\rangle\right|^2 = 2J_f+1
\end{equation}

\noindent were used to go from the expression with the sum over the magnetic substates to the final expression in Eq.\,(\ref{eq:tdcsm_s+}).

Generally, for a transition from an initial state $J_i$ to a final state $J_f$, the following identity, connecting nucleon removal and addition, holds

\begin{equation}
\left|
\langle J_f\|a^\dagger_j\|J_i\rangle
\right|^2 =
\Big|
\langle J_i\|a_j\|J_f\rangle
\Big|^2,
\label{eq:hc}
\end{equation}

\noindent which follows from Hermitian conjugation of spherical tensors like the nucleon removal and addition operators.

Using Eq.\,(\ref{eq:hc}), the following relationship between nucleon-addition and nucleon-removal spectroscopic factors can be derived:

\begin{equation}
    S^+_j(J_i \to J_f) \cdot (2J_{i}+1) = S_j^-(J_f \to J_i) \cdot (2J_{f}+1). 
    \label{eq:connection}
\end{equation}

For our purposes, we identify the initial state $|i,J_iM_i\rangle$ in Eqs.\,(\ref{eq:tdcsm_s+}) and (\ref{eq:connection}) as the ground state of \nuc{49}{Ti} and the final state $|f,J_fM_f\rangle$ as an excited state in \nuc{50}{Ti}.  With this choice, we can convert the neutron-removal spectroscopic factors calculated by the TDCSM to neutron-addition spectroscopic factors,

\begin{align}
S_j^+(\nuc{49}{Ti} \to \nuc{50}{Ti}) = \frac{2J_f(\nuc{50}{Ti})+1}{2J_i(\nuc{49}{Ti})+1} S^-_j(\nuc{50}{Ti} \to \nuc{49}{Ti}),
\label{eq:final_rel}
\end{align}

\noindent where $J_i(\nuc{49}{Ti)} = 7/2^-$ for the ground state of \nuc{49}{Ti} and $J_f(\nuc{50}{Ti)}$ corresponds to the total angular momentum of the excited state considered (or calculated) in \nuc{50}{Ti}. Comparing Eq.\,(\ref{eq:final_rel}) to Eq.\,(\ref{eq:specfac}), one can then see that the removal spectroscopic factor $S^-_j(\nuc{50}{Ti} \to \nuc{49}{Ti})$ is equivalent to the spin-weighted spectroscopic factor $S'$ determined from the comparison of experimental cross sections to ADWA calculations.

\begin{acknowledgments}
This work was supported by the U.S. National Science Foundation under Grants No. PHY-2012522 (FSU), No. PHY-2412808 (FSU), No. PHY-2209376 (WMU), and No. PHY-2515056 (WMU), and by the U.S. Department of Energy, Office of Science, Office of Nuclear Physics under Award No. DE-SC0009883 (FSU). This work was also supported by project ELI-RO/DFG/2023\_001 ARNPhot funded by the Institute of Atomic Physics, Romania. Parts of this work were carried out under the contract PN 23.21.01.06 sponsored by the Romanian Ministry of Research, Innovation and Digitalization and partially supported by ELI-RO-RDI-2024-AMAP of the Romanian Government. A target provided by the Center for Accelerator Target Science at Argonne National Laboratory was used in this work.
\end{acknowledgments}

\bibliographystyle{apsrev4-1}
\bibliography{main}

@PREAMBLE{
 "\providecommand{\noopsort}[1]{}" 
 # "\providecommand{\singleletter}[1]{#1}%" 
}

@article{Hay_52V,
  title = {Measurement of ${g}_{9/2}$ strength in the stretched ${8}^{\ensuremath{-}}$ state and other negative parity states via the $^{51}\mathrm{V}(d,p)^{52}\mathrm{V}$ reaction},
  author = {Hay, I. C. S. and Cottle, P. D. and Riley, L. A. and Baby, L. T. and Baker, S. and Conley, A. L. and Esparza, J. and Hanselman, K. and Heinze, M. and Houlihan, D. and Kelly, B. and Kemper, K. W. and McCann, G. W. and Renom, R. and Sandrik, A. and Simms, D. and Spieker, M. and Wiedenh\"over, I.},
  journal = {Phys. Rev. C},
  volume = {109},
  issue = {2},
  pages = {024302},
  numpages = {6},
  year = {2024},
  month = {Feb},
  publisher = {American Physical Society},
  doi = {10.1103/PhysRevC.109.024302},
  url = {https://link.aps.org/doi/10.1103/PhysRevC.109.024302}
}

@article{Riley_53Cr,
  title = {${g}_{9/2}$ neutron strength in the $N=29$ isotones and the $^{52}\mathrm{Cr}(d,p)^{53}\mathrm{Cr}$ reaction},
  author = {Riley, L. A. and Simms, D. T. and Baby, L. T. and Conley, A. L. and Cottle, P. D. and Esparza, J. and Hanselman, K. and Hay, I. C. S. and Heinze, M. and Kelly, B. and Kemper, K. W. and McCann, G. W. and Renom, R. and Spieker, M. and Wiedenh\"over, I.},
  journal = {Phys. Rev. C},
  volume = {108},
  issue = {4},
  pages = {044306},
  numpages = {8},
  year = {2023},
  month = {Oct},
  publisher = {American Physical Society},
  doi = {10.1103/PhysRevC.108.044306},
  url = {https://link.aps.org/doi/10.1103/PhysRevC.108.044306}
}

@article{Riley_55Fe,
  title = {$^{54}\mathrm{Fe}(d,p)^{55}\mathrm{Fe}$ and the evolution of single neutron energies in the $N=29$ isotones},
  author = {Riley, L. A. and Hay, I. C. S. and Baby, L. T. and Conley, A. L. and Cottle, P. D. and Esparza, J. and Hanselman, K. and Kelly, B. and Kemper, K. W. and Macon, K. T. and McCann, G. W. and Quirin, M. W. and Renom, R. and Saunders, R. L. and Spieker, M. and Wiedenh\"over, I.},
  journal = {Phys. Rev. C},
  volume = {106},
  issue = {6},
  pages = {064308},
  numpages = {8},
  year = {2022},
  month = {Dec},
  publisher = {American Physical Society},
  doi = {10.1103/PhysRevC.106.064308},
  url = {https://link.aps.org/doi/10.1103/PhysRevC.106.064308}
}

@article{Riley_51Ti,
  title = {$^{50}\mathrm{Ti}(d,p)^{51}\mathrm{Ti}$: Single-neutron energies in the $N=29$ isotones, and the $N=32$ subshell closure},
  author = {Riley, L. A. and Nebel-Crosson, J. M. and Macon, K. T. and McCann, G. W. and Baby, L. T. and Caussyn, D. and Cottle, P. D. and Esparza, J. and Hanselman, K. and Kemper, K. W. and Temanson, E. and Wiedenh\"over, I.},
  journal = {Phys. Rev. C},
  volume = {103},
  issue = {6},
  pages = {064309},
  numpages = {8},
  year = {2021},
  month = {Jun},
  publisher = {American Physical Society},
  doi = {10.1103/PhysRevC.103.064309},
  url = {https://link.aps.org/doi/10.1103/PhysRevC.103.064309}
}

@article{Wei21a,
  title = {Microscopic Structure of the Low-Energy Electric Dipole Response of $^{120}\mathrm{Sn}$},
  author = {Weinert, M. and Spieker, M. and Potel, G. and Tsoneva, N. and M\"uscher, M. and Wilhelmy, J. and Zilges, A.},
  journal = {Phys. Rev. Lett.},
  volume = {127},
  issue = {24},
  pages = {242501},
  numpages = {7},
  year = {2021},
  month = {Dec},
  publisher = {American Physical Society},
  doi = {10.1103/PhysRevLett.127.242501},
  url = {https://link.aps.org/doi/10.1103/PhysRevLett.127.242501}
}

@article{61Nidp_Mark,
  title = {Experimental study of excited states of $^{62}\mathrm{Ni}$ via one-neutron $(d,p)$ transfer up to the neutron-separation threshold and characteristics of the pygmy dipole resonance states},
  author = {Spieker, M. and Baby, L. T. and Conley, A. L. and Kelly, B. and M\"uscher, M. and Renom, R. and Sch\"uttler, T. and Zilges, A.},
  journal = {Phys. Rev. C},
  volume = {108},
  issue = {1},
  pages = {014311},
  numpages = {14},
  year = {2023},
  month = {Jul},
  publisher = {American Physical Society},
  doi = {10.1103/PhysRevC.108.014311},
  url = {https://link.aps.org/doi/10.1103/PhysRevC.108.014311}
}

@article{CeBrA_NIM,
title = {The CeBrA demonstrator for particle-γ coincidence experiments at the FSU Super-Enge Split-Pole Spectrograph},
journal = {Nuclear Instruments and Methods in Physics Research Section A: Accelerators, Spectrometers, Detectors and Associated Equipment},
volume = {1058},
pages = {168827},
year = {2024},
issn = {0168-9002},
doi = {https://doi.org/10.1016/j.nima.2023.168827},
url = {https://www.sciencedirect.com/science/article/pii/S0168900223008185},
author = {A.L. Conley and B. Kelly and M. Spieker and R. Aggarwal and S. Ajayi and L.T. Baby and S. Baker and C. Benetti and I. Conroy and P.D. Cottle and I.B. D’Amato and P. DeRosa and J. Esparza and S. Genty and K. Hanselman and I. Hay and M. Heinze and D. Houlihan and M.I. Khawaja and P.S. Kielb and A.N. Kuchera and G.W. McCann and A.B. Morelock and E. Lopez-Saavedra and R. Renom and L.A. Riley and G. Ryan and A. Sandrik and V. Sitaraman and E. Temanson and M. Wheeler and C. Wibisono and I. Wiedenhöver}
}

@article{Spi25a,
  title = {Experimental study of $^{53}\mathrm{Cr}$ via the ($d,p\ensuremath{\gamma}$) reaction},
  author = {Spieker, M. and Riley, L. A. and Heinze, M. and Conley, A. L. and Kelly, B. and Cottle, P. D. and Aggarwal, R. and Ajayi, S. and Baby, L. T. and Baker, S. and Conroy, I. and D'Amato, I. B. and Esparza, J. and Genty, S. and Hay, I. and Kemper, K. W. and Khawaja, M. I. and Kielb, P. S. and Kuchera, A. N. and Lopez-Saavedra, E. and Morelock, A. B. and Piekarewicz, J. and Sandrik, A. and Sitaraman, V. and Temanson, E. and Wibisono, C. and Wiedenhoever, I.},
  journal = {Phys. Rev. C},
  volume = {112},
  issue = {6},
  pages = {064331},
  numpages = {17},
  year = {2025},
  month = {Dec},
  publisher = {American Physical Society},
  doi = {10.1103/wp85-rthk},
  url = {https://link.aps.org/doi/10.1103/wp85-rthk}
}

@article{Mass50DataSheet,
title = {Nuclear Data Sheets for A=50},
journal = {Nuclear Data Sheets},
volume = {157},
pages = {1-259},
year = {2019},
issn = {0090-3752},
doi = {https://doi.org/10.1016/j.nds.2019.04.001},
url = {https://www.sciencedirect.com/science/article/pii/S0090375219300262},
author = {Jun Chen and Balraj Singh}
}

@article{Mass51DataSheet,
title = {Nuclear Data Sheets for A=51},
journal = {Nuclear Data Sheets},
volume = {144},
pages = {1-296},
year = {2017},
issn = {0090-3752},
doi = {https://doi.org/10.1016/j.nds.2017.08.002},
url = {https://www.sciencedirect.com/science/article/pii/S0090375217300546},
author = {Jimin Wang and Xiaolong Huang}
}

@article{Barnes1,
  title = {Inelastic Deuteron Scattering and ($d,p$) Reactions from Isotopes of Titanium. III. ${\mathrm{Ti}}^{49}(d,p){\mathrm{Ti}}^{50}$},
  author = {Barnes, P. D. and Bockelman, C. K. and Hansen, Ole and Sperduto, A.},
  journal = {Phys. Rev.},
  volume = {140},
  issue = {1B},
  pages = {B42--B47},
  numpages = {0},
  year = {1965},
  month = {Oct},
  publisher = {American Physical Society},
  doi = {10.1103/PhysRev.140.B42},
  url = {https://link.aps.org/doi/10.1103/PhysRev.140.B42}
}

@article{Barnes2,
  title = {Inelastic Deuteron Scattering and ($d, p$) Reactions from Isotopes of Ti. V. $\mathrm{Ti}(d, {d}^{\ensuremath{'}})$},
  author = {Wilhjelm, P. and Hansen, Ole and Comfort, J. R. and Bockelman, C. K. and Barnes, P. D. and Sperduto, A.},
  journal = {Phys. Rev.},
  volume = {166},
  issue = {4},
  pages = {1121--1131},
  numpages = {0},
  year = {1968},
  month = {Feb},
  publisher = {American Physical Society},
  doi = {10.1103/PhysRev.166.1121},
  url = {https://link.aps.org/doi/10.1103/PhysRev.166.1121}
}

@article{50Ti_dpgamma,
doi = {10.1088/0305-4616/10/6/016},
url = {https://dx.doi.org/10.1088/0305-4616/10/6/016},
year = {1984},
month = {jun},
publisher = {},
volume = {10},
number = {6},
pages = {833},
author = {P Sona and P A Mando and N Taccetti},
title = {Decay scheme and lifetimes of excited levels in 50Ti},
journal = {Journal of Physics G: Nuclear Physics}
}

@article{50Ti_tp,
title = {A study of the (t,p) reactions leading to 48Ti and 50Ti},
journal = {Nuclear Physics A},
volume = {92},
number = {2},
pages = {422-432},
year = {1967},
issn = {0375-9474},
doi = {https://doi.org/10.1016/0375-9474(67)90227-8},
url = {https://www.sciencedirect.com/science/article/pii/0375947467902278},
author = {S. Hinds and R. Middleton}
}

@article{Ball_49Ti_dp_study,
title = {The level structure of 49Ti},
journal = {Nuclear Physics A},
volume = {183},
number = {3},
pages = {472-496},
year = {1972},
issn = {0375-9474},
doi = {https://doi.org/10.1016/0375-9474(72)90351-X},
url = {https://www.sciencedirect.com/science/article/pii/037594747290351X},
author = {A.E. Ball and G. Brown and A. Denning and R.N. Glover}
}

@article{fresco,
title = {Coupled reaction channels calculations in nuclear physics},
journal = {Computer Physics Reports},
volume = {7},
number = {4},
pages = {167-212},
year = {1988},
issn = {0167-7977},
doi = {https://doi.org/10.1016/0167-7977(88)90005-6},
url = {https://www.sciencedirect.com/science/article/pii/0167797788900056},
author = {Ian J. Thompson}
}

@article{Wales_Johnson,
title = {Deuteron break-up effects in (p, d) reactions at 65 MeV},
journal = {Nuclear Physics A},
volume = {274},
number = {1},
pages = {168-176},
year = {1976},
issn = {0375-9474},
doi = {https://doi.org/10.1016/0375-9474(76)90234-7},
url = {https://www.sciencedirect.com/science/article/pii/0375947476902347},
author = {G.L. Wales and R.C. Johnson}
}

@article{Koning_OMPS,
title = {Local and global nucleon optical models from 1 keV to 200 MeV},
journal = {Nuclear Physics A},
volume = {713},
number = {3},
pages = {231-310},
year = {2003},
issn = {0375-9474},
doi = {https://doi.org/10.1016/S0375-9474(02)01321-0},
url = {https://www.sciencedirect.com/science/article/pii/S0375947402013210},
author = {A.J. Koning and J.P. Delaroche}
}

@article{Kay_A15_SF,
  title = {Quenching of Single-Particle Strength in $A=15$ Nuclei},
  author = {Kay, B. P. and Tang, T. L. and Tolstukhin, I. A. and Roderick, G. B. and Mitchell, A. J. and Ayyad, Y. and Bennett, S. A. and Chen, J. and Chipps, K. A. and Crawford, H. L. and Freeman, S. J. and Garrett, K. and Gott, M. D. and Hall, M. R. and Hoffman, C. R. and Jayatissa, H. and Macchiavelli, A. O. and MacGregor, P. T. and Sharp, D. K. and Wilson, G. L.},
  journal = {Phys. Rev. Lett.},
  volume = {129},
  issue = {15},
  pages = {152501},
  numpages = {6},
  year = {2022},
  month = {Oct},
  publisher = {American Physical Society},
  doi = {10.1103/PhysRevLett.129.152501},
  url = {https://link.aps.org/doi/10.1103/PhysRevLett.129.152501}
}

@article{SumRule_TransferReactions,
  title = {Test of Sum Rules in Nucleon Transfer Reactions},
  author = {Schiffer, J. P. and Hoffman, C. R. and Kay, B. P. and Clark, J. A. and Deibel, C. M. and Freeman, S. J. and Howard, A. M. and Mitchell, A. J. and Parker, P. D. and Sharp, D. K. and Thomas, J. S.},
  journal = {Phys. Rev. Lett.},
  volume = {108},
  issue = {2},
  pages = {022501},
  numpages = {5},
  year = {2012},
  month = {Jan},
  publisher = {American Physical Society},
  doi = {10.1103/PhysRevLett.108.022501},
  url = {https://link.aps.org/doi/10.1103/PhysRevLett.108.022501}
}

@article{Muk10a,
  title = {Unitary correlation in nuclear reaction theory: Separation of nuclear reactions and spectroscopic factors},
  author = {Mukhamedzhanov, A. M. and Kadyrov, A. S.},
  journal = {Phys. Rev. C},
  volume = {82},
  issue = {5},
  pages = {051601},
  numpages = {5},
  year = {2010},
  month = {Nov},
  publisher = {American Physical Society},
  doi = {10.1103/PhysRevC.82.051601},
  url = {https://link.aps.org/doi/10.1103/PhysRevC.82.051601}
}

@book{Bohrbook, 
author = {Bohr, Aage and Mottelson, Ben R},
title = {Nuclear Structure},
publisher = {World Scientific Publishing Company},
year = {1998},
doi = {10.1142/3530},
address = {},
edition   = {},
URL = {https://www.worldscientific.com/doi/abs/10.1142/3530},
eprint = {https://www.worldscientific.com/doi/pdf/10.1142/3530}
}

@book{Soloviev, title={Theory of Atomic Nuclei: Quasiparticles and Phonons}, publisher={CRC Press}, author={Soloviev, V.G.}, year={1992}, edition={1$^{\mathrm{st}}$ edition}, doi={https://doi.org/10.1201/9781003062868}}

@article{spi24a,
author={Spieker, M. and Almaraz-Calderon, S.}, 
title = {Nuclear structure and direct reaction studies in particle-$\gamma$ coincidence experiments at the FSU John D. Fox superconducting linear accelerator laboratory},
journal={Frontiers in Physics},
volume={12},
year={2024},
pages={1511394},
url={https://www.frontiersin.org/journals/physics/articles/10.3389/fphy.2024.1511394},
doi={10.3389/fphy.2024.1511394},
issn={2296-424X}
}

@article{Tsoneva2016,
title = {Energy–density functional plus quasiparticle–phonon model theory as a powerful tool for nuclear structure and astrophysics},
journal = {Physics of Atomic Nuclei},
volume = {79},
number = {6},
pages = {885-903},
year = {2016},
issn = {0167-7977},
doi = {https://doi.org/10.1134/S1063778816060247},
url = {},
author = {N. Tsoneva and H. Lenske}
}

@article{Sol81a,
title = {Fragmentation of two-quasiparticle states in spherical nuclei},
journal = {Nuclear Physics A},
volume = {370},
number = {1},
pages = {13-29},
year = {1981},
issn = {0375-9474},
doi = {https://doi.org/10.1016/0375-9474(81)90752-1},
url = {https://www.sciencedirect.com/science/article/pii/0375947481907521},
author = {V.G. Soloviev and O. Stoyanova and V.V. Voronov}
}

@article{Lit19a,
  title = {Toward an accurate strongly coupled many-body theory within the equation-of-motion framework},
  author = {Litvinova, Elena and Schuck, Peter},
  journal = {Phys. Rev. C},
  volume = {100},
  issue = {6},
  pages = {064320},
  numpages = {26},
  year = {2019},
  month = {Dec},
  publisher = {American Physical Society},
  doi = {10.1103/PhysRevC.100.064320},
  url = {https://link.aps.org/doi/10.1103/PhysRevC.100.064320}
}

@article{Lit23a,
    author = {E. Litvinova},
    title = {On the dynamical kernels of fermionic equations of motion in strongly-correlated media},
    journal = {European Physical Journal A},
    year = {2023},
    volume = {59},
    pages = {291},
    doi = {https://doi.org/10.1140/epja/s10050-023-01198-y}
}

@article{Lit22a,
  title = {Microscopic response theory for strongly coupled superfluid fermionic systems},
  author = {Litvinova, Elena and Zhang, Yinu},
  journal = {Phys. Rev. C},
  volume = {106},
  issue = {6},
  pages = {064316},
  numpages = {15},
  year = {2022},
  month = {Dec},
  publisher = {American Physical Society},
  doi = {10.1103/PhysRevC.106.064316},
  url = {https://link.aps.org/doi/10.1103/PhysRevC.106.064316}
}

@article{Vol09a,
  title = {Time-dependent approach to the continuum shell model},
  author = {Volya, Alexander},
  journal = {Phys. Rev. C},
  volume = {79},
  issue = {4},
  pages = {044308},
  numpages = {19},
  year = {2009},
  month = {Apr},
  publisher = {American Physical Society},
  doi = {10.1103/PhysRevC.79.044308},
  url = {https://link.aps.org/doi/10.1103/PhysRevC.79.044308}
}

@article{Vol14a,
    author = {Volya, A. and Zelevinsky, V.},
    title = {Continuum shell model and nuclear physics at the edge of stability},
    journal = {Phys. Atom. Nuclei},
    pages = {969},
    year = {2014},
    url = {https://doi.org/10.1134/S1063778814070163},
    doi = {10.1134/S1063778814070163}
}

@article{Lub19a,
  title = {Structure of $^{38}\mathrm{Cl}$ and the quest for a comprehensive shell model interaction},
  author = {Lubna, R. S. and Kravvaris, K. and Tabor, S. L. and Tripathi, Vandana and Volya, A. and Rubino, E. and Allmond, J. M. and Abromeit, B. and Baby, L. T. and Hensley, T. C.},
  journal = {Phys. Rev. C},
  volume = {100},
  issue = {3},
  pages = {034308},
  numpages = {12},
  year = {2019},
  month = {Sep},
  publisher = {American Physical Society},
  doi = {10.1103/PhysRevC.100.034308},
  url = {https://link.aps.org/doi/10.1103/PhysRevC.100.034308}
}

@article{Lub20a,
  title = {Evolution of the $N=20$ and 28 shell gaps and two-particle-two-hole states in the FSU interaction},
  author = {Lubna, R. S. and Kravvaris, K. and Tabor, S. L. and Tripathi, Vandana and Rubino, E. and Volya, A.},
  journal = {Phys. Rev. Res.},
  volume = {2},
  issue = {4},
  pages = {043342},
  numpages = {11},
  year = {2020},
  month = {Dec},
  publisher = {American Physical Society},
  doi = {10.1103/PhysRevResearch.2.043342},
  url = {https://link.aps.org/doi/10.1103/PhysRevResearch.2.043342}
}

@Article{Ring1996,
title = {Relativistic mean field theory in finite nuclei},
journal = {Progress in Particle and Nuclear Physics},
volume = {37},
pages = {193-263},
year = {1996},
issn = {0146-6410},
doi = {https://doi.org/10.1016/0146-6410(96)00054-3},
url = {https://www.sciencedirect.com/science/article/pii/0146641096000543},
author = {P. Ring}
}

@article{Kay13a,
  title = {Quenching of Cross Sections in Nucleon Transfer Reactions},
  author = {Kay, B. P. and Schiffer, J. P. and Freeman, S. J.},
  journal = {Phys. Rev. Lett.},
  volume = {111},
  issue = {4},
  pages = {042502},
  numpages = {5},
  year = {2013},
  month = {Jul},
  publisher = {American Physical Society},
  doi = {10.1103/PhysRevLett.111.042502},
  url = {https://link.aps.org/doi/10.1103/PhysRevLett.111.042502}
}

@article{Tos21a,
  title = {Updated systematics of intermediate-energy single-nucleon removal cross sections},
  author = {Tostevin, J. A. and Gade, A.},
  journal = {Phys. Rev. C},
  volume = {103},
  issue = {5},
  pages = {054610},
  numpages = {7},
  year = {2021},
  month = {May},
  publisher = {American Physical Society},
  doi = {10.1103/PhysRevC.103.054610},
  url = {https://link.aps.org/doi/10.1103/PhysRevC.103.054610}
}

@article{Aum21a,
title = {Quenching of single-particle strength from direct reactions with stable and rare-isotope beams},
journal = {Progress in Particle and Nuclear Physics},
volume = {118},
pages = {103847},
year = {2021},
issn = {0146-6410},
doi = {https://doi.org/10.1016/j.ppnp.2021.103847},
url = {https://www.sciencedirect.com/science/article/pii/S0146641021000016},
author = {T. Aumann and C. Barbieri and D. Bazin and C.A. Bertulani and A. Bonaccorso and W.H. Dickhoff and A. Gade and M. Gómez-Ramos and B.P. Kay and A.M. Moro and T. Nakamura and A. Obertelli and K. Ogata and S. Paschalis and T. Uesaka}
}

@ARTICLE{Mac25a,
    
AUTHOR={Macchiavelli, Augusto O.  and Paschalis, Stefanos  and Petri, Marina },        
TITLE={Some aspects of the quenching of single-particle strength in atomic nuclei},        
JOURNAL={Frontiers in Physics},        
VOLUME={Volume 13 - 2025},
YEAR={2025},
URL={https://www.frontiersin.org/journals/physics/articles/10.3389/fphy.2025.1530428},
DOI={10.3389/fphy.2025.1530428},
ISSN={2296-424X}}

@article{Jia25a,
title = {Quenching of single-particle strength inferred from nucleon-removal transfer reactions on $^{15}$C},
journal = {Physics Letters B},
volume = {868},
pages = {139789},
year = {2025},
issn = {0370-2693},
doi = {https://doi.org/10.1016/j.physletb.2025.139789},
url = {https://www.sciencedirect.com/science/article/pii/S0370269325005507},
author = {Y.C. Jiang and J. Chen and B.P. Kay and C.R. Hoffman and T.L. Tang and I.A. Tolstukhin and M.R. Xie and J.G. Li and N. Michel and M.L. Avila and Y. Ayyad and D. Bazin and S. Bennett and J.A. Clark and S.J. Freeman and H. Jayatissa and G. Li and W.P. Liu and J.L. Lou and A. Munoz-Ramos and C. Müller-Gatermann and T. Nathan and D. Santiago-Gonzalez and D.K. Sharp and Y.P. Shen and A.H. Wuosmaa and C.X. Yuan}
}

@article{Ata18a,
  title = {Quasifree ($p$, $2p$) Reactions on Oxygen Isotopes: Observation of Isospin Independence of the Reduced Single-Particle Strength},
  author = {Atar, L. and Paschalis, S. and Barbieri, C. and Bertulani, C. A. and D\'{\i}az Fern\'andez, P. and Holl, M. and Najafi, M. A. and Panin, V. and Alvarez-Pol, H. and Aumann, T. and Avdeichikov, V. and Beceiro-Novo, S. and Bemmerer, D. and Benlliure, J. and Boillos, J. M. and Boretzky, K. and Borge, M. J. G. and Caama\~no, M. and Caesar, C. and Casarejos, E. and Catford, W. and Cederkall, J. and Chartier, M. and Chulkov, L. and Cortina-Gil, D. and Cravo, E. and Crespo, R. and Dillmann, I. and Elekes, Z. and Enders, J. and Ershova, O. and Estrade, A. and Farinon, F. and Fraile, L. M. and Freer, M. and Galaviz Redondo, D. and Geissel, H. and Gernh\"auser, R. and Golubev, P. and G\"obel, K. and Hagdahl, J. and Heftrich, T. and Heil, M. and Heine, M. and Heinz, A. and Henriques, A. and Hufnagel, A. and Ignatov, A. and Johansson, H. T. and Jonson, B. and Kahlbow, J. and Kalantar-Nayestanaki, N. and Kanungo, R. and Kelic-Heil, A. and Knyazev, A. and Kr\"oll, T. and Kurz, N. and Labiche, M. and Langer, C. and Le Bleis, T. and Lemmon, R. and Lindberg, S. and Machado, J. and Marganiec-Galkazka, J. and Movsesyan, A. and Nacher, E. and Nikolskii, E. Y. and Nilsson, T. and Nociforo, C. and Perea, A. and Petri, M. and Pietri, S. and Plag, R. and Reifarth, R. and Ribeiro, G. and Rigollet, C. and Rossi, D. M. and R\"oder, M. and Savran, D. and Scheit, H. and Simon, H. and Sorlin, O. and Syndikus, I. and Taylor, J. T. and Tengblad, O. and Thies, R. and Togano, Y. and Vandebrouck, M. and Velho, P. and Volkov, V. and Wagner, A. and Wamers, F. and Weick, H. and Wheldon, C. and Wilson, G. L. and Winfield, J. S. and Woods, P. and Yakorev, D. and Zhukov, M. and Zilges, A. and Zuber, K.},
  collaboration = {R$^{3}$B Collaboration},
  journal = {Phys. Rev. Lett.},
  volume = {120},
  issue = {5},
  pages = {052501},
  numpages = {7},
  year = {2018},
  month = {Jan},
  publisher = {American Physical Society},
  doi = {10.1103/PhysRevLett.120.052501},
  url = {https://link.aps.org/doi/10.1103/PhysRevLett.120.052501}
}

@article{Poh23a,
  title = {Multiple Mechanisms in Proton-Induced Nucleon Removal at $\ensuremath{\sim}100\text{ }\text{ }\mathrm{MeV}/\text{Nucleon}$},
  author = {Pohl, T. and Sun, Y. L. and Obertelli, A. and Lee, J. and G\'omez-Ramos, M. and Ogata, K. and Yoshida, K. and Cai, B. S. and Yuan, C. X. and Brown, B. A. and Baba, H. and Beaumel, D. and Corsi, A. and Gao, J. and Gibelin, J. and Gillibert, A. and Hahn, K. I. and Isobe, T. and Kim, D. and Kondo, Y. and Kobayashi, T. and Kubota, Y. and Li, P. and Liang, P. and Liu, H. N. and Liu, J. and Lokotko, T. and Marqu\'es, F. M. and Matsuda, Y. and Motobayashi, T. and Nakamura, T. and Orr, N. A. and Otsu, H. and Panin, V. and Park, S. Y. and Sakaguchi, S. and Sasano, M. and Sato, H. and Sakurai, H. and Shimizu, Y. and Stefanescu, A. I. and Stuhl, L. and Suzuki, D. and Togano, Y. and Tudor, D. and Uesaka, T. and Wang, H. and Xu, X. and Yang, Z. H. and Yoneda, K. and Zenihiro, J.},
  journal = {Phys. Rev. Lett.},
  volume = {130},
  issue = {17},
  pages = {172501},
  numpages = {8},
  year = {2023},
  month = {Apr},
  publisher = {American Physical Society},
  doi = {10.1103/PhysRevLett.130.172501},
  url = {https://link.aps.org/doi/10.1103/PhysRevLett.130.172501}
}

@ARTICLE{Heb25a,
    
AUTHOR={Hebborn, C.  and Nunes, F. M. },          
TITLE={Systematic study of the propagation of uncertainties to transfer observables},       
JOURNAL={Frontiers in Physics},      
VOLUME={Volume 13 - 2025},
YEAR={2025},
URL={https://www.frontiersin.org/journals/physics/articles/10.3389/fphy.2025.1525170},
DOI={10.3389/fphy.2025.1525170}, 
ISSN={2296-424X}}

@article{Heb23a,
  title = {New Perspectives on Spectroscopic Factor Quenching from Reactions},
  author = {Hebborn, C. and Nunes, F. M. and Lovell, A. E.},
  journal = {Phys. Rev. Lett.},
  volume = {131},
  issue = {21},
  pages = {212503},
  numpages = {6},
  year = {2023},
  month = {Nov},
  publisher = {American Physical Society},
  doi = {10.1103/PhysRevLett.131.212503},
  url = {https://link.aps.org/doi/10.1103/PhysRevLett.131.212503}
}

@article{Bar09a,
  title = {Role of Long-Range Correlations in the Quenching of Spectroscopic Factors},
  author = {Barbieri, C.},
  journal = {Phys. Rev. Lett.},
  volume = {103},
  issue = {20},
  pages = {202502},
  numpages = {4},
  year = {2009},
  month = {Nov},
  publisher = {American Physical Society},
  doi = {10.1103/PhysRevLett.103.202502},
  url = {https://link.aps.org/doi/10.1103/PhysRevLett.103.202502}
}

@misc{ensdf,
howpublished = {{National Nuclear Data Center, information extracted from the Evaluated Nuclear Structure Data File (ENSDF), \url{https://www.nndc.bnl.gov/ensdf/}}},
year = {2026}
}

@book{satchler,
  author		= {Satchler, G.R.},
  title			= {{Direct Nuclear Reactions}},
  address		= "New {Y}ork",
  publisher		= {Clarendum Press, Oxford University Press},
  year			= {1983}
}

@article{Fur02a,
title = {Are occupation numbers observable?},
journal = {Physics Letters B},
volume = {531},
number = {3},
pages = {203-208},
year = {2002},
issn = {0370-2693},
doi = {https://doi.org/10.1016/S0370-2693(01)01504-0},
url = {https://www.sciencedirect.com/science/article/pii/S0370269301015040},
author = {R.J. Furnstahl and H.-W. Hammer}
}

@article{Fur10a,
doi = {10.1088/0954-3899/37/6/064005},
url = {https://doi.org/10.1088/0954-3899/37/6/064005},
year = {2010},
month = {mar},
publisher = {},
volume = {37},
number = {6},
pages = {064005},
author = {Furnstahl, R J and Schwenk, A},
title = {How should one formulate, extract and interpret ‘non-observables’ for nuclei?},
journal = {Journal of Physics G: Nuclear and Particle Physics}
}

@article{Spi25b,
	author = {Spieker, M.},
	title = {Experimental studies of the microscopic structure of the pygmy dipole resonance and comments on a possible pygmy quadrupole resonance},
	DOI= "10.1140/epja/s10050-025-01669-4",
	url= "https://doi.org/10.1140/epja/s10050-025-01669-4",
	journal = {Eur. Phys. J. A},
	year = 2025,
	volume = 61,
	number = 8,
	pages = "197",
}

@article{Cha15a,
title = {Empirical pairing gaps, shell effects, and di-neutron spatial correlation in neutron-rich nuclei},
journal = {Nuclear Physics A},
volume = {940},
pages = {210-226},
year = {2015},
issn = {0375-9474},
doi = {https://doi.org/10.1016/j.nuclphysa.2015.04.010},
url = {https://www.sciencedirect.com/science/article/pii/S0375947415001062},
author = {S.A. Changizi and Chong Qi and R. Wyss}
}

@article{Kel26a,
  title = {Detailed View at Magnetic Dipole Strengths: The Case of Semimagic ${}^{50}\mathrm{Ti}$},
  author = {Kelly, B. and Spieker, M. and Friman-Gayer, U. and Baby, L. T. and Beck, T. and Conley, A. L. and Finch, S. W. and Isaak, J. and Krishichayan and Litvinova, E. and Pai, H. and Pietralla, N. and Savran, D. and Tornow, W. and Tsoneva, N. and Volya, A. and Werner, V.},
  journal = {Phys. Rev. Lett.},
  volume = {136},
  issue = {8},
  pages = {082502},
  numpages = {7},
  year = {2026},
  month = {Feb},
  publisher = {American Physical Society},
  doi = {10.1103/82y9-svrd},
  url = {https://link.aps.org/doi/10.1103/82y9-svrd}
}

@article{Dug12a,
  title = {Ab initio approach to effective single-particle energies in doubly closed shell nuclei},
  author = {Duguet, T. and Hagen, G.},
  journal = {Phys. Rev. C},
  volume = {85},
  issue = {3},
  pages = {034330},
  numpages = {13},
  year = {2012},
  month = {Mar},
  publisher = {American Physical Society},
  doi = {10.1103/PhysRevC.85.034330},
  url = {https://link.aps.org/doi/10.1103/PhysRevC.85.034330}
}

@article{Dug15a,
  title = {Nonobservable nature of the nuclear shell structure: Meaning, illustrations, and consequences},
  author = {Duguet, T. and Hergert, H. and Holt, J. D. and Som\`a, V.},
  journal = {Phys. Rev. C},
  volume = {92},
  issue = {3},
  pages = {034313},
  numpages = {15},
  year = {2015},
  month = {Sep},
  publisher = {American Physical Society},
  doi = {10.1103/PhysRevC.92.034313},
  url = {https://link.aps.org/doi/10.1103/PhysRevC.92.034313}
}

@article{Din26a,
  title = {From Spin to Pseudospin Symmetry: The Origin of Magic Numbers in Nuclear Structure},
  author = {Ding, C. R. and Wang, C. C. and Yao, J. M. and Hergert, H. and Liang, H. Z. and Bogner, S. K.},
  journal = {Phys. Rev. Lett.},
  volume = {136},
  issue = {5},
  pages = {052501},
  numpages = {7},
  year = {2026},
  month = {Feb},
  publisher = {American Physical Society},
  doi = {10.1103/8lzc-j1lx},
  url = {https://link.aps.org/doi/10.1103/8lzc-j1lx}
}

@article{Lap93a,
title = {Quasi-elastic electron scattering off nuclei},
journal = {Nuclear Physics A},
volume = {553},
pages = {297-308},
year = {1993},
issn = {0375-9474},
doi = {https://doi.org/10.1016/0375-9474(93)90630-G},
url = {https://www.sciencedirect.com/science/article/pii/037594749390630G},
author = {L. Lapikás}
}

@book{Dickhoff2005,
author = {Dickhoff, Willem H and Van Neck, Dimitri},
title = {Many-Body Theory Exposed!},
publisher = {WORLD SCIENTIFIC},
year = {2005},
doi = {10.1142/5804},
address = {},
edition   = {},
URL = {https://www.worldscientific.com/doi/abs/10.1142/5804},
eprint = {https://www.worldscientific.com/doi/pdf/10.1142/5804}
}

@Article{Litvinova2021a,
  author  = {Litvinova, Elena and Zhang, Yinu},
  journal = {Physical Review C},
  title   = {Many-body theory for quasiparticle states in superfluid fermionic systems},
  year    = {2021},
  number  = {4},
  pages   = {044303},
  volume  = {104},
  doi     = {10.1103/PhysRevC.104.044303},
}

@Article{Bortignon1981a,
  author       = {Bortignon, P. F. and Broglia, R. A.},
  title        = {{Role of the nuclear surface in a unified description of the damping of single-particle states and giant resonances}},
  journal      = {Nuclear Physics},
  year         = {1981},
  volume       = {A371},
  pages        = {405-429},
  doi          = {10.1016/0375-9474(81)90055-5},
  slaccitation = {%%CITATION = NUPHA,A371,405;%%},
}

@Article{MahauxBortignonBrogliaEtAl1985,
title = {Dynamics of the shell model},
journal = {Physics Reports},
volume = {120},
number = {1},
pages = {1-274},
year = {1985},
issn = {0370-1573},
doi = {https://doi.org/10.1016/0370-1573(85)90100-0},
url = {https://www.sciencedirect.com/science/article/pii/0370157385901000},
author = {C. Mahaux and P.F Bortignon and R.A Broglia and C.H Dasso}
}

@Article{LitvinovaRing2006,
  title = {Covariant theory of particle-vibrational coupling and its effect on the single-particle spectrum},
  author = {Litvinova, E. and Ring, P.},
  journal = {Phys. Rev. C},
  volume = {73},
  issue = {4},
  pages = {044328},
  numpages = {11},
  year = {2006},
  month = {Apr},
  publisher = {American Physical Society},
  doi = {10.1103/PhysRevC.73.044328},
  url = {https://link.aps.org/doi/10.1103/PhysRevC.73.044328}
}

@Article{Vaquero2020,
  title = {Fragmentation of Single-Particle Strength around the Doubly Magic Nucleus  $^{132}\mathrm{Sn}$ and the Position of the $0{f}_{5/2}$ Proton-Hole State in $^{131}\mathrm{In}$},
  author = {Vaquero, V. and Jungclaus, A. and Aumann, T. and Tscheuschner, J. and Litvinova, E. V. and Tostevin, J. A. and Baba, H. and Ahn, D. S. and Avigo, R. and Boretzky, K. and Bracco, A. and Caesar, C. and Camera, F. and Chen, S. and Derya, V. and Doornenbal, P. and Endres, J. and Fukuda, N. and Garg, U. and Giaz, A. and Harakeh, M. N. and Heil, M. and Horvat, A. and Ieki, K. and Imai, N. and Inabe, N. and Kalantar-Nayestanaki, N. and Kobayashi, N. and Kondo, Y. and Koyama, S. and Kubo, T. and Martel, I. and Matsushita, M. and Million, B. and Motobayashi, T. and Nakamura, T. and Nakatsuka, N. and Nishimura, M. and Nishimura, S. and Ota, S. and Otsu, H. and Ozaki, T. and Petri, M. and Reifarth, R. and Rodr\'{\i}guez-S\'anchez, J. L. and Rossi, D. and Saito, A. T. and Sakurai, H. and Savran, D. and Scheit, H. and Schindler, F. and Schrock, P. and Semmler, D. and Shiga, Y. and Shikata, M. and Shimizu, Y. and Simon, H. and Steppenbeck, D. and Suzuki, H. and Sumikama, T. and Symochko, D. and Syndikus, I. and Takeda, H. and Takeuchi, S. and Taniuchi, R. and Togano, Y. and Tsubota, J. and Wang, H. and Wieland, O. and Yoneda, K. and Zenihiro, J. and Zilges, A.},
  journal = {Phys. Rev. Lett.},
  volume = {124},
  issue = {2},
  pages = {022501},
  numpages = {6},
  year = {2020},
  month = {Jan},
  publisher = {American Physical Society},
  doi = {10.1103/PhysRevLett.124.022501},
  url = {https://link.aps.org/doi/10.1103/PhysRevLett.124.022501}
}

@Article{VanderSluys1993,
  author       = {Van der Sluys, V. and Van Neck, D. and Waroquier, M. and Ryckebusch, J.},
  title        = {{Fragmentation of single-particle strength in spherical open-shell nuclei: Application to the spectral functions in 142 Nd}},
  journal      = {Nuclear Physics},
  year         = {1993},
  volume       = {A551},
  pages        = {210-240},
  doi          = {10.1016/0375-9474(93)90479-H},
  slaccitation = {%%CITATION = NUPHA,A551,210;%%},
}

@Article{Mishev2010,
  author  = {Mishev, S. and Voronov, V. V.},
  journal = {Physical Review C},
  title   = {An Extended Approximation for the Lowest-lying States in Odd-mass Nuclei},
  year    = {2010},
  pages   = {064312},
  volume  = {82},
  doi     = {10.1103/PhysRevC.82.064312},
}

@Article{Litvinova2012,
  title = {Quasiparticle-vibration coupling in a relativistic framework: Shell structure of $Z=120$ isotopes},
  author = {Litvinova, Elena},
  journal = {Phys. Rev. C},
  volume = {85},
  issue = {2},
  pages = {021303},
  numpages = {6},
  year = {2012},
  month = {Feb},
  publisher = {American Physical Society},
  doi = {10.1103/PhysRevC.85.021303},
  url = {https://link.aps.org/doi/10.1103/PhysRevC.85.021303}
}

@Article{Afanasjev2015,
  title = {Impact of collective vibrations on quasiparticle states of open-shell odd-mass nuclei and possible interference with the tensor force},
  author = {Afanasjev, A. V. and Litvinova, E.},
  journal = {Phys. Rev. C},
  volume = {92},
  issue = {4},
  pages = {044317},
  numpages = {7},
  year = {2015},
  month = {Oct},
  publisher = {American Physical Society},
  doi = {10.1103/PhysRevC.92.044317},
  url = {https://link.aps.org/doi/10.1103/PhysRevC.92.044317}
}

@Article{Malov1976,
  author  = {Malov, L. A. and Soloviev, V. G.},
  journal = {Nuclear Physics A},
  title   = {Fragmentation of single-particle states and neutron strength functions in deformed nuclei},
  year    = {1976},
  pages   = {87--107},
  volume  = {270},
  doi     = {10.1016/0375-9474(76)90129-9},
}

@Article{Zhang2022,
  author  = {Zhang, Yinu and Bjel\v{c}i\'c, Antonio and Nik\v{s}i\'c, Tamara and Litvinova, Elena and Ring, Peter and Schuck, Peter},
  journal = {Physical Review C},
  title   = {Many-body approach to superfluid nuclei in axial geometry},
  year    = {2022},
  number  = {4},
  pages   = {044326},
  volume  = {105},
}

@Inbook{Elbek1969,
author={Elbek, Bent
and Tj{\o}m, Per Olav},
editor={Baranger, Michel
and Vogt, Erich},
title={Single Nucleon Transfer in Deformed Nuclei},
bookTitle={Advances in Nuclear Physics},
year="1969",
publisher="Springer US",
address="Boston, MA",
pages="259--323",
isbn="978-1-4757-9018-4",
doi="10.1007/978-1-4757-9018-4_4",
url="https://doi.org/10.1007/978-1-4757-9018-4_4"
}

@book{RingSchuck,
    author = {Ring, P. and Schuck, P.},
    title = {The Nuclear Many-Body Problem},
    publisher = {Springer Berlin, Heidelberg},
    year = {2004}
}

@Inbook{Kunz1993,
author={Kunz, P. D.
and Rost, E.},
editor = {"Langanke, K.
and Maruhn, J. A.
and Koonin, S. E.},
title={The Distorted-Wave Born Approximation},
bookTitle={Computational Nuclear Physics 2: Nuclear Reactions},
year="1993",
publisher={Springer New York},
address={New York, NY},
pages="88--107",
isbn="978-1-4613-9335-1",
doi="10.1007/978-1-4613-9335-1_5",
url="https://doi.org/10.1007/978-1-4613-9335-1_5"
}

@book{Bertulani, 
author = {Bertulani, Carlos A.},
title = {Nuclear Physics in a Nutshell},
publisher = {Princeton University Press},
year = {2007},
isbn = {9780691125053},
URL = {https://press.princeton.edu/books/hardcover/9780691125053/nuclear-physics-in-a-nutshell?srsltid=AfmBOoon7rppY88PEefqs1whIq6i9DpFbi4s1fuoCUS41f212i_fSy6G}
}

@article{Sch13a,
  title = {Valence nucleon populations in the Ni isotopes},
  author = {Schiffer, J. P. and Hoffman, C. R. and Kay, B. P. and Clark, J. A. and Deibel, C. M. and Freeman, S. J. and Honma, M. and Howard, A. M. and Mitchell, A. J. and Otsuka, T. and Parker, P. D. and Sharp, D. K. and Thomas, J. S.},
  journal = {Phys. Rev. C},
  volume = {87},
  issue = {3},
  pages = {034306},
  numpages = {15},
  year = {2013},
  month = {Mar},
  publisher = {American Physical Society},
  doi = {10.1103/PhysRevC.87.034306},
  url = {https://link.aps.org/doi/10.1103/PhysRevC.87.034306}
}

@article{Fre17a,
  title = {Experimental study of the rearrangements of valence protons and neutrons amongst single-particle orbits during double-$\ensuremath{\beta}$ decay in $^{100}\mathrm{Mo}$},
  author = {Freeman, S. J. and Sharp, D. K. and McAllister, S. A. and Kay, B. P. and Deibel, C. M. and Faestermann, T. and Hertenberger, R. and Mitchell, A. J. and Schiffer, J. P. and Szwec, S. V. and Thomas, J. S. and Wirth, H.-F.},
  journal = {Phys. Rev. C},
  volume = {96},
  issue = {5},
  pages = {054325},
  numpages = {15},
  year = {2017},
  month = {Nov},
  publisher = {American Physical Society},
  doi = {10.1103/PhysRevC.96.054325},
  url = {https://link.aps.org/doi/10.1103/PhysRevC.96.054325}
}

\end{document}